\documentclass[aps, prd, twocolumn, showpacs, showkeys, preprintnumbers, bibnotes, floatfix, longbibliography, notitlepage, nofootinbib, superscriptaddress, superscriptgaddress]{revtex4-2}

\pdfoutput=1
\usepackage[colorlinks=true,breaklinks=true]{hyperref}
\usepackage{amsmath}
\usepackage{amsfonts}
\usepackage{amssymb}
\usepackage{mathrsfs}
\usepackage{graphicx}
\usepackage{color}
\usepackage[utf8]{inputenc}
\usepackage{amsfonts}
\usepackage{graphicx}
\usepackage{booktabs}
\usepackage{multirow}
\usepackage{makecell}
\usepackage{siunitx}
\usepackage{ragged2e}
\usepackage{rotating}
\usepackage{float}
\usepackage[dvipsnames]{xcolor}
\definecolor{darkred}{rgb}{0.5,0,0}
\definecolor{darkblue}{rgb}{0,0,0.5}
\definecolor{firebrick}{rgb}{0.75,0.125,0.125}
\definecolor{darkgreen}{rgb}{0,0.5,0}
\hypersetup{urlcolor=darkblue,
	    citecolor=darkgreen,
	    linkcolor=firebrick}

\usepackage{orcidlink}
\usepackage{lipsum}
\usepackage[normalem]{ulem}
\usepackage{enumitem}
\usepackage[capitalise]{cleveref}
\usepackage{pifont}
\usepackage{subfigure}
\usepackage{physics}
\usepackage{comment}
\usepackage{mwe}

\long\def\exclude#1{}

\newcommand{\ie}{{\it i.e.}}

\newcommand{\eg}{{\it e.g.}}

\newcommand{\cf}{{\it cf.}}

\newcommand{\eq}{Eq.}

\newcommand{\fig}{Fig.}

\newcommand{\Refe}{Ref.}
\newcommand{\Refes}{Refs.}
\newcommand{\equ}[1]{\eq~(\ref{equ:#1})}
\newcommand{\figu}[1]{\fig~\ref{fig:#1}}

\newcommand{\ship}{\textsc{SHiP}} 

\DeclareMathOperator{\arctantwo}{arctan2}

\graphicspath{{figures/}}

\begin{document}

\title{Astrophysical bounds on the high-energy evolution of neutrino mixing}

\author{Mauricio Bustamante}
\email{mbustamante@nbi.ku.dk}
\affiliation{Niels Bohr International Academy, Niels Bohr Institute,\\University of Copenhagen, 2100 Copenhagen, Denmark}

\author{Qinrui Liu}
\email{qinrui\_liu@sfu.ca}
\affiliation{Department of Physics, Simon Fraser University, Burnaby, BC V5A 1S6, Canada}
\affiliation{Arthur B. McDonald Canadian Astroparticle Physics Research Institute, Kingston ON K7L 3N6, Canada}

\author{Gabriela Barenboim}
\email{gabriela.barenboim@uv.es}
\affiliation{Departament de F\'isica Te\'orica and IFIC, Universitat de Val\`encia-CSIC, E-46100, Burjassot, Spain}

\date{April 15, 2026}

\begin{abstract}

While conventional oscillation experiments measure neutrino mixing parameters with high precision, these measurements are strictly confined to sub-TeV scales. At higher energies, renormalization-group effects can cause these parameters to evolve with the transferred momentum, $Q$. High-energy and ultra-high-energy astrophysical neutrinos, spanning TeV to EeV energies, probe high values of $Q$ unreachable by conventional experiments, offering an unprecedented test of high-energy mixing. We use the flavor composition of these neutrinos---the relative proportions of $\nu_e$, $\nu_\mu$, and $\nu_\tau$---to constrain this evolution, both phenomenologically and within dimension-6 Standard Model Effective Field Theory. We account for astrophysical uncertainties---an unavoidable requirement to obtain realistic results, even though this weakens the bounds. Although present IceCube measurements lack the sensitivity to detect this running, we forecast that upcoming multi-detector combinations will place unprecedented bounds on the high-energy evolution of neutrino mixing.

\end{abstract}

\maketitle


\section{Introduction}
\label{sec:introduction}

Neutrino oscillations---the periodic transformation of one neutrino flavor into another during propagation---are measured with high precision. Combined observations of solar, atmospheric, reactor, and accelerator neutrinos have established that mixing among the three active flavors---$\nu_e$, $\nu_\mu$, and $\nu_\tau$---is described by the Pontecorvo--Maki--Nakagawa--Sakata (PMNS) matrix~\cite{Pontecorvo:1957cp, Maki:1962mu}, parametrized by three mixing angles $(\theta_{12}, \theta_{23}, \theta_{13})$ and a CP-violation phase ($\delta_{\rm CP}$).  Today, the mixing angles are known to a precision of 1--3\% [at 68\% confidence level (C.L.)] and the CP-violation phase, to about 16\%~\cite{deSalas:2020pgw, Capozzi:2021fjo, Esteban:2024eli}.  

Yet, the values of the mixing parameters are inferred exclusively from sub-TeV neutrino experiments (see, however, \Refe~\cite{Bustamante:2026aur}).  Thus, a fundamental question remains: \textbf{\textit{are the mixing parameters universal constants, or are they different at low and high neutrino energies?}}  The expectation from quantum field theory is that, indeed, the parameters evolve---or \textit{run}---with the transferred momentum, $Q$, at which neutrino interactions occur, the value of which rises with energy (as is the case for dimensionless couplings). However, no evolution has been observed to date.

Significant evolution of the mixing parameters would modify the $\nu_\alpha \to \nu_\beta$ oscillation probabilities ($\alpha, \beta = e, \mu, \tau$) relative to their standard, no-running expectation, leading to potentially detectable effects in neutrino experiments capable of distinguishing flavors.  While the running---and its effect on the probabilities---is expected to grow more prominent with $Q$, accessing these values requires higher-energy neutrinos, which are scarcer.

From theory, the predicted size and form of the running of the mixing parameters are model-dependent.  In the Standard Model (SM) extended with massive Majorana neutrinos, the mass matrix responsible for granting neutrinos mass is generated by the  dimension-5 Weinberg operator~\cite{Weinberg:1979sa}.  Its Wilson coefficients---and hence the physical mixing parameters---evolve with $Q$ through renormalization group equations (RGEs)~\cite{Babu:1993qv, Peskin:1995ev}.  Within the SM, however, the running is negligible, since it is suppressed by the small tau Yukawa coupling, $y_\tau \approx 0.01$.  But in other well-motivated extensions, such as the Minimal Supersymmetric Standard Model (MSSM)~\cite{Dimopoulos:1981zb, Haber:1984rc, Antusch:2003kp, Antusch:2005gp} or, more generally, the SM Effective Field Theory (SMEFT) supplemented with higher-dimensional operators~\cite{Buchmuller:1985jz, Brivio:2017vri, Jenkins:2017jig, Jenkins:2017dyc}, the running can be significantly enhanced, potentially shifting the mixing parameters by observationally relevant amounts between low- and high-energy scales.  \textbf{\textit{Thus, detecting appreciable RG running of the neutrino mixing parameters would constitute unmistakable evidence of new physics.}}

From experiment, sub-TeV neutrino data have probed primarily momenta $Q \lesssim$~15~GeV.  Reference~\cite{Babu:2021cxe} (see also \Refes~\cite{Babu:2022non, Ge:2023azz, Ge:2024ibn}) showed that, in this regime, extending the SM with a new, light secluded sector could induce observable effects in present (T2K, NO$\nu$A) and future long-baseline GeV-scale neutrino experiments, including new sources of CP violation, zero-baseline flavor transitions, and apparent CPT violation.  However, if the characteristic energy scale of the RG-inducing new physics is instead heavy---\ie, above the TeV scale---the RG running of the mixing parameters would be undetectable in conventional oscillation experiments.  

In such case, higher neutrino energies would be needed to access higher momenta.  Above the TeV scale, neutrinos are detected via their deep inelastic scattering (DIS) off nucleons~\cite{CTEQ:1993hwr, Conrad:1997ne, Formaggio:2012cpf}.  In DIS, a neutrino of energy $E_\nu$ can transfer, as a maximum, a momentum of $\sqrt{2 E_\nu m_N}$, where $m_N \approx 1$~GeV is the nucleon mass.  For instance, atmospheric neutrinos with energies of up to $E_\nu \sim$~100~TeV may probe momenta as high as  450~GeV.  In reality, however, the \textit{average} value of $Q$ in the DIS of 100-TeV neutrinos is significantly smaller---around 7~GeV---because the parton distribution functions (PDFs) of the nucleon favor lower values of $Q$ (more precisely, lower values of the Bjorken-$x$ parameter).

To overcome this, we turn to even higher-energy neutrinos of astrophysical origin---the most energetic ones known, with TeV--EeV energies.  They provide a unique window into the high-$Q$ regime---as proposed by \Refe~\cite{Bustamante:2010bf} and then revisited by \Refes~\cite{Babu:2021cxe, Mir:2025fae}---and into new physics in general~\cite{Ahlers:2018mkf, Ackermann:2019cxh, Arguelles:2019rbn, Ackermann:2022rqc, Arguelles:2022tki, MammenAbraham:2022xoc}.  High-energy astrophysical neutrinos with energies in the TeV--10~PeV range are detected regularly by neutrino telescopes IceCube~\cite{IceCube:2013low}, KM3NeT~\cite{KM3Net:2016zxf}, and Baikal-GVD~\cite{Baikal-GVD:2025rhg}, soon to be joined by more detectors~\cite{Ackermann:2022rqc, MammenAbraham:2022xoc, Guepin:2022qpl}.  Ultra-high-energy (UHE) neutrinos above 100~PeV will be targeted by next-generation dedicated telescopes~\cite{Ackermann:2022rqc, MammenAbraham:2022xoc, Guepin:2022qpl}, while the first UHE neutrino has already been observed by KM3NeT~\cite{KM3NeT:2025npi}. 

The most energetic among these neutrinos may access momenta of tens of TeV.  Yet, as for atmospheric neutrinos, the PDFs pull the average $Q$ down. This, combined with the increasing scarcity of neutrinos of growing energy, places the average $Q$ accessible by high- and ultra-high-energy astrophysical neutrinos at 20--40~GeV (\figu{main_results}).  This range represents roughly an order-of-magnitude improvement in the average $Q$ available to probe the high-energy evolution of neutrino mixing.

\section{Synopsis}

\textbf{\textit{We probe the RG running of neutrino mixing by searching for its imprint on the \emph{flavor composition} of high-energy astrophysical neutrinos}}---the relative fractions of $\nu_e$, $\nu_\mu$, and $\nu_\tau$ arriving at Earth~\cite{Rachen:1998fd, Kashti:2005qa, Bustamante:2015waa, Song:2020nfh}---which reflects RG-induced changes in the flavor-transition probabilities.  The flavor composition is a versatile probe of astrophysics~\cite{Rachen:1998fd, Athar:2000yw, Crocker:2001zs, Barenboim:2003jm, Beacom:2003nh, Beacom:2004jb, Kashti:2005qa, Mena:2006eq, Kachelriess:2006ksy, Lipari:2007su, Esmaili:2009dz, Choubey:2009jq, Hummer:2010ai, Winter:2013cla, Palladino:2015zua, Bustamante:2015waa, Biehl:2016psj, Bustamante:2019sdb, Ackermann:2019ows, Bustamante:2020bxp, Song:2020nfh, AlvesBatista:2021eeu, Liu:2023lxz, Bhattacharya:2023mmp, Telalovic:2023tcb, Dev:2023znd} and fundamental physics~\cite{Beacom:2002vi, Barenboim:2003jm, Beacom:2003nh, Beacom:2003eu, Beacom:2003zg, Serpico:2005bs, Mena:2006eq, Lipari:2007su, Pakvasa:2007dc, Esmaili:2009dz, Choubey:2009jq, Esmaili:2009fk, Bhattacharya:2009tx, Bhattacharya:2010xj, Bustamante:2010nq, Mehta:2011qb, Baerwald:2012kc, Fu:2012zr, Pakvasa:2012db, Chatterjee:2013tza, Xu:2014via, Aeikens:2014yga, Arguelles:2015dca, Bustamante:2015waa, Pagliaroli:2015rca, Shoemaker:2015qul, deSalas:2016svi, Gonzalez-Garcia:2016gpq, Bustamante:2016ciw, Rasmussen:2017ert, Dey:2017ede, Bustamante:2018mzu, Farzan:2018pnk, Ahlers:2018yom, Brdar:2018tce, Palladino:2019pid, Ackermann:2019cxh, Arguelles:2019rbn, Ahlers:2020miq, Karmakar:2020yzn, Fiorillo:2020gsb, Song:2020nfh, AlvesBatista:2021eeu, Arguelles:2022tki, MammenAbraham:2022xoc, Telalovic:2023tcb, Liu:2024wmk, Telalovic:2025xor}.  While short terrestrial baselines restrict conventional oscillation experiments primarily to two-flavor transitions, astrophysical neutrinos undergo full three-flavor mixing over cosmological distances, enabling tests of this paradigm~\cite{Beacom:2003zg, Bhattacharjee:2005nh, Serpico:2005sz, Serpico:2005bs, Balaji:2006wi, Xing:2006xd, Meloni:2006gv, Winter:2006ce, Rodejohann:2006qq, Blum:2007ie, Hwang:2007na, Pakvasa:2007dc, Choubey:2008di, Maltoni:2008jr, Xing:2008fg, Esmaili:2009dz, Meloni:2012nk, Lai:2013isa, Chatterjee:2013tza, Bustamante:2026aur}.

\begin{figure}[t!]
 \centering
 \includegraphics[width=\columnwidth]{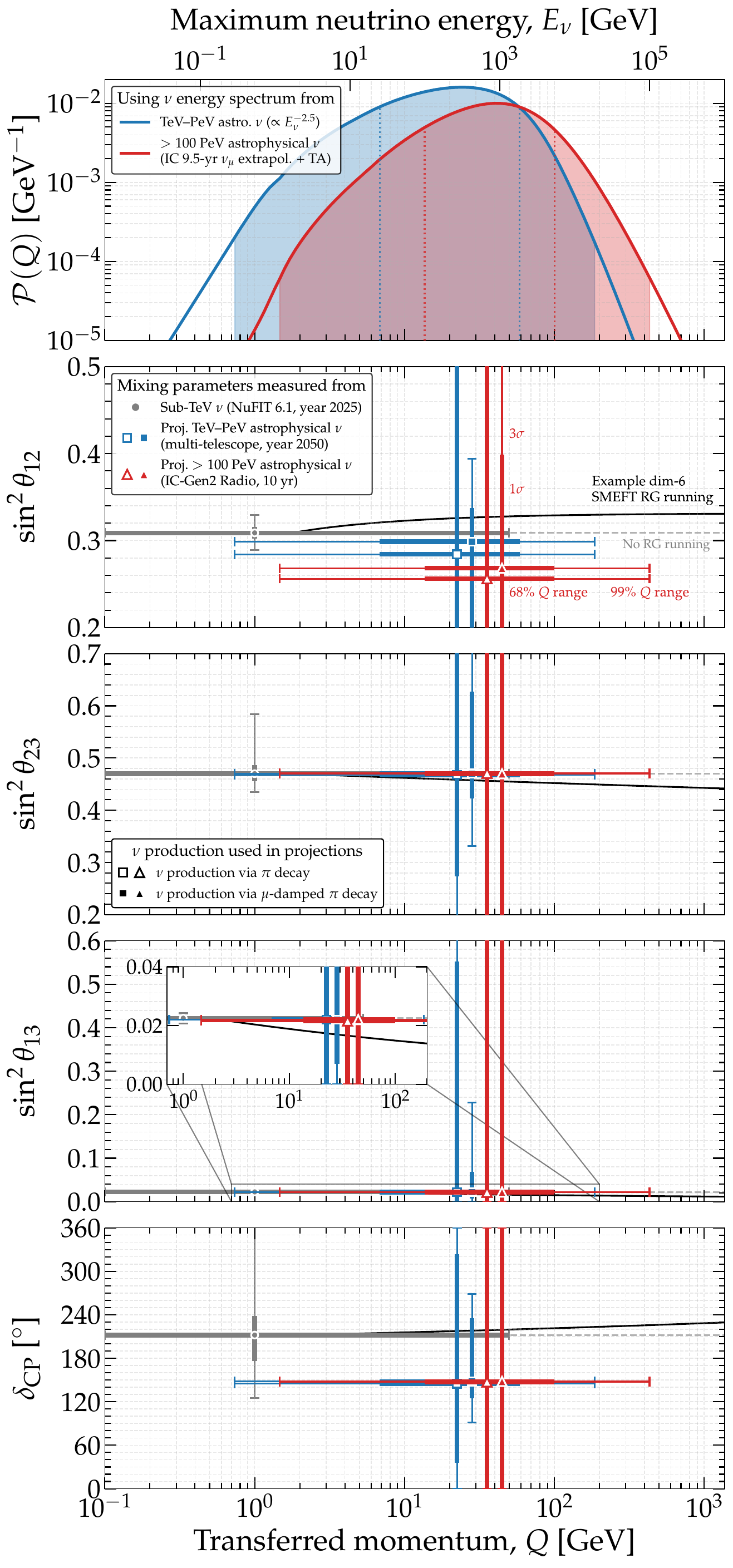}
 \vspace*{-0.55cm}
 \caption{\textbf{Evolution of neutrino mixing parameters with transferred momentum, $Q$.}  \textit{Top panel:} Probability distribution of $Q$ in the deep inelastic scattering of high-energy (TeV--PeV) and ultra-high-energy (UHE, $> 100$~PeV) astrophysical neutrinos on nucleons.  The maximum neutrino energy is $E_\nu = Q^2/(2 m_N)$, where $m_N \approx 1$~GeV is the nucleon mass. \textit{Lower panels:} Variation of the mixing parameters with $Q$. Present measurements reflect the NuFIT 6.1 global oscillation fit~\cite{Esteban:2024eli}. Projections are for TeV--PeV multi-detector observations and UHE detection in the radio array of IceCube-Gen2, assuming astrophysical neutrino production via full or muon-damped pion decay. Evolution under a specific SMEFT RG running scheme is shown for illustration (Sec.~\ref{sec:rg_running-smeft}).}
 \label{fig:main_results}
\end{figure}

We execute a two-pronged analysis: first, treating the high-$Q$ mixing parameters $(\theta_{12}^\prime, \theta_{23}^\prime, \theta_{13}^\prime, \delta_{\rm CP}^\prime)$ as free phenomenological quantities to remain agnostic to specific RG scenarios; and second, constraining RG-inducing dimension-6 SMEFT coefficients. 

Figure~\ref{fig:main_results} summarizes our main results. We adopt present IceCube TeV--PeV flavor measurements~\cite{Abbasi:2025fjc} and project the sensitivity of multi-detector combinations. Present data offer no sensitivity, but \textbf{\textit{future detectors will constrain $\theta_{23}^\prime$ and $\theta_{13}^\prime$ by 2040.}}  While our projections yield limited precision compared to conventional oscillation experiments---owing primarily to the unknown flavor composition with which neutrinos are produced at their sources---they are meaningful. Extending observations to the UHE regime yields weaker constraints due to larger flavor uncertainties.

The rest of this paper is organized as follows. Section~\ref{sec:flavor_comp} introduces the formalism of standard and RG-modified flavor transitions, the production of high-energy astrophysical neutrinos, and our predictions for their flavor composition at Earth. Section~\ref{sec:rg_running} details the RG running of neutrino mixing parameters, with a focus on the SMEFT and the momentum distributions accessible to neutrino telescopes. Section~\ref{sec:statistics} outlines our statistical procedure used to constrain both the generic high-$Q$ mixing parameters and the specific SMEFT coefficients. Section~\ref{sec:results} presents our main results, including present limits, future TeV--PeV multi-detector combinations, and UHE radio array projections. Finally, Section~\ref{sec:summary} summarizes our findings and outlook. Appendices~\ref{app:rge_approximations}--\ref{app:smeft_regions_single_parameter} contain additional derivations and detailed results. 


\section{Flavor composition in high-energy astrophysical neutrinos}
\label{sec:flavor_comp}

\textit{We establish the theoretical and experimental framework for predicting and measuring the flavor composition of high-energy astrophysical neutrinos. We detail how the standard PMNS matrix maps the initial flavor composition at the source---canonically produced via full or muon-damped pion decay---to the observable flavor composition at Earth. We introduce the phenomenological framework for modifying these flavor-transition probabilities to account for potential high-$Q$ renormalization-group running at the detection scale. By juxtaposing these theoretical predictions against the detection capabilities of current and next-generation neutrino telescopes, we demonstrate the premise of this work: while standard three-flavor mixing confines the expected flavor composition at Earth to a remarkably narrow and rigid band, high-energy modifications to the mixing parameters can break this restriction, driving the flavor composition into expansive new regions accessible to future experiments.}


\subsection{Standard flavor transitions}
\label{sec:flavor_comp-osc_std}

A neutrino of a given flavor, $\nu_\alpha$, is a superposition of mass eigenstates $\nu_i$, \ie, 
\begin{equation}
 \nu_\alpha = \sum_{i=1}^3 U_{\alpha i}^\ast \nu_i \;,
\end{equation}
where $U_{\alpha i}$ are elements of the PMNS matrix.  In its standard parametrization~\cite{ParticleDataGroup:2024cfk}, the matrix depends on three mixing angles ($\theta_{12}$, $\theta_{23}$, $\theta_{13}$), and one CP-violation phase ($\delta_{\rm CP}$), whose values are determined experimentally.  

Neutrinos \textit{oscillate}: while propagating, the probability of detecting them as having a certain flavor, including one different from their original one, varies periodically as a function of energy and distance traveled.  The oscillation length of neutrinos of energy $E_\nu$ is $4\pi E_\nu / \Delta m^2$, where $\Delta m^2 \approx $~10$^{-4}$--10$^{-3}$~eV$^2$ is the squared-mass difference between the mass eigenstates.  For high-energy astrophysical neutrinos with $E_\nu \gtrsim 1$~TeV, the oscillation length is tiny compared to their cosmological-scale baselines of hundreds to Mpc to a few Gpc to Earth.  Further, over these vast distances, the different mass-eigenstate wave packets physically separate due to their different group velocities, leading to a complete loss of coherence.  This, combined with the spread in baselines due to the distribution of astrophysical sources and the limited energy resolution of neutrino telescopes, precludes sensitivity to the rapid oscillation in the flavor-transition probabilities.

Instead, we are sensitive to the average probabilities.  The standard probability of a neutrino produced with flavor $\alpha$ being detected with flavor $\beta$ is
\begin{equation}
    P_{\alpha\beta}^{\mathrm{std}} = \sum_{i=1}^{3} |U_{\alpha i}|^2 |U_{\beta i}|^2 \;.
    \label{equ:prob_standard}
\end{equation}
This standard formulation implicitly assumes that the mixing matrix at the source is identical to the mixing matrix at the detector.  The values of the mixing parameters are determined by global fits to data from oscillation experiments with sub-TeV neutrino energies~\cite{deSalas:2020pgw, Capozzi:2021fjo, Esteban:2024eli}.


\subsection{Flavor transitions with modified high-$Q$ mixing parameters}
\label{sec:flavor_comp-osc_mod}

Because neutrino production occurs primarily via pion decay (Sec.~\ref{sec:producing-high-energy-neutrinos}), the momentum scale accessible at production is governed by the mass of the pion (\ie, $Q = m_\pi \approx 140$~MeV). Hence, the mixing parameters at production are well-approximated---as in Sec.~\ref{sec:flavor_comp-osc_std}---by the standard mixing parameters measured in conventional sub-TeV oscillation experiments. 

However, high-energy astrophysical neutrinos detected at Earth typically interact via deep inelastic scattering on nucleons ($\nu N$ DIS), where the \textit{maximum kinematically allowed} transferred momentum can be significantly larger than in conventional oscillation experiments, \ie, $Q \approx $~TeV--PeV vs.~GeV. (Although, as we explain below the \textit{most likely} momenta for high-energy astrophysical neutrinos are in the 10--100~GeV range.) Thus, if the neutrino mixing parameters are subject to RG running, the mixing parameters at detection may show signs of this and significantly deviate from those at production. 

To account for this effect, we introduce a distinct, modified detection-scale matrix $U^\prime(Q)$ parametrized by high-$Q$ mixing angles ($\theta_{12}^\prime, \theta_{23}^\prime, \theta_{13}^\prime$) and a CP-violation phase ($\delta_{\mathrm{CP}}^\prime$). The mixing matrix at production, $U$, remains the PMNS matrix. Thus, the flavor-transition probability is modified to project the propagated mass eigenstates onto the altered high-$Q$ flavor basis at the detector, \ie,
\begin{equation}
 P_{\alpha\beta} = \sum_{i=1}^{3} |U_{\alpha i}|^2 |U^\prime_{\beta i}|^2 \;.
 \label{equ:prob_modified}
\end{equation}
Because $U$ and $U^\prime$ are unitary, the sum of probabilities equals unity, \ie, $\sum_\alpha P_{\alpha\beta} = 1$.  The goal of our work is to extract the values of the altered mixing parameters from the detection of high-energy astrophysical neutrinos.

The magnitude and direction of the changes to the mixing parameters depend on the value of $Q$ and on the specific RG running scenario adopted, of which there are multiple possibilities.  Figure~\ref{fig:main_results} shows one hand-picked example that introduces RG running via a generic dimension-6 operator in the SMEFT formalism, which we expand upon later, in Sec.~\ref{sec:rg_running-smeft}.

An exhaustive exploration of the different viable RG running schemes that could affect high-energy astrophysical neutrinos is not the goal of our work.  Instead, our goal is to assess the detectability of high-$Q$ modifications to neutrino mixing that apply to all possible RG running schemes, establishing realistic benchmarks for which schemes are experimentally testable.  Later, we report on the minimum detectable size of these modifications.  Our sweeping approach is motivated by the limited precision we find available in high-energy astrophysical neutrinos to test RG running, which suits model-independent searches better than model-dependent ones. 

Thus, in the first part of our analysis we make the simplifying assumption that the mixing matrix at detection, $U^\prime$, while potentially different from $U$, is independent of $Q$.  In other words, we extract constant values of $\theta_{12}^\prime$, $\theta_{23}^\prime$, $\theta_{13}^\prime$, and $\delta_{\rm CP}^\prime$, effectively representing their $Q$-averaged values [formally defined later in \equ{flavor_ratios_q_avg}], without ascribing any specific RG running to them.  In practice, we compute $P_{\alpha\beta}$ in \equ{prob_modified} by varying the modified mixing parameters away from their standard low-energy values. The standard parameters in $U$ are fixed to their currently allowed values from the NuFIT 6.1 global analysis~\cite{Esteban:2024eli}, accounting for their experimental uncertainty.

\begin{figure*}[t]
    \centering
    \includegraphics[width=\textwidth]{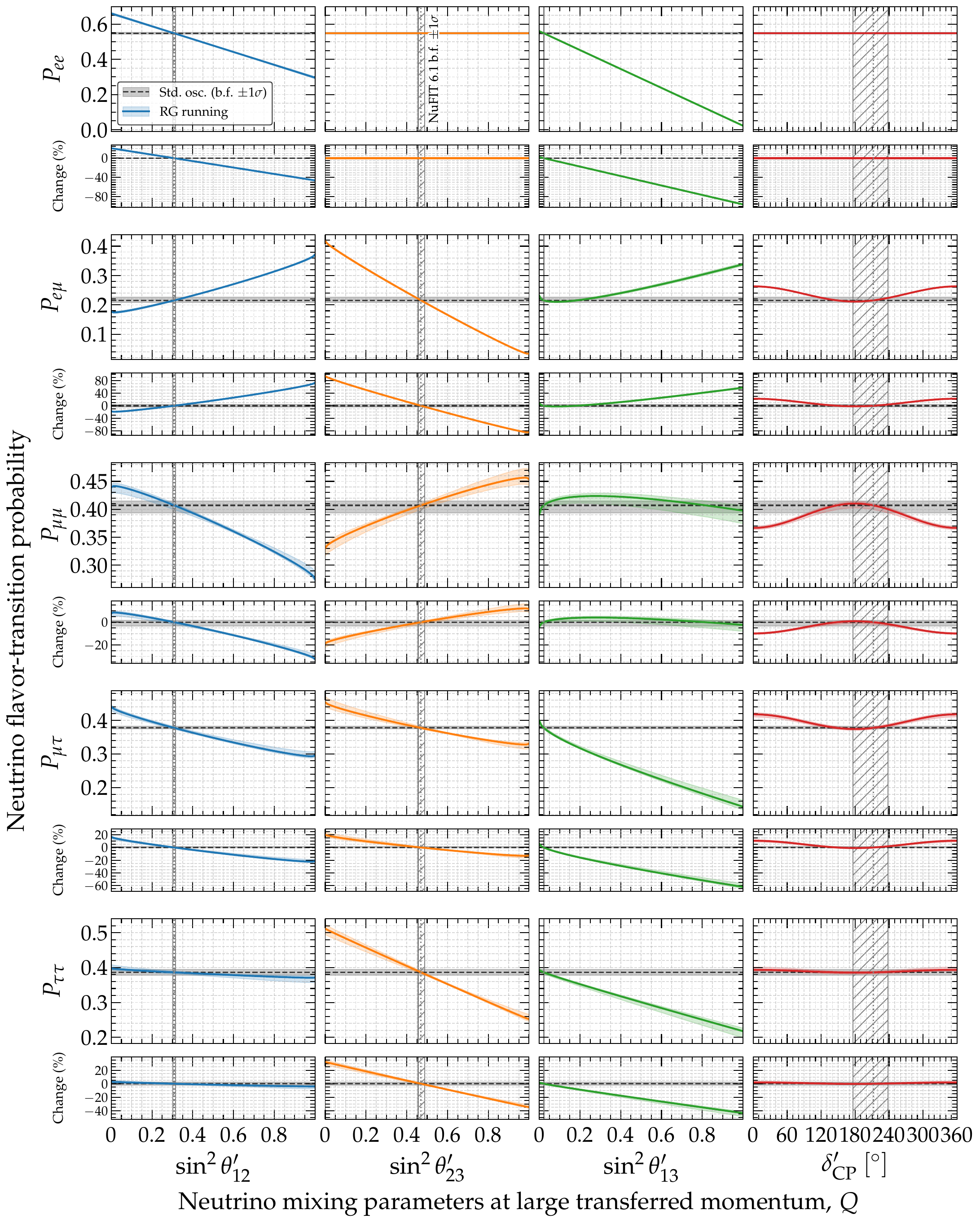}
    \caption{\textbf{Neutrino flavor-transition probabilities as a function of the modified high-$Q$ mixing parameters.}  Rows display $P_{ee}$, $P_{e\mu}$, $P_{\mu\mu}$, $P_{\mu\tau}$, and $P_{\tau\tau}$ computed via \equ{prob_modified}. Columns show independent variations of the high-energy parameters $\sin^2\theta_{12}^\prime$, $\sin^2\theta_{23}^\prime$, $\sin^2\theta_{13}^\prime$, and $\delta_{\mathrm{CP}}^\prime$, with the remaining parameters fixed to their standard low-energy values. We compare these RG-modified probabilities against the standard case [\equ{prob_standard}]. Probability bands denote $1\sigma$ uncertainties. Vertical bands indicate NuFIT 6.1 best-fit (``b.f.'') and $1\sigma$ intervals for the varied parameter, assuming normal mass ordering and including Super-Kamiokande data. Sub-panels quantify the relative change, $(P_{\alpha\beta} - P_{\alpha\beta}^{\mathrm{std}}) / P_{\alpha\beta}^{\mathrm{std}}$. See Sec.~\ref{sec:flavor_comp-osc_mod} for details.}
    \label{fig:probabilities}
\end{figure*}

Figure~\ref{fig:probabilities} illustrates the dependence of the flavor-transition probabilities on the modified mixing parameters.  We focus on the independent diagonal channels ($P_{ee}$, $P_{\mu\mu}$, $P_{\tau\tau}$) and the symmetric off-diagonal channels ($P_{e\mu}$, $P_{\mu\tau}$).  The figure shows that modifications of the probabilities of up to tens of percent are possible, with different flavor-transition channels exhibiting distinct sensitivity to specific high-$Q$ mixing angles.  

Broadly stated, large departures of $\theta_{12}^\prime$, $\theta_{23}^\prime$, and $\theta_{13}^\prime$ from their standard counterparts $\theta_{12}$, $\theta_{23}$, and $\theta_{13}$ can all impact the probabilities.  Large values of $\theta_{13}^\prime$ have an especially strong effect given that its standard counterpart is small, $\theta_{13} \approx 8^\circ$.  Later, we find that the effect of varying $\theta_{12}^\prime$ is unfortunately obscured by the uncertainty on the flavor composition with which neutrinos are produced, mitigating our sensitivity to this angle. The impact of $\delta_{\rm CP}^\prime$ remains as marginal as that of its standard counterpart, $\delta_{\rm CP}$.  These observations prefigure the predominant sensitivity to $\theta_{23}^\prime$ and $\theta_{13}^\prime$ that we find later. 


\subsection{Producing high-energy neutrinos}
\label{sec:producing-high-energy-neutrinos}

In standard astrophysical scenarios, high-energy neutrinos are expected to be produced when high-energy protons---accelerated within astrophysical environments such as active galactic nuclei, gamma-ray bursts, or starburst galaxies---interact with ambient matter ($pp$ interactions)~\cite{Margolis:1977wt, Stecker:1978ah, Kelner:2006tc} or radiation ($p\gamma$ interactions)~\cite{Stecker:1978ah, Mucke:1999yb, Kelner:2008ke, Hummer:2010vx}. These interactions produce charged and neutral pions. The neutral pions decay into gamma rays ($\pi^0 \to \gamma\gamma$), while the charged pions decay to yield neutrinos.

The primary neutrino production channel is the decay of charged pions and their daughter muons, \ie, $\pi^+ \to \mu^+ + \nu_\mu$ followed by $\mu^+ \to e^+ + \nu_e + \bar{\nu}_\mu$, and their charge-conjugated processes. At the astrophysical sources, before any oscillations occur, this full decay chain yields an initial flavor composition, defined as $(f_{e}, f_{\mu}, f_{\tau})_{\rm S}$, where $f_{\alpha, \rm S}$ is the proportion of $\nu_\alpha + \bar{\nu}_\alpha$ produced, of approximately $\left( \frac{1}{3}, \frac{2}{3}, 0 \right)_{\rm S}$. This is the nominal expectation for high-energy astrophysical neutrino production; we refer to it as the \textit{full pion-decay} scenario. 

However, the conditions at the source can significantly alter this initial composition. If the source harbors a strong magnetic field, the intermediate muons produced in pion decay may cool substantially via synchrotron radiation before they decay~\cite{Kashti:2005qa, Kachelriess:2007tr, Lipari:2007su}. In this case, the high-energy neutrino flux is dominated by the initial pion decay, yielding a flavor composition of $(0, 1, 0)_{\rm S}$. This \textit{muon-damped} scenario is expected to become dominant at the highest energies (see, \eg, \Refes~\cite{Winter:2013cla, Bustamante:2020bxp}), where the muon lifetime is sufficiently time-dilated to ensure severe energy losses prior to decay.  

(A third, less common standard scenario arises from the beta-decay of free neutrons, which can be produced either from the photo-dissociation of accelerated heavy nuclei or from the $p\gamma$ interactions themselves~\cite{Anchordoqui:2003vc}. The decay $n \to p + e^- + \bar{\nu}_e$ yields a pure electron anti-neutrino flux, resulting in a flavor composition of $(1, 0, 0)_{\rm S}$. See Appendix~\ref{app:analytical_approximation_smeft-flavor_distance_beta_decay} for why this scenario is less likely and why we do not consider it in our main results.)

Because of the high energy threshold required for tau-lepton production, standard astrophysical sources are expected to produce a negligible fraction of $\nu_\tau$~\cite{Farzan:2021gbx}. Thus, across all standard production mechanisms, $f_{\tau, \rm S} \approx 0$. While exotic scenarios---\eg, decay of heavy dark matter~\cite{Feldstein:2013kka, Esmaili:2013gha, Bhattacharya:2014vwa}, interactions involving leptoquarks~\cite{Barger:2013pla}, or sterile neutrinos~\cite{Arguelles:2019tum, Ahlers:2020miq}---can generate appreciable $f_{\tau, \rm S}$, they lie outside the standard astrophysical paradigm. Consequently, the initial flavor composition of high-energy astrophysical neutrinos is generally bounded to mixtures of $\nu_e$ and $\nu_\mu$, parametrized as $(f_{e, {\rm S}}, 1 - f_{e, {\rm S}}, 0)_{\rm S}$.  Neutrino interactions with matter inside the sources are unlikely to modify these flavor ratios before exiting them~\cite{Mena:2006eq, Razzaque:2009kq, Sahu:2010ap, Varela:2014mma, Xiao:2015gea} (see, however, \Refe~\cite{Dev:2023znd}).


\subsection{Predicting the flavor composition at Earth}
\label{sec:flavor_ratios-theory}

Using the flavor-transition probabilities $P_{\alpha\beta}$ in \equ{prob_modified}, the expected flavor ratios at Earth are a combination of the flavor ratios at the sources, $f_{\alpha, {\rm S}}$, \ie,
\begin{equation}
 f_{\beta, \oplus}
 (\boldsymbol{\theta}, \boldsymbol{\theta}^\prime)
 =
 \sum_{\alpha \in \{e, \mu, \tau\}} 
 P_{\alpha\beta} 
 (\boldsymbol{\theta}, \boldsymbol{\theta}^\prime)
 f_{\alpha, {\rm S}} \;.
 \label{equ:flavor_ratios}
\end{equation}
Here, the standard mixing parameters, $\boldsymbol{\theta} \equiv \{ \theta_{12}, \theta_{23}, \theta_{13}, \delta_{\rm CP} \}$, determine the PMNS matrix $U$, and the modified mixing parameters, $\boldsymbol{\theta}^\prime \equiv \{ \theta_{12}^\prime, \theta_{23}^\prime, \theta_{13}^\prime, \delta_{\rm CP}^\prime \}$, determine the high-$Q$ mixing matrix $U^\prime$.  The nominal expectation from neutrino production via full pion decay and standard mixing [\ie, $U^\prime = U$, with mixing parameters given by their best-fit values (Tables~\ref{tab:results_mix_params_tev_pev}, \ref{tab:results_mix_params_uhe})] is about $\left( \frac{1}{3}, \frac{1}{3}, \frac{1}{3} \right)_\oplus$.

\begin{figure*}[htpb]
    \centering
    \includegraphics[width=\textwidth]{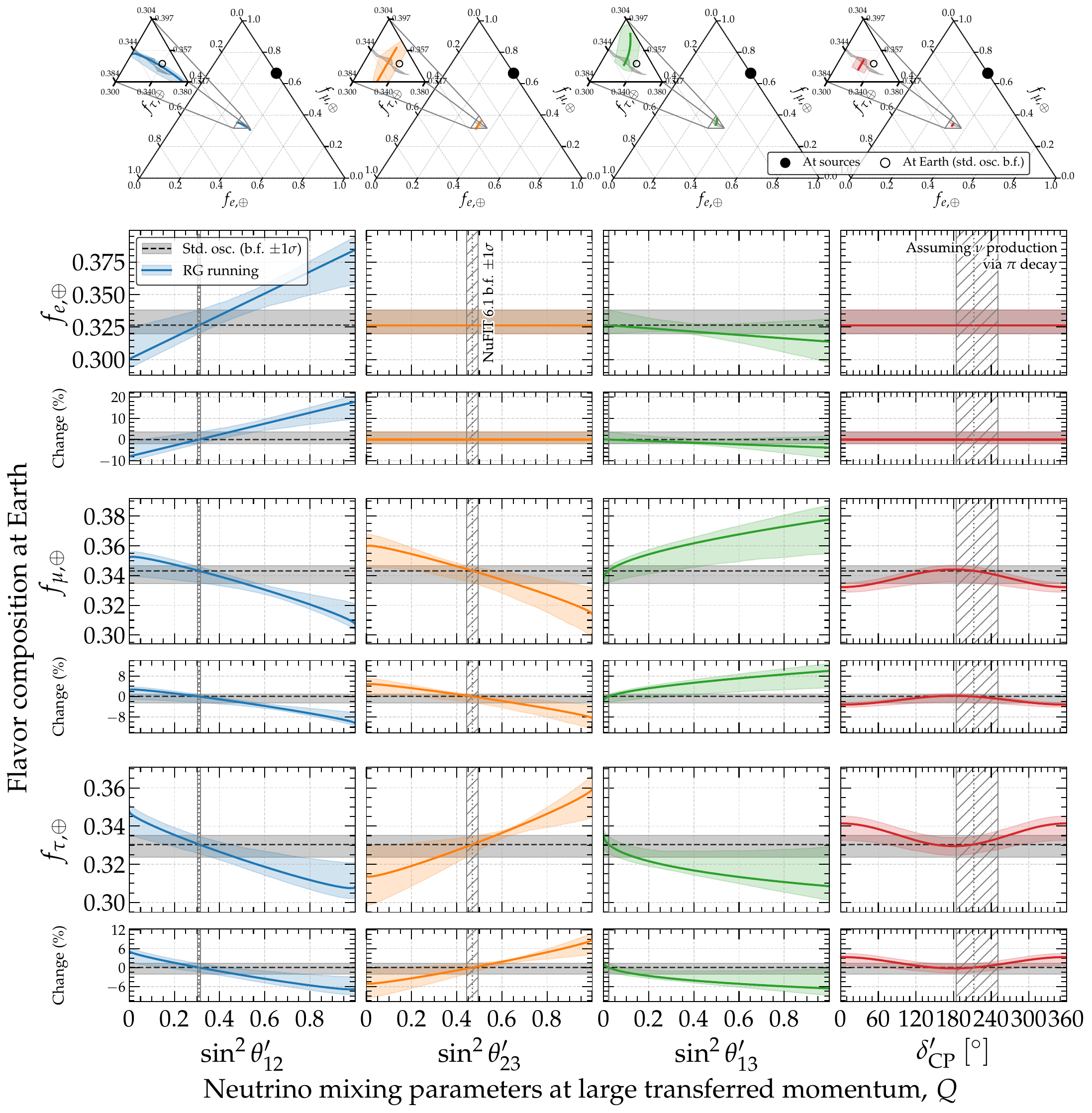}
    \caption{\textbf{Flavor composition at of high-energy astrophysical neutrinos at Earth as a function of the modified high-$Q$ mixing parameters.}  Neutrino production is assumed to occur via full pion decay, $\ie$, the flavor composition at the sources is $\left( \frac{1}{3}, \frac{2}{3}, 0\right )_{\rm S}$. The top row shows the trajectory of the flavor composition, $(f_e, f_\mu, f_\tau)_\oplus$, spanned by varying a single high-energy parameter, compared against the region allowed by standard oscillations.  Underneath, rows display $f_{e, \oplus}$, $f_{\mu, \oplus}$, and $f_{\tau, \oplus}$ individually, computed via \equ{flavor_ratios}.  Columns show independent variations of the high-energy parameters $\sin^2\theta_{12}^\prime$, $\sin^2\theta_{23}^\prime$, $\sin^2\theta_{13}^\prime$, and $\delta_{\mathrm{CP}}^\prime$, with the remaining parameters fixed to their standard low-energy values. We compare these RG-modified ratios against the standard case, where $U^\prime = U$ in \equ{prob_standard}. Bands denote $1\sigma$ uncertainties.  Vertical bands indicate NuFIT 6.1 best-fit (``b.f.'') and $1\sigma$ intervals for the varied parameter, assuming normal mass ordering and including Super-Kamiokande data, as in \figu{probabilities}. Sub-panels quantify the relative change, $(f_{\alpha, \oplus} - f_{\alpha, \oplus}^{\rm std}) / f_{\alpha, \oplus}^{\rm std}$. See Sec.~\ref{sec:flavor_ratios-theory} for details and \figu{flavor_ratios_muon} in Appendix~\ref{app:flavor_composition_muon_damped} for the case of neutrino production via muon-damped pion decay.}
    \label{fig:flavor_ratios_pion}
\end{figure*}

Figure \ref{fig:flavor_ratios_pion} illustrates how the flavor composition at Earth responds to isolated variations in the high-energy mixing parameters, assuming neutrino production via full pion decay.
The top panels compare the $1\sigma$-allowed modified flavor regions accessible by varying $\mathbf{\theta}$ and $\mathbf{\theta}^\prime$ vs.~the regions accessible via standard mixing by varying $\boldsymbol{\theta}$ only~\cite{Bustamante:2015waa, Song:2020nfh}.  The rows underneath show the variation of the individual flavor ratios.

Figure \ref{fig:flavor_ratios_pion} reveals that, broadly, even large departures of $\theta_{12}^\prime$, $\theta_{23}^\prime$, and $\theta_{13}^\prime$ from their standard counterparts $\theta_{12}$, $\theta_{23}$, and $\theta_{13}$ have only a mild impact on the flavor composition at Earth, typically of $\lesssim 10\%$.  This contrasts with the behavior of the probabilities in \figu{probabilities}, where deviations were as large as tens of percent.  This low sensitivity is particular to our choice of neutrino production via full pion decay; we elaborate on this in Sec.~\ref{sec:flavor_ratios-impact_sources}.

Regardless, we can draw physical insight from \figu{flavor_ratios_pion}:
\begin{enumerate}
 \item
  The variation of $\sin^2\theta_{12}^\prime$ primarily modulates $f_{e, \oplus}$. As $\theta_{12}^\prime$ increases, so does $f_{e, \oplus}$, the remaining flavor content split almost symmetrically between the muon and tau flavors, due to $\theta_{23} \approx 45^\circ$. 
  This results in a nearly linear trajectory in the flavor triangle pointing away from the pure-electron corner. 
 \item
  The parameter $\sin^2\theta_{23}^\prime$ governs the $\mu$-$\tau$ symmetry at Earth. In standard oscillations, $\theta_{23}$ is close to maximal ($45^\circ$), which leads to roughly equal fractions of $\mu$ and $\tau$ flavors regardless of the initial composition. As seen in the sub-panels of \figu{flavor_ratios_pion}, altering $\theta_{23}^\prime$ breaks this degeneracy, creating a distinct anti-correlation: increasing $\theta_{23}^\prime$ suppresses $f_{\mu, \oplus}$ while enhancing $f_{\tau, \oplus}$, with $f_{e, \oplus}$ remaining largely decoupled. The flavor trajectory thus runs almost parallel to the $f_{\tau, \oplus}$ axis. If next-generation neutrino telescopes measure a significant $\mu$-$\tau$ asymmetry, it could be interpreted as evidence of RG evolution affecting the 2-3 sector.
 \item 
  The effects of $\sin^2\theta_{13}^\prime$ and $\delta_{\rm CP}^\prime$ are mathematically more intricate but phenomenologically sub-dominant. Because standard $\theta_{13}$ is small, its variation primarily enhances the mixing of $f_{e, \oplus}$ with the other flavors beyond standard mixing. The phase, $\delta_{\rm CP}^\prime$, introduces only a mild modulation effect.
\end{enumerate}

This behavior is present also for other choices of the flavor composition at the sources.  Importantly, however, the magnitude of the effects varies depending on the choice. In particular, \figu{flavor_ratios_muon} in Appendix~\ref{app:flavor_composition_muon_damped} shows that, for neutrino production via muon-damped pion decay, the variation of the high-energy mixing parameters can induce much larger changes in $f_{\alpha, \oplus}$, of up to 100\%.  We elaborate on why next.


\subsection{Why is neutrino production via full pion decay less sensitive to RG running?}
\label{sec:flavor_ratios-impact_sources}

The relative robustness of the flavor ratios at Earth under the nominal expectation of production via full pion decay, $\left( \frac{1}{3}, \frac{2}{3}, 0 \right)_{\rm S}$, against high-$Q$ mixing-parameter variations, as compared to the muon-damped case, $(0:1:0)_{\rm S}$, originates from the unitarity of the mixing matrices and the averaging of different flavor-transition probabilities.

In the muon-damped scenario, the flavor ratios at Earth map exclusively to the muon transition probabilities, $f_{\beta, \oplus} = P_{\mu\beta}$. Consequently, any high-$Q$ modification to the mixing angles is transferred undiluted to the flavor ratios at Earth.  In contrast, the flavor composition at Earth under full pion-decay production is a weighted superposition: $f_{\beta, \oplus} = \frac{1}{3} P_{e\beta} + \frac{2}{3} P_{\mu\beta}$. Because total probability is conserved (\ie, $\sum_\alpha P_{\alpha\beta} = 1$), an RG-induced enhancement in one transition channel (\eg, $P_{\mu e}$) is typically accompanied by a suppression in another (\eg, $P_{ee}$). When evaluating the electron flavor fraction at Earth, $f_{e, \oplus} = \frac{1}{3} P_{ee} + \frac{2}{3} P_{\mu e}$, for instance, this inverse correlation leads to a partial cancellation of the parametric dependence. This weighted sum dilutes the impact of high-$Q$ modifications on the flavor ratios.

From a geometric perspective in flavor-composition space, this damping effect under full pion decay is a consequence of the proximity of its prediction at Earth to the flavor-democratic point of $\left( \frac{1}{3}, \frac{1}{3}, \frac{1}{3} \right)_\oplus$. Under standard mixing, where $\theta_{23}$ is nearly maximal and $\theta_{12}$ is large, the initial $\left( \frac{1}{3}, \frac{2}{3}, 0 \right)_{\rm S}$ composition  diffuses into a nearly flavor-democratic composition at Earth. Once a completely decohered quantum system reaches a highly mixed state, extreme deformations of the underlying dynamics governing its evolution are required to displace it significantly from equilibrium. 

This principle generalizes broadly across astrophysical production mechanisms: sources with ``purer'' initial flavor composition evade this unitarity-driven dilution and provide superior sensitivity to high-energy mixing.  For instance, a pure $\bar{\nu}_e$ source driven by neutron decay, $(1, 0, 0)_{\rm S}$, yields $f_{\beta, \oplus} = P_{e\beta}$, exhibiting an extreme, unbuffered sensitivity to variations in $\theta_{12}^\prime$ that mirrors the sensitivity of the muon-damped case. 

Later, when extracting the high-$Q$ mixing parameters from flavor-composition measurements, our statistical procedure profiles over all possible flavor compositions at the sources, considering all of them as equally probable as a reflection of our extant ignorance on what is the real neutrino production mechanism.  While some of these initial compositions provide high sensitivity to the high-$Q$ mixing parameters, others do not. Our net sensitivity to the high-$Q$ mixing parameters is thus the result of exploring across this entire space of possible flavor compositions at the sources.


\subsection{Measuring the flavor composition}
\label{sec:flavor_comp-measuring}

\textbf{\textit{TeV--PeV neutrino detection.---}}In water- and ice-based optical-Cherenkov high-energy neutrino telescopes~\cite{Markov:1961tyz} optimized for TeV--PeV neutrino detection, like IceCube~\cite{IceCube:2013low}, KM3NeT~\cite{KM3Net:2016zxf}, and Baikal-GVD~\cite{Baikal-GVD:2025rhg} neutrinos are detected via their deep inelastic scattering off nucleons ($\nu N$ DIS)~\cite{CTEQ:1993hwr, Conrad:1997ne, Formaggio:2012cpf}, which can be either charged-current (CC)---if mediated by a $W$ boson---or neutral-current---if mediated by a $Z$ boson.  

In CC interactions, the final state contains an energetic charged lepton of the same flavor as the incoming neutrino, \ie, an electron, muon, or tau.  In NC interactions, the final state lepton is instead a neutrino of the same flavor as the incoming one.  In both cases, the final-state lepton receives, on average, 80\% of the energy of the original neutrino.  The remaining 20\% is carried by the final-state hadrons created by the breaking-up and ensuing hadronization of the interacting nucleon.  

In CC and NC interactions the final-state charged particles emit Cherenkov light, which propagates through the transparent medium of the detector (ice in IceCube, water in KM3NeT and Baikal-GVD), becoming attenuated in the process, and eventually being detected by an array of photomultipliers.  From the spatial and time profiles of the detected light, analyses infer the energy, arrival direction, and flavor of the interacting neutrino.

Optical Cherenkov neutrino telescopes classify events into three primary topologies:
\begin{itemize}
 \item \textbf{Cascades:} These are generated predominantly by the CC interactions of $\nu_e$ and $\nu_\tau$, which produce localized particle showers. NC interactions from all flavors also produce showers, but, because the NC cross section is smaller, and because a higher-energy, less abundant incident neutrino is required to produce an NC shower of the same deposited energy as a CC shower, the NC contribution is subdominant. However, because showers from CC $\nu_e$, CC $\nu_\tau$, and NC interactions look nearly identical at a given energy, there is a degeneracy in measuring the $\nu_e$ and $\nu_\tau$ flavor ratios.
 \item \textbf{Tracks:} These are primarily generated by CC $\nu_\mu$ interactions, which produce an energetic final-state muon capable of traveling several kilometers, leaving a long, visible track of Cherenkov light alongside the hadronic shower at the interaction vertex. Tracks can also arise from CC $\nu_\tau$ interactions when the final-state tau promptly decays into a muon (an $\approx 18\%$ branching ratio) and the production and decay showers cannot be spatially resolved.
 \item \textbf{Double cascades:} This topology is a unique signature of CC $\nu_\tau$ interactions. It consists of a first shower from the initial neutrino-nucleon DIS, and, if the resulting tau is sufficiently energetic to travel a resolvable distance before decaying, a second shower from its subsequent decay.
\end{itemize}

Because of these inherent degeneracies---particularly between $\nu_e$ and $\nu_\tau$ cascades---identifying the flavor of a neutrino on an event-by-event basis is effectively unfeasible.  
Instead, the flavor fractions $f_{\oplus, \alpha}$ are reconstructed collectively using a statistical ensemble of events. These analyses traditionally rely on ``starting events'', where the neutrino interaction vertex occurs within the instrumented volume and all three event topologies can be distinguished. However, such analyses are statistically limited by the low event rate of astrophysical neutrinos (\eg, about 10 neutrinos per ${\rm km}^3$ per year above 60 TeV~\cite{IceCube:2020wum}).  So far, flavor measurements have been carried out only on IceCube data, either by the IceCube Collaboration~\cite{IceCube:2015rro, IceCube:2015gsk, IceCube:2018pgc, IceCube:2020fpi, Abbasi:2025fjc} or externally to it~\cite{Mena:2014sja, Palomares-Ruiz:2015mka, Palladino:2015zua}.

To overcome these statistical limitations, flavor measurements can be significantly improved by including through-going tracks---events where a $\nu_\mu$ interacts outside the instrumented volume, producing a muon that traverses the detector. While these events only constrain the $\nu_\mu$ fraction, they are more numerous; when combined with starting events, they appreciably tighten the overall flavor composition fits, as shown in \Refe~\cite{IceCube:2015gsk}. 

Future analyses may mitigate the prevailing $\nu_e$-$\nu_\tau$ degeneracy by differentiating between $\nu_e$-induced electromagnetic showers and the largely hadronic showers induced by $\nu_\tau$. This separation relies on late-time Cherenkov-light ``echoes'' from the decay of low-energy muons and the capture of neutrons~\cite{Li:2016kra}, the latter of which has shown promising preliminary results on IceCube data~\cite{Farrag:2023jut, Dutta:2025qgk} (see \Refe~\cite{Steuer:2017tca} for earlier results). We do not, however, include echoes in our projections.

\smallskip

\textbf{\textit{Present TeV--PeV flavor measurements.---}}For the present flavor-composition measurements, we adopt the recent 11.4-year IceCube Medium Energy Starting Events (MESE) measurement~\cite{Abbasi:2025fjc} (\figu{ternary_theory}, right panel), the first reporting non-zero content of all flavors at 68\% C.L. MESE events have energies from 1~TeV to 10~PeV and high astrophysical purity~\cite{IceCube:2014rwe, IceCube:2025tgp}.  The right panel of \figu{ternary_theory} shows our approximation to this MESE measurement.  The current measurement is broad enough to encompass both the standard mixing band and a significant fraction of the RG-allowed space, severely limiting our present ability to distinguish between the two regimes. However, projected combinations of the observations by multiple neutrino telescopes will boost sensitivity.

\smallskip

\textbf{\textit{Future TeV--PeV flavor measurements.---}}For our projected flavor-composition measurements by optical-Cherenkov neutrino telescopes, we adopt the same projections as \Refe~\cite{Bustamante:2026aur}, based on multi-detector combinations.  These are, in turn, based on the methods introduced in \Refe~\cite{Liu:2023flr} to infer flavor-composition measurements from a combination of High-Energy Starting Events~\cite{Schonert:2008is, IceCube:2013low, Gaisser:2014bja, IceCube:2014stg, Arguelles:2018awr, IceCube:2020wum} (HESE, tracks, cascades, and double cascades above 60~TeV)---sensitive to all flavors---and through-going muons~\cite{IceCube:2019cia, IceCube:2021xar}.  We forecast multi-telescope detection in existing IceCube, Baikal-GVD~\cite{Baikal-GVD:2025rhg}, and KM3NeT~\cite{KM3Net:2016zxf}, plus in future telescopes~\cite{MammenAbraham:2022xoc, Ackermann:2022rqc} P-ONE~\cite{P-ONE:2020ljt}, IceCube-Gen2~\cite{IceCube-Gen2:2020qha}, NEON~\cite{Zhang:2024slv}, TRIDENT~\cite{TRIDENT:2022hql}, and HUNT~\cite{Huang:2023mzt}---up to 30 times the size of IceCube---by scaling IceCube event rates by the detector size, as in \Refe~\cite{Schumacher:2025qca} (also \Refes~\cite{Song:2020nfh, Fiorillo:2022rft, Telalovic:2023tcb, Liu:2023flr, Schumacher:2025qca}).  For details, see \Refes~\cite{Liu:2023flr, Bustamante:2026aur}.

The right panel of \figu{ternary_theory} shows our projections (at the 68\%~C.L.; for the full flavor likelihood, see the Suppl.~Mat.~of \Refe~\cite{Bustamante:2026aur}). By 2040, combined multi-decade exposures from km$^3$-scale detectors (IceCube, Baikal-GVD, and KM3NeT) will shrink the observational uncertainties dramatically. By 2050, the integration of tens-of-km$^3$ next-generation facilities (IceCube-Gen2, P-ONE, and HUNT) will yield precision contours capable of isolating specific source mechanisms. 

Figure~\ref{fig:ternary_theory} shows the measurements are tightest on the muon flavor fraction, $f_{\mu,\oplus}$, owing to the large number of through-going muon tracks present in the simulated data used to derive the measurements.  The 2040 flavor-composition contour remains elongated along the approximate $f_{e, \oplus}$-$f_{\tau, \oplus}$ axis, reflecting the experimental difficulty in separating these two flavors in a HESE sample.  By 2050, the innate tau-identification capabilities in the HESE sample---via double cascades---and immense event statistics shrink this contour into a tight, nearly symmetric region, constraining all three flavors comparably well.

The ultimate physical insight derived from these projections is clear: because the standard mixing prediction is so extraordinarily narrow, it constitutes an exceptionally pristine null hypothesis. If the observed high-statistics flavor composition ultimately converges outside this thin standard band, it will provide an astrophysics-independent signature of new high-energy dynamics affecting neutrino mixing.

\smallskip

\textbf{\textit{Future UHE flavor measurements.---}}In the UHE regime (\ie, energies over 100~PeV), the rapidly decreasing astrophysical neutrino flux necessitates detection volumes vastly larger than what is feasible for optical-Cherenkov telescopes. Future UHE observatories~\cite{Ackermann:2022rqc, MammenAbraham:2022xoc}, like the planned radio array of IceCube-Gen2~\cite{IceCube-Gen2:2020qha}, will overcome this (see, \eg, \Refe~\cite{Valera:2022wmu}) by detecting the coherent radio-frequency emission---the Askaryan effect---produced by neutrino-induced particle showers propagating through the ice~\cite{Schroder:2016hrv, Barwick:2022vqt}.  (Other techniques based on high-altitude fluorescence and Cherenkov-light detection are also being explored~\cite{Otte:2018uxj, POEMMA:2020ykm}.)

Reference~\cite{Coleman:2024scd} showed how flavor discrimination in radio detectors differs from that in optical telescopes. Because radio arrays observe the brief, intense radio flash emitted by the primary interaction rather than imaging kilometer-long tracks, they rely on the longitudinal development of the shower, which imprints itself on the angular distribution and frequency spectrum of the radio pulse. In CC $\nu_e$ interactions, the final-state electron from the DIS event initiates a purely electromagnetic shower. At UHE energies, the Landau-Pomeranchuk-Migdal (LPM) effect significantly elongates this electromagnetic cascade. Conversely, the identification of CC $\nu_\mu$ and $\nu_\tau$ interactions is achieved through their multiple catastrophic energy losses. As the high-energy final-state muon or tau propagates through the ice, it undergoes repeated, discrete energy losses that generate secondary showers, which can be detected as coincident radio pulses by multiple radio stations. 

This stark difference in detection signatures allows radio arrays to cleanly separate CC $\nu_e$ interactions from the rest, effectively isolating the electron neutrino fraction, $f_{e, \oplus}$, from the combined non-electron fraction, $f_{\mu, \oplus} + f_{\tau, \oplus}$. For our projected UHE flavor-composition measurements, we adopt the sensitivities forecasted for the IceCube-Gen2 radio array from \Refe~\cite{Coleman:2024scd}. Because distinguishing $\nu_\mu$ from $\nu_\tau$ based on these multi-station catastrophic loss signatures is exceptionally challenging, the resulting UHE flavor contours are elongated along the $f_{\mu,\oplus}$-$f_{\tau,\oplus}$ axis.  Further, the lower neutrino flux expected at ultra-high energies (see Fig.~2 in \Refe~\cite{Valera:2022wmu}), compared to the flux at TeV--PeV energies, widens the UHE flavor contours compared to their TeV--PeV counterparts.

Despite this, the precision on $f_{e, \oplus}$ provides a critical constraint. These projected UHE measurements will offer an independent, complementary probe of the flavor composition at the highest accessible energies. This is precisely the regime where new-physics effects---\eg, Lorentz-invariance violation, novel neutrino interactions---are expected to grow with energy, providing an unprecedented test of the standard mixing paradigm.  Below, we extract constraints on the high-$Q$ mixing parameters from these UHE projections.

(The capabilities of the radio array of IceCube-Gen2 could be complemented by giant surface radio arrays like the Giant Radio Array for Neutrino Detection (GRAND)~\cite{GRAND:2018iaj}.  Reference~\cite{Testagrossa:2023ukh} showed how combining the dedicated sensitivity of GRAND to Earth-skimming $\nu_\tau$ in-air interactions~\cite{Huege:2016veh, Schroder:2016hrv} with all-flavor in-ice observations by IceCube-Gen2 radio array could break the $\nu_\mu$-$\nu_\tau$ degeneracy, yielding tightened constraints on the UHE flavor composition.  However, in our projections in this work we use exclusively the stand-alone flavor-measurement capabilities of the IceCube-Gen2 radio array described above.)


\subsection{Flavor regions at Earth accessible by varying the high-$Q$ mixing parameters}
\label{sec:flavor_comp-regions_earth}

\begin{figure*}[t!]
    \centering
    \includegraphics[width=0.497\textwidth]{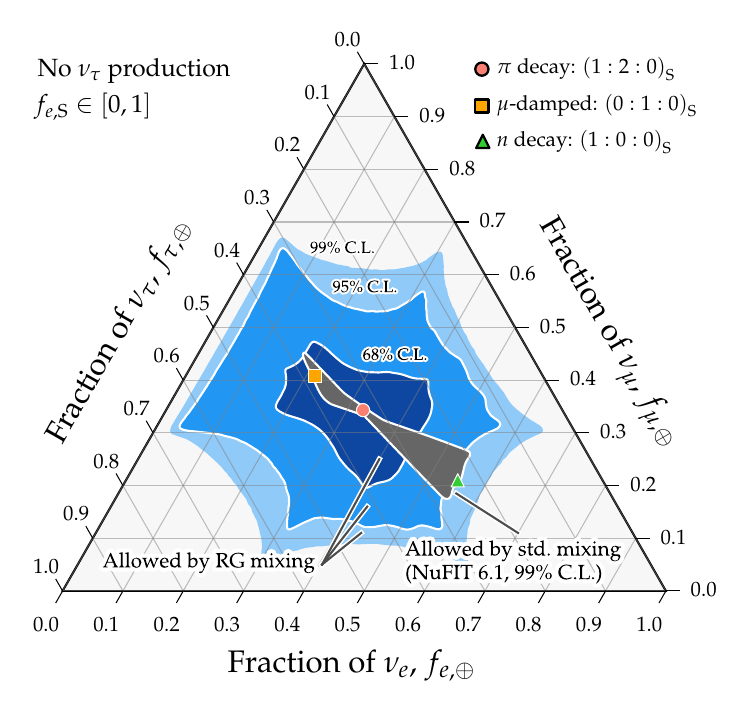}
    \includegraphics[width=0.497\textwidth]{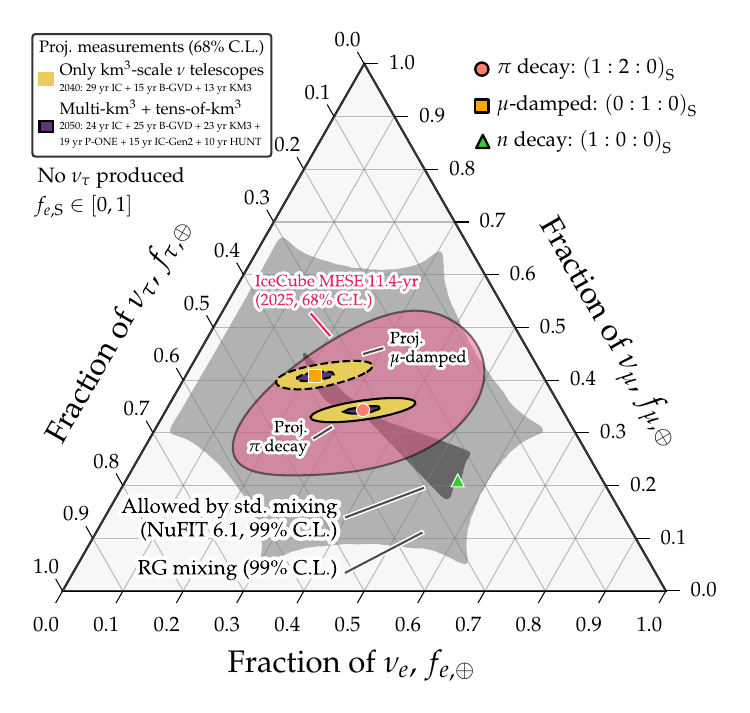}
    \caption{\textbf{Standard and RG-modified allowed regions of flavor composition at Earth.} The region of flavor composition accessible by standard mixing is obtained by varying the standard mixing parameters within their allowed intervals from NuFIT 6.1 (99\% C.L.) and the flavor composition at the sources $f_{e,S} \in [0,1]$ (assuming $f_{\tau,S} = 0$).  The region accessible by RG-modified mixing is obtained by varying, in addition, the high-$Q$ mixing parameters within their physically allowed regions.  \textit{Left:} The RG-modified allowed regions at 68\%, 95\%, and 99\% C.L.~regions.  \textit{Right:} The 99\% C.L. RG-allowed theoretical space juxtaposed with current and projected experimental sensitivities. Current 68\% C.L. constraints from IceCube (11.4-yr MESE) are shown in gray. Projected 68\% C.L. contours for 2040 (combining IceCube, Baikal-GVD, and KM3NeT) and 2050 (adding IceCube-Gen2, P-ONE, and HUNT) are shown for benchmark pion-decay and muon-damped source scenarios. 
    }
    \label{fig:ternary_theory}
\end{figure*}

To fully map the discovery potential of astrophysical flavor ratios, we must generalize beyond fixed benchmark choices of the flavor compositions the sources. While neutrino production mechanisms like full and muon-damped pion decay are well-motivated, astrophysical environments may produce arbitrary mixtures of $\nu_e$ and $\nu_\mu$. 

Figure~\ref{fig:ternary_theory} explores this generalized landscape by allowing the initial electron neutrino fraction to vary freely, $f_{e, {\rm S}} \in [0,1]$, while maintaining the standard assumption of negligible $\nu_\tau$ production ($f_{\tau, {\rm S}} = 0$).  Because astrophysical neutrino production occurs at low $Q$ (see \figu{main_results}), the projection of source flavors onto the propagated mass eigenstates is governed strictly by the standard, unvaried PMNS matrix, which inherently possesses strong $\mu$-$\tau$ symmetry because $\theta_{23} \approx 45^\circ$. Consequently, a pure $\nu_\tau$ source would populate the mass eigenstates in nearly identical proportions to a pure $\nu_\mu$ source.  Allowing for $\nu_\tau$ production ($f_{\tau, {\rm S}} > 0$) would barely expand the accessible flavor space at Earth (see \Refe~\cite{Bustamante:2015waa}), regardless of how severely RG running breaks $\mu$-$\tau$ symmetry at the high-$Q$ detection scale.  (Nevertheless, letting $f_{\tau, {\rm S}}$ float does impact our parameter constraints; see Appendix~\ref{app:results_with_nu_tau_production}.)

Figure~\ref{fig:ternary_theory} shows that, under standard mixing, even an unconstrained initial mixture of $\nu_e$ and $\nu_\mu$ maps to a remarkably restrictive band at Earth.  This is the \textit{theoretically palatable region} introduced in \Refe~\cite{Bustamante:2015waa} (see also \Refe~\cite{Song:2020nfh}).  This small band is a consequence of the high precision with which the standard mixing angles are known~\cite{Esteban:2024eli}.  To capture known experimental correlations between standard mixing parameters in \figu{ternary_theory} and in our statistical analysis later (Sec.~\ref{sec:statistics}), we sample their values from the NuFIT 6.1~\cite{Esteban:2024eli} two-dimensional $\Delta\chi^2$ profiles, specifically, from the pairwise profiles of $\delta_{\mathrm{CP}}$ and $\sin^2\theta_{23}$, and of $\sin^2\theta_{12}$ and $\sin^2\theta_{13}$.

However, under RG running, and allowing for the high-$Q$ mixing parameters to float freely, this rigidity is shattered.  As shown in the left panel of Fig.~\ref{fig:ternary_theory}, varying over all possible values of the high-$Q$ parameters and $f_{e, {\rm S}} \in [0, 1]$ drastically inflates the accessible flavor space.  While the allowed RG-modified region at 68\% C.L.~does not deviate far from the theoretically palatable region, the full 99\% C.L.~RG-modified footprint encompasses the majority of the flavor space.  This expansion confirms that RG running naturally accesses regions of flavor composition at Earth that are inaccessible by standard mixing.

Appendix~\ref{app:allowed_flavor_regions} shows that the RG-allowed flavor region at Earth is spanned dominantly by the variation of $\theta_{12}^\prime$ and $\theta_{23}^\prime$, and sub-dominantly by that of $\theta_{13}^\prime$ and $\delta_{\rm CP}^\prime$.  However, the sensitivity to these different parameters depends on exactly where on the flavor triangle the flavor measurement falls.  This is reflected later (Sec.~\ref{sec:results}) in our projections assuming measurements based on neutrino production via full and muon-damped pion decay.


\section{Renormalization group running in neutrinos}
\label{sec:rg_running}

\textit{We model the $Q$-dependent evolution of neutrino mixing parameters, contrasting the negligible running in the SM with the potentially large, observable effects introduced by beyond-Standard-Model (BSM) frameworks like the MSSM and particularly dimension-6 operators in the SMEFT. By folding the momentum-dependent RGEs over the realistic $Q$-distributions of deep inelastic scattering for TeV--PeV and UHE astrophysical neutrinos, we find that their steeply falling energy spectra heavily weight the interactions toward moderate momenta $Q = $~10--100~GeV, dampening the observable effects of the running. Individual SMEFT coefficients impact the mixing angles hierarchically: $\theta_{12}$ is uniquely sensitive due to its being governed by the small solar mass-squared splitting ($\Delta m^2_{21}$), while the running of $\theta_{13}$ and $\theta_{23}$ is suppressed by the larger atmospheric splitting ($\Delta m^2_{31}$).}


\subsection{RG running of the neutrino mass matrix}

So far, we have treated the high-$Q$ mixing parameters  as phenomenological quantities that we vary independently of each other.  Now we turn to predicting their running from low to high $Q$ scales---from production to detection---explicitly. This is achieved by computing the RG running of the effective neutrino mass matrix between these two scales. Below, first we present an overview of RG running in neutrinos generically, and then we show examples within specific running schemes.

In the Standard Model (SM) extended with massive Majorana neutrinos, the masses and mixing parameters are generated after electroweak symmetry breaking via the unique dimension-5 Weinberg operator~\cite{Weinberg:1979sa},
\begin{equation}
 \mathcal{L}^{(5)}
 =
 \frac{c_5}{\Lambda} (\overline{L^c} \tilde{H})(\tilde{H}^T L) + \text{h.c.} \;,
 \label{equ:weinberg_operator}
\end{equation}
where $L$ is the left-handed lepton doublet, $H$ is the Higgs doublet, and $\Lambda$ is the scale of new physics.  After electroweak symmetry breaking, the Higgs field acquires a vacuum expectation value (VEV) $v \approx 246$~GeV, reducing this operator to the effective Majorana mass matrix for the active neutrinos, $M_\nu = - [v^2 / ( 2 \Lambda )] c_5$ in the flavor basis, where $c_5$ is a symmetric matrix of dimensionless Wilson coefficients in flavor space.  Diagonalizing the mass matrix using the PMNS mixing matrix, $U$, rotates the mass matrix into the mass basis, revealing the masses of the neutrino mass eigenstates $\nu_1$, $\nu_2$, and $\nu_3$ as its eigenvalues, \ie, $U^\dagger M_\nu U = \text{diag}(m_1, m_2, m_3)$.

The evolution of the mass matrix with the momentum scale $Q$, is governed by a set of coupled non-linear RGEs. We restrict our analysis to one-loop RGEs. Two-loop corrections are suppressed by an additional loop factor of $1/(16\pi^2)$ and yield negligible shifts that fall well below the current experimental sensitivities for high-energy astrophysical neutrino flavor composition (Sec.~\ref{sec:flavor_comp-measuring}).

The generic one-loop RGE for $M_\nu$ is determined by its beta function, $\beta \equiv \frac{d M_\nu}{dt}$~\cite{Chankowski:1993tx, Babu:1993qv, Casas:1999tg, Antusch:2003kp, Antusch:2005gp}, 
\begin{equation}
 16\pi^2 \frac{d M_\nu}{dt}
 =
 \alpha M_\nu + P^T M_\nu + M_\nu P \;,
 \label{equ:rg_equation}
\end{equation}
where $t = \ln(Q/Q_0)$ is the integration variable, and $Q_0$ is the momentum at which initial boundary conditions are defined before the RG running begins.  In our numerical results below, we set $Q_0 = 1$~GeV.

In \equ{rg_equation}, the scalar coefficient $\alpha$ encapsulates all flavor-universal corrections. These arise from gauge-boson loops and the Higgs field renormalization (which is dominated by the top-quark Yukawa and the Higgs self-coupling). Because these interactions do not distinguish between lepton generations, they scale the entire mass matrix uniformly and only affect the absolute neutrino mass scale, not the mixing.  Thus, $\alpha$ governs the RG running of the squared-mass differences, $\Delta m_{ij}^2$.

In contrast, the matrix $P$ in \equ{rg_equation} captures the flavor-dependent  renormalization of the lepton doublets. This term is driven directly by the charged-lepton Yukawa couplings ($P \propto Y_e^\dagger Y_e$). Because the charged-lepton masses are highly hierarchical ($m_e \ll m_\mu \ll m_\tau$), $P$ breaks the flavor symmetry. It is precisely this non-universal $P$ term that induces flavor transitions, driving the differentiated running of the neutrino mixing angles and the CP-violation phase across momentum scales.  

The shape of $P$ depends on the RG running scheme we adopt.  While our first set of results on the high-$Q$ mixing parameters ($\theta_{12}$, $\theta_{23}$, and $\theta_{13}$), $\delta_{\rm CP}$) is independent of any specific scheme, our second set of results is on the specific coefficients that control RG running within dimension-6 SMEFT.  Thus, below we motivate our study of the RG running of the neutrino mixing parameters by presenting three illustrative schemes: the SM, the MSSM, and the aforementioned dimension-6 SMEFT.

Regardless of the specific high-energy framework adopted, integrating the RGEs requires fixing boundary conditions on the mixing parameters. In all subsequent numerical evaluations, we anchor the initial neutrino mixing angles ($\theta_{12}$, $\theta_{23}$, and $\theta_{13}$), the CP-violation phase ($\delta_{\rm CP}$), and the mass-squared differences ($\Delta m_{21}^2$, $\Delta m_{31}^2$) at the reference scale $Q_0 = 1$~GeV to the central best-fit values provided by the NuFIT 6.1 global fit. Because these NuFIT parameters are derived from comprehensive fits to sub-TeV neutrino oscillation data, using them as our low-$Q$ boundary conditions ensures that the RG-modified flavor transitions we compute later for TeV--PeV astrophysical neutrinos remain strictly consistent with current constraints.


\subsection{Mixing-parameter extraction}

At each value of $Q$ evaluated when solving the beta function, the running mixing parameters are extracted by diagonalizing the Hermitian matrix $H = M_\nu^\dagger M_\nu$, \ie,
\begin{equation}
 H = \tilde{U} M_{\text{diag}}^2 \tilde{U}^\dagger \;,
\end{equation}
where $M_{\text{diag}}^2 = \text{diag}(m_1^2, m_2^2, m_3^2)$ contains the running squared-mass eigenvalues, and $\tilde{U}$ is a generic unitary complex matrix parametrized in the standard PMNS form.  The squared-mass differences are computed directly from these eigenvalues.

The mixing angles are extracted from the elements of the PMNS matrix $\tilde{U}$ via
\begin{eqnarray}
 \sin^2 \theta_{13} &=& |\tilde{U}_{e3}|^2  \;, \\
 \sin^2 \theta_{12} &=& \frac{|\tilde{U}_{e2}|^2}{1 - |\tilde{U}_{e3}|^2} \\
 \sin^2 \theta_{23} &=& \frac{|\tilde{U}_{\mu 3}|^2}{1 - |\tilde{U}_{e3}|^2} \;.
\end{eqnarray}
The CP-violation phase is extracted from the rephasing-invariant quartic product $C \equiv \tilde{U}_{e3} \tilde{U}_{\mu 2} \tilde{U}_{e2}^* \tilde{U}_{\mu 3}^*$ via
\begin{equation}
 \delta_{\rm CP}
 =
 \arctantwo \left[\text{Im}(C), \text{Re}(C)\right] \;.
\end{equation}
In our numerical results, we verify the continuity of the phase evolution by tracking the Jarlskog invariant, $J = c_{12} s_{12} c_{23} s_{23} c_{13}^2 s_{13} \cos \delta_{\rm CP}$ (with $c_{ij} \equiv \cos \theta_{ij}$, $s_{ij} \equiv \sin \theta_{ij}$), across the range of $Q$ values to ensure no unphysical discrete jumps occur.

Although later (Sec.~\ref{sec:rg_running-impact_smeft_coefficients}) we show approximate analytical expressions for the RG running of the mixing parameters to extract physical insight---in the spirit of the seminal work in \Refes~\cite{Casas:1999tg, Antusch:2003kp}---when producing our results we compute the RG evolution strictly numerically.


\subsection{RG running in the Standard Model}
\label{sec:rg_running-sm}

In the SM, the flavor-universal trace is
\begin{equation}
 \alpha_{\rm SM} = -3g_2^2 + 2\lambda + 6y_t^2 \;,
\end{equation}
where $g_2$ is the $SU(2)_L$ weak isospin gauge coupling constant (which represents the $W$ and $Z$ bosons running in the one-loop diagrams connecting to Higgs and lepton lines), $\lambda$ is the Higgs self-coupling constant (which accompanies the $(H^\dagger H)^2$ term in the SM Higgs potential), and $y_t \approx 1$ is the top-quark Yukawa coupling, which dominates the determination of $\alpha_{\rm SM}$.

The flavor-dependent matrix in \equ{rg_equation} is dominated by the charged-lepton Yukawa matrix $Y_e$, \ie
\begin{equation}
 P_{\rm SM} \approx -\frac{3}{2} Y_e^\dagger Y_e \;.
\end{equation}
Because $Y_e = \text{diag}(y_e, y_\mu, y_\tau)$ is highly hierarchical, the high-$Q$ running ($Q > m_\tau \approx 1.776$ GeV) is nearly entirely driven by the tau Yukawa coupling ($y_\tau \approx 0.01$). 

When $Q$ drops below the tau-mass threshold, $m_\tau$, the tau lepton decouples from the one-loop corrections governing the RG running.  At this threshold, the tau  Yukawa contribution is removed from $P_\text{SM}$. The running is then handed over to the muon Yukawa coupling ($y_\mu \approx 6 \times 10^{-4}$). Because the rate of change is proportional to the square of the Yukawa couplings, this transition suppresses RG running by orders of magnitude. 

Figure~\ref{fig:rge_running_comparison} shows the SM RG running of the mixing parameters.  It reveals that, even above $m_\tau$, the SM running is inherently constrained by the smallness of the tau Yukawa coupling. Consequently, for a normal neutrino mass ordering with a strictly hierarchical spectrum ($m_1 \ll m_2 \ll m_3$), the resulting SM running of the mixing angles across the entire 0.1--100 GeV range is negligible.  In contrast, the RG running of the neutrino masses can be significant.  This is because it is driven by  $\alpha_{\rm SM}$ in \equ{rg_equation}, which is dominated by the top-quark Yukawa coupling ($y_t = 1$), roughly a hundred times higher than the tau Yukawa coupling.


\subsection{RG running in the MSSM}
\label{sec:rg_running-mssm}

To induce large RG running in the neutrino mixing angles  between low-$Q$ and high-$Q$ scales, we must consider new-physics frameworks.  We consider first the MSSM.

Unlike the SM, the MSSM~\cite{Dimopoulos:1981zb, Haber:1984rc} requires two Higgs doublets. The doublet $H_u$, with VEV $v_u$, couples to up-type quarks, while the doublet $H_d$, with VEV $v_d$, couples to down-type quarks and charged leptons. Because the total electroweak VEV $v$ is shared such that $v_d = v \cos\beta$, where $\tan\beta \equiv v_u / v_d$, the charged leptons acquire their masses from a suppressed VEV when $\tan\beta$ is large.

Consequently, to reproduce the observed physical lepton masses, the effective Yukawa couplings of the charged leptons ($\ell = e, \mu, \tau$) must be significantly enhanced by a factor of $1/\cos\beta$, yielding
\begin{equation}
 y_{\ell}^{\text{eff}} \simeq y_{\ell} \sqrt{1 + \tan^2\beta} \;.
 \label{equ:yukawa_mssm}
\end{equation}
Above the supersymmetry-breaking scale $M_\text{SUSY}$ [of at least 1--10~TeV in light of bounds from the Large Hadron Collider (LHC)], the beta function for the effective neutrino mass matrix retains the same form as in the SM, \equ{rg_equation}.  However, the loop coefficients $\alpha$ and $P$ are modified by the presence of supersymmetric partners and the second Higgs doublet. When $Q < M_{\text{SUSY}}$, the heavy supersymmetric partners and the extended Higgs sector are integrated out.  The beta function abruptly transitions back to its SM form, slowing down the running of the mixing angles back to its negligible SM regime.

Crucially for the running of the mixing angles, the flavor-dependent matrix $P_{\text{MSSM}} \simeq (Y_e^{\text{eff}})^\dagger Y_e^{\text{eff}}$ absorbs the $\tan\beta$ enhancement from \equ{yukawa_mssm}.  As a result, for large values of $\tan\beta$ (\eg, $\tan\beta \sim 50$), the tau contribution to $P_{\text{MSSM}}$ is amplified by $\sim$$\tan^2\beta$ relative to the SM, accelerating the RG running of the mixing angles.  However, LHC bounds from heavy Higgs searches (\eg, $H/A \to \tau^+\tau^-$) and rare meson decays ($B_s \to \mu^+\mu^-$) severely constrain large $\tan\beta \gtrsim 40$. Consequently, such dramatic running within the MSSM is only viable today if the supersymmetric scale is pushed to higher energies.  

Figure~\ref{fig:rge_running_comparison} shows the MSSM RG running of the mixing parameters when choosing a moderate illustrative value of $\tan \beta = 10$ and a SUSY mass scale of $M_\text{SUSY} = 1$~TeV.  We assume normal neutrino mass ordering, with $m_1 = 0.05$~eV the lightest mass; the other two masses are computed as $m_2 = \sqrt{m_1^2 + \Delta m_{21}^2}$ and $m_3 = \sqrt{m_1^2 + \Delta m_{31}^2}$.  The accelerated RG running starts once $Q > M_\text{SUSY}$.  The running of $\theta_{12}$ is visibly enhanced---though still within its present $3\sigma$ globally allowed range--- while $\theta_{23}$ and $\theta_{13}$ remain relatively stable. 

This disparity arises because the RG evolution of the mixing angles is inversely proportional to the mass-squared splittings, $\Delta m^2_{ij}$. The running of $\theta_{12}$ is primarily governed by the solar mass splitting ($\Delta m^2_{21} \simeq 7.5 \times 10^{-5} \text{ eV}^2$), whereas the running of $\theta_{23}$ and $\theta_{13}$ is suppressed by the larger atmospheric splitting ($\Delta m^2_{31} \simeq 2.5 \times 10^{-3} \text{ eV}^2$). Although the $\tan\beta$ enhancement amplifies the overall running rate by two orders of magnitude relative to the SM, this is only sufficient to overcome the $\Delta m^2$ suppression for $\theta_{12}$, leaving the other angles largely unaffected.  (This is strictly true under a hierarchical mass spectrum, which we assume for \figu{rge_running_comparison}. Under a quasi-degenerate spectrum, where $m_1 \approx m_2 \approx m_3$, the other angles can also be significantly affected.) 


\subsection{RG running in the SMEFT}
\label{sec:rg_running-smeft}

Beyond specific supersymmetric extensions of the SM, the RG evolution can be modified generically using SMEFT. In this framework, the effects of heavy new physics residing at a scale $\Lambda \gg v$ (where $v$ is the electroweak scale) are parametrized by an infinite tower of higher-dimensional operators added to the SM Lagrangian~\cite{Buchmuller:1985jz, Brivio:2017vri, Jenkins:2017jig, Jenkins:2017dyc}: $\mathcal{L}_{\rm SMEFT} = \mathcal{L}_{\rm SM} + \sum_{d=5}^{\infty} \frac{1}{\Lambda^{n-4}} \mathcal{L}^{(d)}$. 

The lowest-order addition, the dimension-5 Weinberg operator [\equ{weinberg_operator}], violates lepton number ($\Delta L = 2$) and generates Majorana neutrino masses; however, its phenomenological imprint on flavor transitions is strictly kinematic, yielding standard oscillation phases that scale as $\sim \Delta m^2 L / E$, which strictly decohere and vanish for high-energy astrophysical neutrinos (Sec.~\ref{sec:flavor_comp-osc_std}). Odd-dimensional operators ($d=5, 7, \dots$) generically violate lepton or baryon number and are highly suppressed. 

Conversely, dimension-6 operators ($\mathcal{L}^{(6)}$)~\cite{Buchmuller:1985jz, Grzadkowski:2010es} represent the leading-order, lepton- and baryon-number-conserving corrections that provide non-standard interactions~\cite{Wolfenstein:1977ue, Grossman:1995wx, Proceedings:2019qno} and lepton-flavor-violating effects. Because their interaction cross sections scale as $\sim (E/\Lambda)^2$ and their induced propagation phase shifts scale as $\sim E/\Lambda^2$, they may dominate the anomalous BSM phenomenology at the high energies probed by neutrino telescopes. While higher-order lepton-number-conserving operators exist (\eg, dimension-8), their effects are suppressed by an overwhelming factor of $\mathcal{O}(v^4/\Lambda^4)$ or $\mathcal{O}(E^4/\Lambda^4)$. Assuming the underlying EFT expansion remains valid ($E, v \ll \Lambda$), dimension-8 and higher operators are entirely negligible compared to dimension-6 ones. For this reason, our analysis focuses exclusively on dimension-6 SMEFT operators. Specifically, we use the Warsaw basis~\cite{Grzadkowski:2010es}, which provides a complete, non-redundant set of independent dimension-6 operators.

In the exact one-loop matching using the Warsaw basis, the beta function for the effective neutrino mass matrix, \equ{rg_equation}, is modified by dimension-6 operators, notably the $SU(2)_L$ triplet operator $O_{H\ell}^{(3)} = (H^\dagger i \overset{\leftrightarrow}{D}_\mu^I H)(\bar{\ell}_p \tau^I \gamma^\mu \ell_r)$, where the superscript $I \in \{ 1, 2, 3 \}$ denotes contraction with the Pauli matrices. The modified beta function becomes
\begin{eqnarray}
 16\pi^2 \frac{d M_\nu}{dt}
 &=&
 \alpha_{\rm SM} M_\nu
 +
 P_{\rm SM}^T M_\nu
 +
 M_\nu P_{\rm SM} 
 \nonumber \\ 
 && 
 +~
 2\left(C_{H\ell}^{(3)T} M_\nu
 +
 M_\nu C_{H\ell}^{(3)}\right) \;,
 \label{equ:rg_equation_smeft}
\end{eqnarray}
where $C_{H\ell}^{(3)}$ is the Hermitian Wilson coefficient matrix for the dimension-6 insertion,
\begin{equation}
 C_{H\ell}^{(3)}
 =
 c_{\rm SMEFT}
 \begin{pmatrix}
  C_{11} & C_{12} & C_{13} \\ 
  C_{12}^\ast & C_{22} & C_{23} \\ 
  C_{13}^\ast & C_{23}^\ast & C_{33}
 \end{pmatrix} \;,
 \label{equ:wilson_coeff_matrix_general}
\end{equation}
where $c_{\rm SMEFT} \equiv (v / \Lambda_{\rm SMEFT})^2$.
Following standard EFT power-counting up to $\mathcal{O}(1/\Lambda_{\rm SMEFT}^2)$ the background loop contributions are restricted to their unperturbed SM values, with the flavor-dependent matrix and the scalar trace remaining as $P_{\rm SM}$ and $\alpha_{\rm SM}$.  (Although, \equ{wilson_coeff_matrix_general} is written in the flavor basis, we write the SMEFT coefficients as $C_{ij}$, with $i, j = 1,2,3$, to follow literature convention, with the tacit understanding that these indices refer to $e$, $\mu$, and $\tau$.)

\begin{figure}[t!]
 \centering
 \includegraphics[width=\columnwidth]{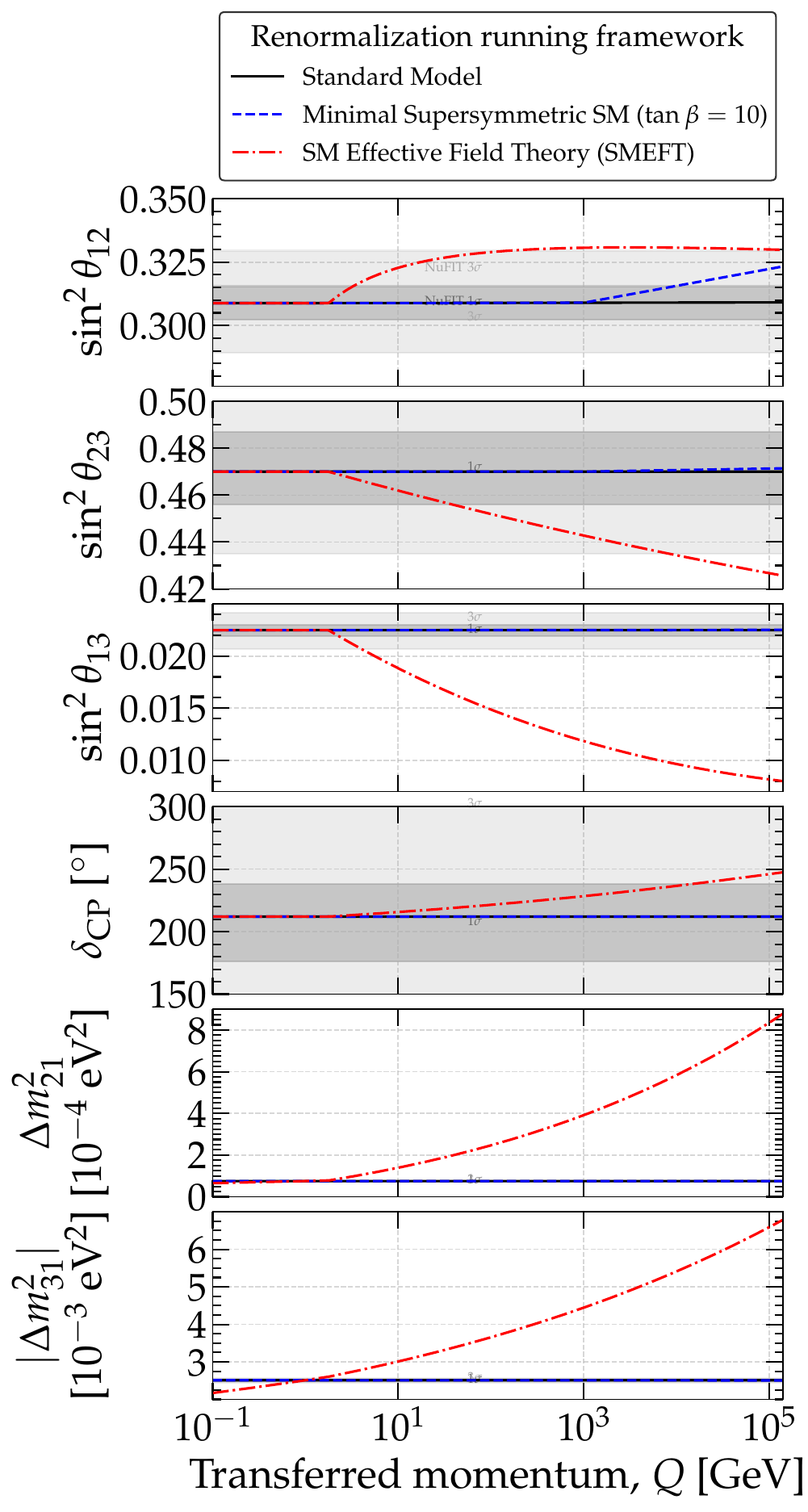}
 \caption{\textbf{Renormalization group (RG) evolution of the neutrino mixing parameters} The mixing angles ($\theta_{12}$, $\theta_{23}$, $\theta_{13}$) and CP-violation phase ($\delta_{\rm CP}$) evolve as a function of the transferred momentum, $Q$, depending on the RG scheme chosen. The Standard Model (SM) expectation exhibits negligible running (Sec.~\ref{sec:rg_running-sm}). The Minimal Supersymmetric Standard Model (MSSM) with $\tan\beta = 10$ (Sec.~\ref{sec:rg_running-mssm}) exhibits limited evolution, while the phenomenological dimension-6 SMEFT exhibits appreciable evolution that makes it possible to place constraints on the RG-modified mixing parameters using high-energy astrophysical neutrino measurements.}
 \label{fig:rge_running_comparison}
\end{figure}

Figure~\ref{fig:rge_running_comparison} (also \figu{main_results}) illustrates the SMEFT RG running of the mixing parameters using hand-picked illustrative values of the couplings, 
\begin{eqnarray}
 C_{11} &=& 0.1 \;, \\
 C_{12} &=& 0.1 + 0.15i \;, \\
 C_{13} &=& 0.01 \;,  \label{equ:wilson_coeff_matrix} \\
 C_{22} &=& 0.11 \;, \\
 C_{23} &=& 0.01 - 0.02i \;, \\
 C_{33} &=& 0.2 \;,
\end{eqnarray}
and $c_{\rm SMEFT} = 3/2$ as baseline value in our examples. The non-zero complex off-diagonal elements in $C_{H\ell}^{(3)}$ for $Q > m_\tau$ introduce flavor-violating interactions that drive the RG running of the mixing angles and the CP-violation phase.  Below the tau-mass threshold, tau-specific dimension-6 operators are integrated out alongside the SM tau lepton, and the flavor-dependent running is handed over to the muon Yukawa coupling.  However, owing to its smallness, the RG running of the mixing angles effectively ceases.

The profound contrast in the RG evolution between the MSSM and the SMEFT frameworks arises from four fundamental differences in the tensor structures and magnitudes of their respective beta functions:
\begin{enumerate}
 \item 
  The underlying flavor structures dictate entirely different mixing dynamics: in the MSSM, the active driver $P_{\rm MSSM} \propto \text{diag}(0, y_\mu^2, y_\tau^2)$ is  diagonal in the flavor basis. Consequently, the term $P^T M_\nu + M_\nu P$ in the beta function solely scales existing matrix elements multiplicatively ($d M_{ij} / dt \propto M_{ij}$); this preserves the existing flavor topology and remains subject to the kinematic suppression of the mass gaps $\Delta m_{31}^2$ and $\Delta m_{32}^2$. Conversely, the SMEFT insertion matrix $C_{H\ell}^{(3)}$ is dense. The additive matrix multiplication $C^T M_\nu + M_\nu C$ in \equ{rg_equation_smeft} breaks lepton flavor conservation, generating cross-talk between lepton generations that overcomes the atmospheric mass-splitting suppression and forces all mixing angles to run appreciably.
 \item
  While the MSSM driver is strictly real and relies entirely on the low-energy boundary condition of $\delta_{\rm CP}$ (manually set by us to the current global best fit) to evolve CP-violation effects, the $C_{H\ell}^{(3)}$ matrix introduces explicit imaginary components. These act as an independent source of CP violation, appreciably driving $\delta_{\rm CP}$ away from its low-energy value. 
 \item
  The bare magnitude of the SMEFT driver is significantly larger: even with a $\tan\beta = 10$ enhancement, the dominant MSSM coupling remains modest [$(y_\tau^{\rm eff})^2 \simeq 0.01$], whereas our tuned SMEFT parameters are $\mathcal{O}(0.1\text{--}1.0)$, providing enhancement that is orders-of-magnitude more powerful. 
 \item
  This structural disparity fundamentally alters the trajectory of the mass-squared differences, $\Delta m_{ij}^2$. In the MSSM, the absolute mass running is dominated by the flavor-universal scalar trace ($\alpha_{\rm SM} M_\nu$), which scales all mass eigenvalues uniformly and largely preserves the hierarchical mass splittings. In contrast, in the SMEFT framework, the large, non-universal diagonal elements of $C_{H\ell}^{(3)}$ (\eg, $C_{11} \neq C_{33}$) impose distinct, flavor-dependent shifts on the individual mass eigenvalues, causing $\Delta m_{ij}^2$ to splinter and diverge beyond their sub-TeV allowed ranges at high momenta.
\end{enumerate}

In what follows, when presenting RG running results explicitly, we will focus exclusively on the SMEFT, representative of a more general framework than MSSM.


\subsection{Momentum distribution accessible with high-energy astrophysical neutrinos}
\label{sec:rg_running-momentum_dist}

In the $\nu N$ DIS by which high-energy astrophysical neutrinos are detected (Sec.~\ref{sec:flavor_comp-measuring}), the transferred momentum can be significantly larger than in conventional oscillation experiments, \ie, $Q \approx $~TeV--PeV vs.~GeV, as stated in Sec.~\ref{sec:flavor_comp-osc_mod}. Such values could induce significant RG running of the mixing parameters, as in \figu{rge_running_comparison}.

In reality, however, we do not have experimental access to the value of the momentum transferred in the detection of high-energy astrophysical neutrinos in high-energy neutrino telescopes.  Instead, in any particular $\nu N$ DIS of a neutrino of energy $E_\nu$, the transferred momentum takes a random value $Q \leq \sqrt{2 m_N x y E_\nu}$, where $m_N \approx 1$~GeV is the nucleon mass, $0 \leq x \leq 1$ is the Bjorken scaling parameter, and $0 \leq y \leq 1$ is the inelasticity.  Because the values of $x$ and $y$ are random---though not uniformly distributed---so are the values of $Q$.

The top panel of \figu{main_results} (also Figs.~\ref{fig:flavor_smeft_rg_vs_Q} and \ref{fig:combined_pq}) shows the probability distributions of transferred momentum, $\mathcal{P}(Q)$, for high-energy and UHE astrophysical neutrinos.  We compute them by convolving the doubly differential DIS cross section, $d^2\sigma/dxdy$, and the neutrino energy spectrum, $\Phi_\nu$.  We compute the former using the CT18NNLO PDFs~\cite{Hou:2019efy}.  For the latter, we marginalize over energies using two distinct astrophysical flux models: a standard TeV--PeV power-law spectrum $\Phi_\nu \propto E_\nu^{-2.5}$, and a representative UHE flux prediction.  When computing $\mathcal{P}(Q)$, we weigh the contributions of NC and CC interactions of neutrinos and anti-neutrinos.  Appendix~\ref{app:momentum_distribution} contains full details of our calculation.

After building the $\mathcal{P}(Q)$ distributions, we extract from them the most likely values of $Q$ and the 68\% and 99\% highest-posterior-density containment intervals, which we report in Figs.~\ref{fig:main_results} and \ref{fig:rge_running_comparison} as representative of the probed range of $Q$ values.  Because the neutrino fluxes fall steeply with energy, the distributions are dominated by moderate values of $Q$ of tens of GeV, rather than by the TeV--PeV range at the tail of the distributions.  \textbf{\textit{Thus, the steeply falling energy distribution of high-energy astrophysical neutrinos dampens---but does not preclude---their potential to probe large RG running of the mixing parameters.}}

In practice, then, we are sensitive to the $Q$-averaged values of the mixing parameters [and, later, of the flavor fractions at Earth, \equ{flavor_ratios_q_avg}], which we compute as follows.  Given a flux of high-energy astrophysical neutrinos spanning the energy range $[E_\nu^{\rm min}, E_\nu^{\rm max}]$, its associated $Q$-distribution $\mathcal{P}(Q)$, and a specific RG running scheme with parameter values $\boldsymbol{\rho}$, the experimentally relevant mixing parameter for neutrino telescopes is
\begin{equation}
 \tilde{\theta}_{ij}
 (\boldsymbol{\theta}, \boldsymbol{\rho})
 =
 \int_0^{Q_{\rm max}}
 dQ~
 \theta_{ij}
 (\boldsymbol{\theta}, 
 \boldsymbol{\rho}, Q)
 \mathcal{P}(Q) \;,
 \label{equ:flavor_ratios_q_avg}
\end{equation}
where $Q_{\rm max} = \sqrt{2 E_\nu^{\rm max}}$ (after setting $x = y = 1$) is the maximum possible momentum accessible with this neutrino flux, and $\theta_{ij}$ in the integrand is obtained by evolving the RG equations to the scale $Q$.  We illustrate the behavior of these $Q$-averaged parameters below.


\subsection{The impact of each SMEFT coefficient}
\label{sec:rg_running-impact_smeft_coefficients}

\begin{figure*}[t]
 \centering
 \includegraphics[width=\textwidth]{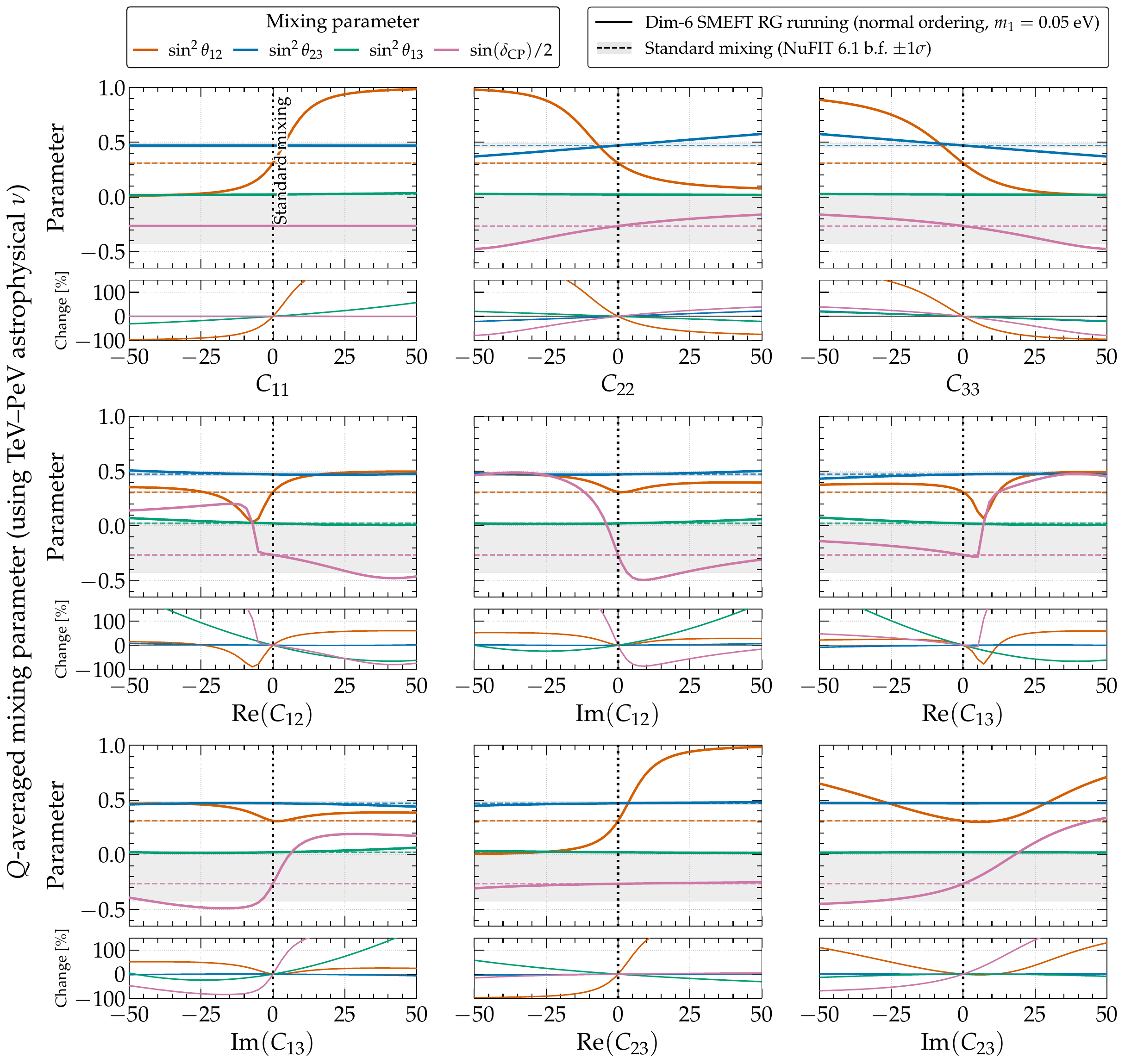}
 \caption{\textbf{ Evolution of the $Q$-averaged neutrino mixing parameters with the SMEFT couplings.}  The physical parameters are integrated over the momentum distribution accessible by a typical TeV--PeV astrophysical neutrino energy spectrum ($\propto E_\nu^{-2.5}$), with sub-panels detailing the relative change from standard mixing expectations.  The pronounced volatility of the solar sector ($\theta_{12}$, $\delta_{\rm CP}$), contrasted against the comparative rigidity of the reactor and atmospheric angles, illustrates how the Standard Model mass hierarchy intrinsically throttles anomalous flavor transitions.  In this figure, we fix $c_{\text{SMEFT}} = 1$ and $\Lambda_\text{SMEFT} = 1$~TeV.  See Sec.~\ref{sec:rg_running-impact_smeft_coefficients} for details.}
 \label{fig:mixing_params_rg_running_smeft}
\end{figure*}

Figure~\ref{fig:mixing_params_rg_running_smeft}  shows the $Q$-averaged values of the mixing parameters computed within the dimension-6 SMEFT scheme.  To isolate the effect of each of the SMEFT coefficients, $C_{ij}$ in \equ{wilson_coeff_matrix_general}, we vary one at a time, fixing the remaining coefficients to zero.  The response of the mixing parameters to the different SMEFT coefficients in \figu{mixing_params_rg_running_smeft} governs the constraints we infer on them later.

To understand the behavior in \figu{mixing_params_rg_running_smeft}, we resort to approximate analytical expressions of the parameter RG running. We derive them below (with the full explicit expansions provided in Appendix~\ref{app:rge_approximations}), including the explicit dependence on the mass hierarchy, leading-order operator expansions, and first-order $\mathcal{O}(s_{13})$ corrections.

\smallskip

\textbf{\textit{Analytical framework.---}}Taking a cue from our full SMEFT beta function, \equ{rg_equation_smeft}, we see that the non-SM RG evolution of the effective neutrino mass matrix $M$ under SMEFT potential is governed by
\begin{equation}
 \frac{dM}{dt} 
 =
 \kappa \left( C_{Hl}^{(3)} M + M (C_{Hl}^{(3)})^\dagger \right) \;,
 \label{equ:beta_function_smeft_simple}
\end{equation}
where $C_{Hl}^{(3)}$ is the Hermitian matrix of dimension-6 Wilson coefficients given in \equ{wilson_coeff_matrix_general}, and $\kappa = \frac{1}{16\pi^2} \frac{v^2}{\Lambda_\text{SMEFT}^2} c_{\text{SMEFT}}$.  

To extract approximate analytical expressions for the running of the mixing parameters, we project the source term on the right-hand side of \equ{beta_function_smeft_simple}, which is written in the flavor basis, into the mass basis by rotating it with the running mixing matrix $\tilde{U}$.  Defining the mass-basis SMEFT matrix as $\tilde{C}_{Hl}^{(3)} = \tilde{U}^\dagger C_{Hl}^{(3)} \tilde{U}$, we relate the off-diagonal elements of the mass matrix (which drive the RG running of the mixing angles) to the anti-Hermitian generator of the running mixing matrix, $T = \tilde{U}^\dagger \dot{\tilde{U}}$. 

Equating these terms yields the essential proportionality for the mixing evolution, 
\begin{equation}
 T_{ij}
 =
 \kappa \frac{m_i^2 + m_j^2}{\Delta m_{ij}^2} \tilde{C}_{ij} \;.
 \label{equ:Tij}
\end{equation}
The standard parametrization of $\tilde{U}$ maps the real  and imaginary components of $T$ to the derivatives of the mixing angles ($\dot{\theta}_{ij}$) and the CP-violation phase ($\dot{\delta}_{\text{CP}}$). 

\smallskip

\textbf{\textit{The impact of mass hierarchy.---}}The structure of $T_{ij}$ reveals that the running of any mixing parameter is inversely proportional to the mass-squared splittings. In the SM with normal mass ordering, the atmospheric splitting ($|\Delta m_{31}^2|$) is roughly thirty times larger than the solar splitting ($\Delta m_{21}^2$). Consequently, in analogy to the MSSM case, the 1-2 sector possesses a smaller denominator in $T_{ij}$, making $\theta_{12}$ about thirty times more sensitive to running than $\theta_{13}$ or $\theta_{23}$. 

To produce our numerical results below, we adopt the normal mass ordering and the illustrative value of $m_1 = 0.05$~eV, as before, placing our system in the hierarchical mass regime and inducing differentiated RG running of the different mixing parameters. However, were the lightest neutrino heavier (\ie, $m_1 \gg \sqrt{\Delta m_{ij}^2}$), the system enters a quasi-degenerate regime. Here, the numerator $(m_i^2 + m_j^2) \approx 2m_1^2$ in \equ{Tij} becomes large while the mass splittings remain fixed, universally amplifying the RG running across all angles and breaking the standard hierarchical suppression.  

\smallskip

\textbf{\textit{Coefficient-by-coefficient approximations.---}}By expanding $\tilde{C}_{ij}$ using the standard PMNS parametrization, we can isolate exactly how each flavor-basis operator $C_{ij}$ drives the mixing parameters. Below, we expand to leading order, ignoring corrections of $\mathcal{O}(s_{13})$ and smaller.
\begin{description}
 \item[Solar angle, $\theta_{12}$] 
  The evolution of the solar angle is driven by the real part of $T_{12}$,
  \begin{equation}
   \frac{d\theta_{12}}{dt} \approx \kappa \frac{m_1^2 + m_2^2}{\Delta m_{21}^2} \text{Re}(\tilde{C}_{12}) \;,
   \label{equ:th12_evol_gen}
  \end{equation}
  which, expanded to leading order, yields
  \begin{align}
   \dot{\theta}_{12} \approx \kappa \frac{m_1^2 + m_2^2}{\Delta m_{21}^2} \Bigg[ &\frac{1}{2}\sin 2\theta_{12} \left( C_{11} - s_{23}^2 C_{33} - c_{23}^2 C_{22} \right) \nonumber \\
   &+ \cos 2\theta_{12} \left( c_{23} \text{Re}(C_{12}) - s_{23} \text{Re}(C_{13}) \right) \nonumber \\
   &+ \frac{1}{2} \sin 2\theta_{12} \sin 2\theta_{23} \text{Re}(C_{23}) \Bigg] \;.
   \label{equ:th12_evol_gen}
  \end{align}
  The solar angle is uniquely vulnerable. Not only is it amplified by the small $\Delta m_{21}^2$ denominator, but the flavor-diagonal $C_{11}$ coefficient drives it at leading order without any suppression. This explains why the presence of a nonzero $C_{11}$ rapidly evolves $\theta_{12}$ and, therefore, disrupts the $\nu_\mu \to \nu_e$ transition.
 \item[Reactor angle, $\theta_{13}$]
  The evolution of the reactor angle couples to the 1-3 sector, \ie,
  \begin{equation}
   \frac{d\theta_{13}}{dt} \approx \kappa \frac{m_1^2 + m_3^2}{\Delta m_{31}^2} \text{Re}(\tilde{C}_{13} e^{-i\delta}) \;,
   \label{equ:th13_evol_gen}
  \end{equation}
  which, expanded to leading order, yields
  \begin{align}
   \dot{\theta}_{13} \approx \kappa \frac{m_1^2 + m_3^2}{\Delta m_{31}^2} \Bigg[ &\frac{1}{2} s_{12} \sin 2\theta_{23} \cos\delta_{\rm CP} (C_{33} - C_{22}) \nonumber \\
   &+ c_{12} s_{23} \text{Re}(C_{12} e^{-i\delta_{\rm CP}}) 
   \nonumber \\
   &+ c_{12} c_{23} \text{Re}(C_{13} e^{-i\delta_{\rm CP}}) \nonumber \\
   &- s_{12} \cos 2\theta_{23} \cos\delta_{\rm CP} \text{Re}(C_{23}) 
   \nonumber \\
   &- s_{12} \sin\delta_{\rm CP} \text{Im}(C_{23}) \Bigg] \;.
   \label{equ:th13_evol_gen}
  \end{align}
  The running of $\theta_{13}$ is heavily suppressed by the larger atmospheric mass splitting $\Delta m_{31}^2$. The coefficient $C_{11}$ no longer appears at leading order; it only enters as an $\mathcal{O}(s_{13})$ correction (not shown). Consequently, $\theta_{13}$ is primarily perturbed by the off-diagonal $C_{13}$ and $C_{12}$ coefficients, which are suppressed by multiple mixing angles, thus requiring massive off-diagonal SMEFT injections to produce observable deviations in its value (and, later, in the probabilities and flavor fractions at Earth).
 \item[Atmospheric angle, $\theta_{23}$]
  The evolution of the atmospheric angle couples to the 2-3 sector, \ie,
  \begin{equation}
   \frac{d\theta_{23}}{dt} \approx \kappa \frac{m_2^2 + m_3^2}{\Delta m_{32}^2} \text{Re}(\tilde{C}_{23}) \;,
   \label{equ:th23_evol_gen}
  \end{equation}
  which, expanded to leading order, yields
  \begin{align}
   \dot{\theta}_{23} \approx \kappa \frac{m_2^2 + m_3^2}{\Delta m_{32}^2} \Bigg[ &\frac{1}{2} c_{12} \sin 2\theta_{23} (C_{22} - C_{33}) \nonumber \\
   &+ c_{12} \cos 2\theta_{23} \text{Re}(C_{23}) \nonumber \\
   &+ s_{12} s_{23} \text{Re}(C_{12}) 
   \nonumber \\
   & + s_{12} c_{23} \text{Re}(C_{13}) \Bigg] \;.
   \label{equ:th23_evol_gen}
  \end{align}
  Similar to $\theta_{13}$, the atmospheric running is strongly suppressed by the heavy $\Delta m_{32}^2 \approx \Delta m_{31}^2$ denominator. At leading order, $\theta_{23}$ is driven by the $\mu$-$\tau$ diagonal asymmetry ($C_{22} - C_{33}$) and the explicit $\mu$-$\tau$ mixing coefficient $C_{23}$. Because standard mixing is nearly maximal ($\theta_{23} \approx 45^\circ$), the $\cos 2\theta_{23}$ factor suppresses the direct  impact of $C_{23}$.
 \item[CP-violation phase, $\delta_{\text{CP}}$]
  The evolution of the CP-violation phase is notoriously complex as it draws from the imaginary components of multiple sectors. Using the phase-invariant definition of $\delta_{\text{CP}}$, its running is approximately given by
  \begin{align}
   \frac{d\delta_{\text{CP}}}{dt} \approx\;& \frac{\text{Im}(T_{12})}{\sin\theta_{12}\cos\theta_{12}} 
   \nonumber \\
   &- \frac{\text{Im}(T_{13} e^{-i\delta})}{\sin\theta_{13}} + \frac{\text{Im}(T_{23})}{\sin\theta_{23}\cos\theta_{23}} \;,
   \label{equ:dcp_evol_gen}
  \end{align}
  which, expanded to leading order, yields
  \begin{widetext}
  \begin{align}
   \label{equ:dcp_rg_running_smeft}
   \dot{\delta}_{\rm CP} \approx &\frac{\kappa}{\sin 2\theta_{12}} \frac{m_1^2 + m_2^2}{\Delta m_{21}^2} \Big[ c_{23} \text{Im}(C_{12}) - s_{23} \text{Im}(C_{13}) \Big]  \\
   &- \frac{\kappa}{s_{13}} \frac{m_1^2 + m_3^2}{\Delta m_{31}^2} \Bigg\{ \cos\delta_{\rm CP} \Big[ c_{12} s_{23} \text{Im}(C_{12}) + c_{12} c_{23} \text{Im}(C_{13}) - s_{12} \text{Im}(C_{23}) \Big] \nonumber \\
   &\qquad\qquad\qquad - \sin\delta_{\rm CP} \Big[ \frac{1}{2} s_{12} \sin 2\theta_{23} (C_{33} - C_{22}) + c_{12} s_{23} \text{Re}(C_{12}) + c_{12} c_{23} \text{Re}(C_{13}) - s_{12} \cos 2\theta_{23} \text{Re}(C_{23}) \Big] \Bigg\} \nonumber \;.
  \end{align}
  \end{widetext}
  The running of $\delta_{\text{CP}}$ is a competition between two distinct enhancements. The first term on the right-hand side of \equ{dcp_rg_running_smeft} is amplified by the small solar mass splitting ($\Delta m_{21}^2$), making it highly sensitive to the imaginary parts of $C_{12}$. The second term, despite being suppressed by the heavy atmospheric splitting, is inversely proportional to the small reactor angle ($s_{13}^{-1}$), yielding an equally large amplification for $\text{Im}(C_{13})$. However, because the high-energy astrophysical neutrinos fully decohere over macroscopic baselines, the transition probabilities are inherently blind to these CP-violating interference terms (see Figs.~\ref{fig:probabilities} and \ref{fig:flavor_ratios_pion} later).  Any remaining sensitivity stems from sub-dominant $\delta_{\rm CP}$ dependence that survives the decoherence (see, \eg, Eq.~(3) in \Refe~\cite{Song:2020nfh}).
\end{description}


\subsection{The SMEFT RG trajectory in flavor space}
\label{sec:rg_running-trajectory_flavor_space}

\begin{figure}[t!]
 \centering
 \includegraphics[width=\columnwidth]{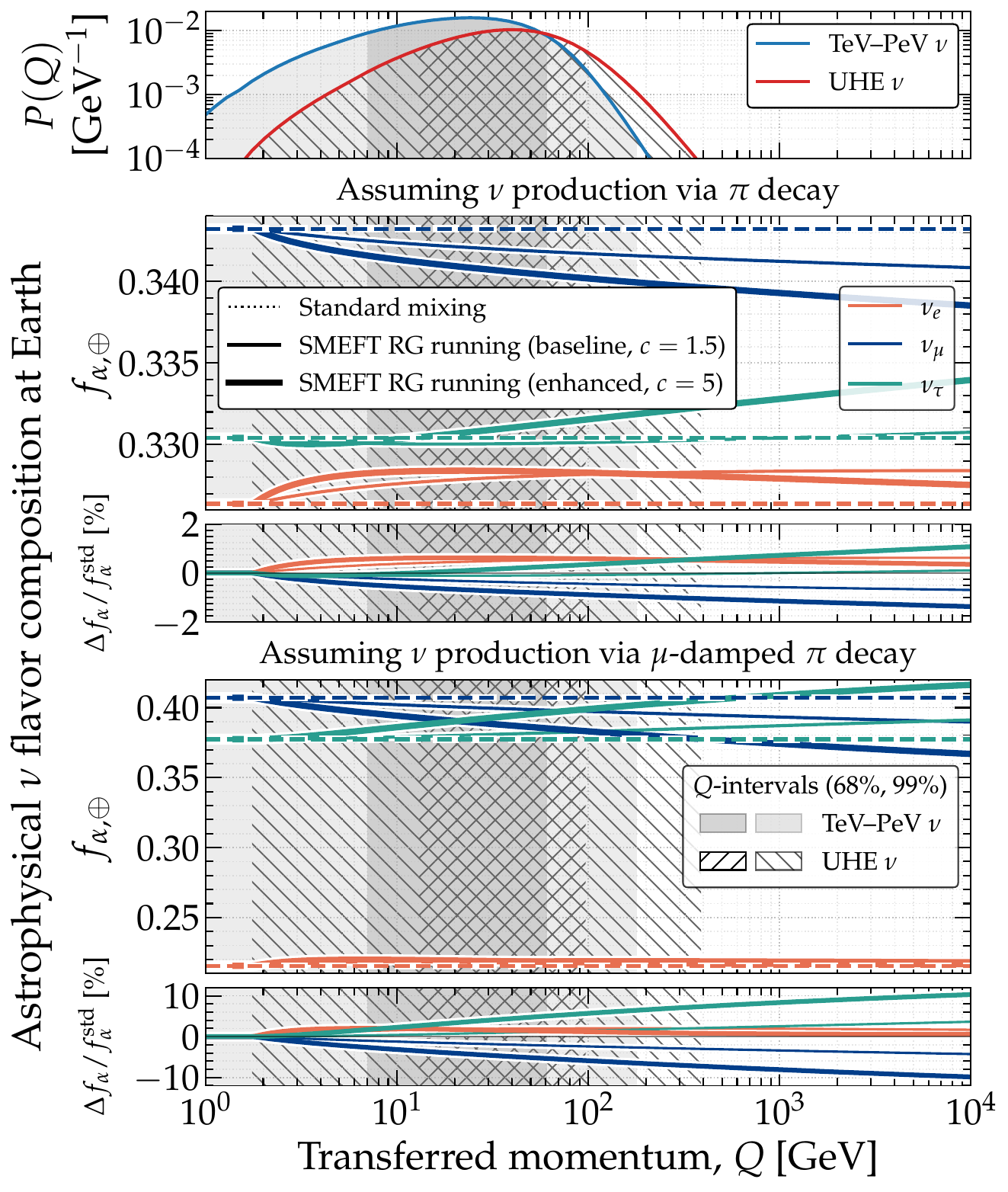}
 \caption{\textbf{Example renormalization-group running of the flavor composition at Earth.} The RG evolution of the flavor fractions $f_{\alpha, \oplus}$ and their relative deviations from standard oscillations ($\Delta f_{\alpha}/f_{\alpha}^{\rm std}$) are shown as a function of the momentum transfer, $Q$, under our example SMEFT scheme (Sec.~\ref{sec:rg_running-smeft}). \textit{Top panel:} Probability distribution of momentum transfer in $\nu N$ DIS for TeV--PeV and UHE astrophysical neutrinos, indicating their respective 68\% and 99\% containment intervals.  \textit{Lower panels:} RG running assuming neutrino production via full and muon-damped pion decay, evaluated with baseline ($c_{\rm SMEFT}=1.5$ in \equ{wilson_coeff_matrix}) and enhanced ($c_{\rm SMEFT}=5.0$) benchmark SMEFT couplings.}
 \label{fig:flavor_smeft_rg_vs_Q}
\end{figure}

\begin{figure}[t!]
 \centering
 \includegraphics[width=\columnwidth]{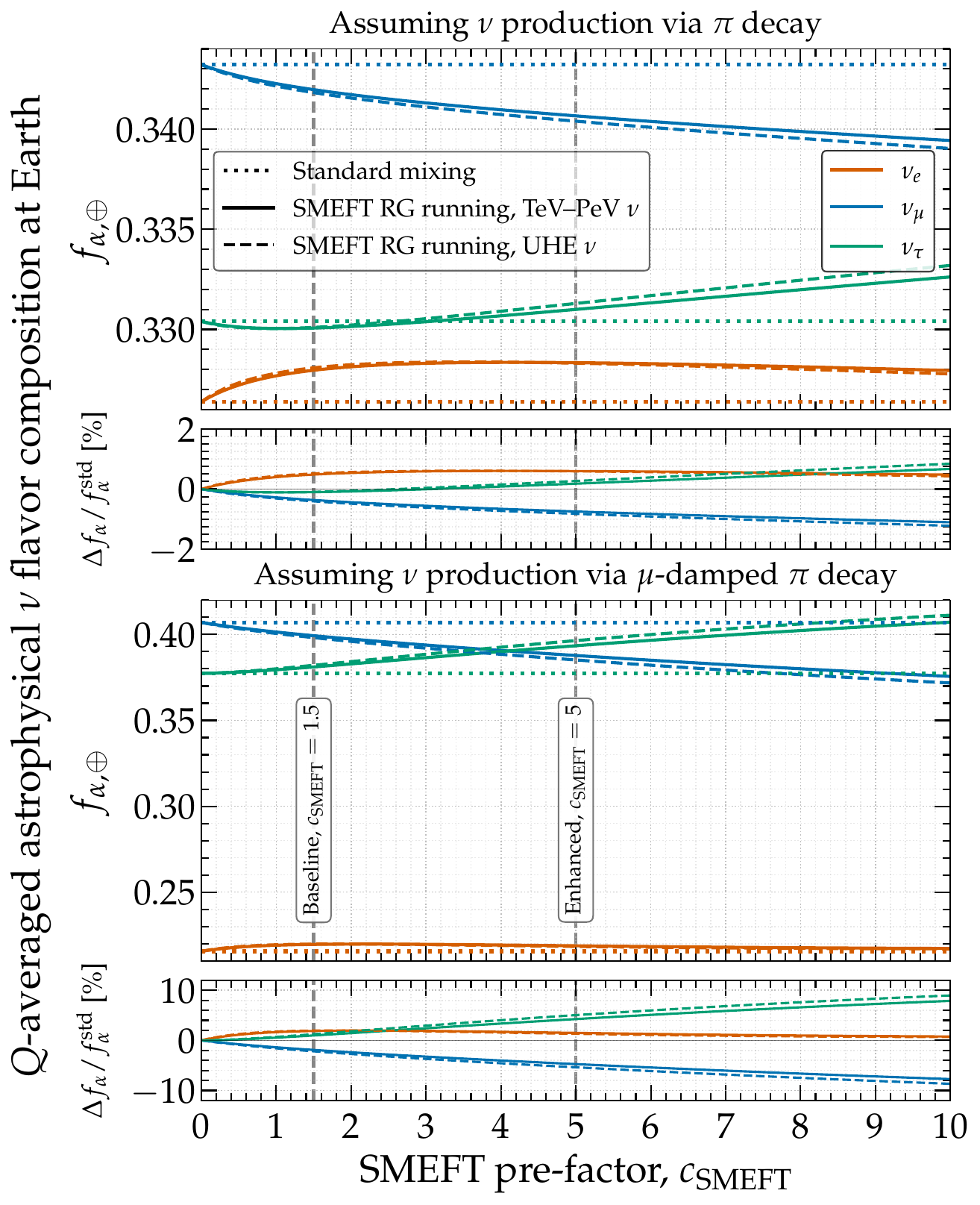}
 \caption{\textbf{Example renormalization-group running of the $Q$-averaged flavor composition at Earth.}  The evolution of the flavor fractions $\tilde{f}_{\alpha, \oplus}$ (\equ{flavor_ratios_q_avg}) and their relative deviations from standard oscillations ($\Delta \tilde{f}_{\alpha} / f_{\alpha}^{\rm std}$) are shown as a function of the prefactor $c_{\rm SMEFT}$ of our example SMEFT scheme  (see Sec.~\ref{sec:rg_running-smeft}).  Neutrino production is assumed to occur via full pion decay (\textit{top two panels}) or muon-damped pion decay (\textit{bottom two panels}). The $Q$-averaged SMEFT RG running is shown for two assumed astrophysical neutrino fluxes: a TeV--PeV flux and a UHE flux. Vertical dashed lines highlight our baseline ($1.5$) and enhanced ($5.0$) $c_{\rm SMEFT}$ benchmarks.  See \figu{flavor_smeft_rg_vs_Q} for the explicit running with $Q$.}
 \label{fig:flavor_smeft_rg_vs_c_smeft}
\end{figure}

Figure~\ref{fig:flavor_smeft_rg_vs_Q} shows the trajectory of the flavor composition at Earth with changing $Q$ under the RG running of the mixing parameters computed using our example SMEFT framework from Figs.~\ref{fig:main_results} and \ref{fig:rge_running_comparison}.  We show the trajectories for two choices of flavor composition at the sources---full pion decay, $\left( \frac{1}{3}, \frac{2}{3}, 0 \right)_{\rm S}$, and muon-damped pion decay, $(0, 1, 0)_{\rm S}$---and for a baseline value of $c_{\rm SMEFT} = 1.5$ from \equ{wilson_coeff_matrix} and an enhanced value of $c_{\rm SMEFT} = 5$.  While the parameter modifications grow with $Q$, the relative size of the modifications remains small to moderate: less than 2\% and 10\% assuming production via full and muon-damped pion decay, respectively.

Because we do not have access to the $Q$ values of individual neutrino detection events, in analogy to the $Q$-averaged mixing parameters (Sec.~\ref{sec:rg_running-impact_smeft_coefficients}), we compute the $Q$-averaged flavor composition,
\begin{equation}
 \tilde{f}_{\alpha, \oplus}
 (\boldsymbol{\theta}, \boldsymbol{\rho})
 =
 \int_0^{Q_{\rm max}}
 dQ~
 f_{\alpha, \oplus}
 [\boldsymbol{\theta}, 
 \boldsymbol{\theta}^\prime(\boldsymbol{\rho}, Q)]
 \mathcal{P}(Q) \;,
 \label{equ:flavor_ratios_q_avg}
\end{equation}
where, as before, $Q_{\rm max} = \sqrt{2 E_\nu^{\rm max}}$ is the maximum possible momentum accessible with this neutrino flux, and $f_{\alpha, \oplus}$ in the integrand is defined in \equ{flavor_ratios}.  

Figure~\ref{fig:flavor_smeft_rg_vs_c_smeft} shows the trajectory of the $Q$-averaged flavor composition---the experimentally accessible quantity---computed within our example SMEFT framework and obtained by varying the prefactor $c_{\rm SMEFT}$ of the Wilson-coefficient matrix, $C_{H\ell}^{(3)}$ in \equ{wilson_coeff_matrix}, while keeping its components fixed.  
The behavior of the $Q$-averaged flavor composition reveals
the experimental challenge of detecting RG running in high-energy astrophysical neutrinos: the flavor shifts are dominated by small $Q$ values, given their steeply falling neutrino energy spectra

Even when using the enhanced value of $c_{\rm SMEFT} = 5$, the relative size of the RG modifications of $\tilde{f}_{\alpha, \oplus}$ is moderate at best: about 1\% and 8\% for production via full and muon-damped pion decay, respectively.  While the exact size and shape of the RG modifications depend on our choices of $c_{\rm SMEFT}$ and on the structure of the SMEFT coefficient matrix, $C_{H\ell}^{(3)}$, the RG modifications of $\tilde{f}_{\alpha, \oplus}$ are small regardless of our specific choices.  \textbf{\textit{This places the detection of the SMEFT RG modifications out of reach of present flavor-composition measurements. However, we show later that it is within reach of our multi-detector flavor measurement projections.}}

\begin{figure*}[t!]
 \centering
 \includegraphics[width=0.497\textwidth]{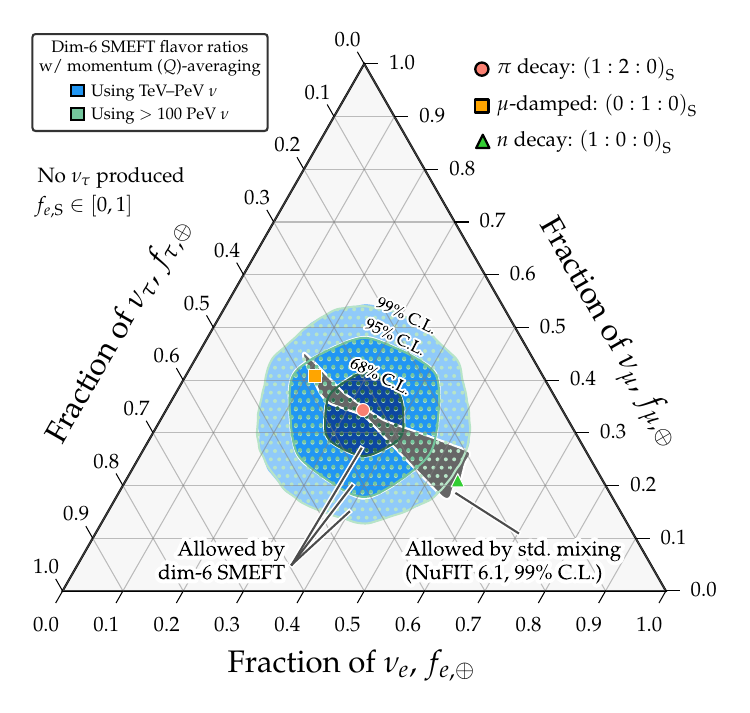}
 \includegraphics[width=0.497\textwidth]{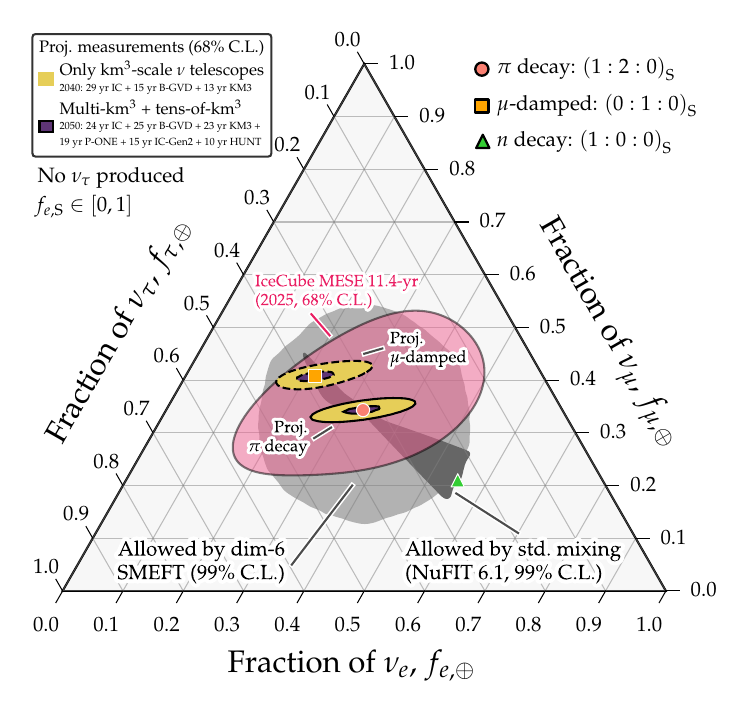}
 \caption{\textbf{Standard and dimension-6-SMEFT allowed regions of flavor composition at Earth.} Similar to \figu{ternary_theory_smeft}, but the RG-regions of flavor composition are now generated by varying simultaneously all of the $C_{ij}$ SMEFT coefficients in \equ{wilson_coeff_matrix_general} (with fixed $c_{\rm SMEFT} = 1$ and $\Lambda_{\rm SMEFT} = 1$~TeV).  The standard mixing parameters are varied within their projected 2040 allowed ranges~\cite{Song:2020nfh}.  The flavor composition at the sources $f_{e,S} \in [0,1]$, assuming $f_{\tau,S} = 0$.  \textit{Left:} The RG-modified allowed regions at 68\%, 95\%, and 99\% C.L.~regions, shown for two choices of $Q$-averaging: using the $Q$-distribution from TeV--PeV neutrinos and from UHE neutrinos [see Figs.~\ref{fig:main_results}, \ref{fig:flavor_smeft_rg_vs_Q}, and \ref{fig:combined_pq}]. \textit{Right:} The 99\% C.L. RG-allowed theoretical space juxtaposed with current and projected experimental sensitivities. Current 68\% C.L. constraints from IceCube (11.4-yr MESE) are shown in gray. Projected 68\% C.L. contours for 2040 (combining IceCube, Baikal-GVD, and KM3NeT) and 2050 (adding IceCube-Gen2, P-ONE, and HUNT) are shown for benchmark pion-decay and muon-damped source scenarios.}
 \label{fig:ternary_theory_smeft}
\end{figure*}

\begin{figure}[t!]
 \centering
 \includegraphics[width=0.497\textwidth]{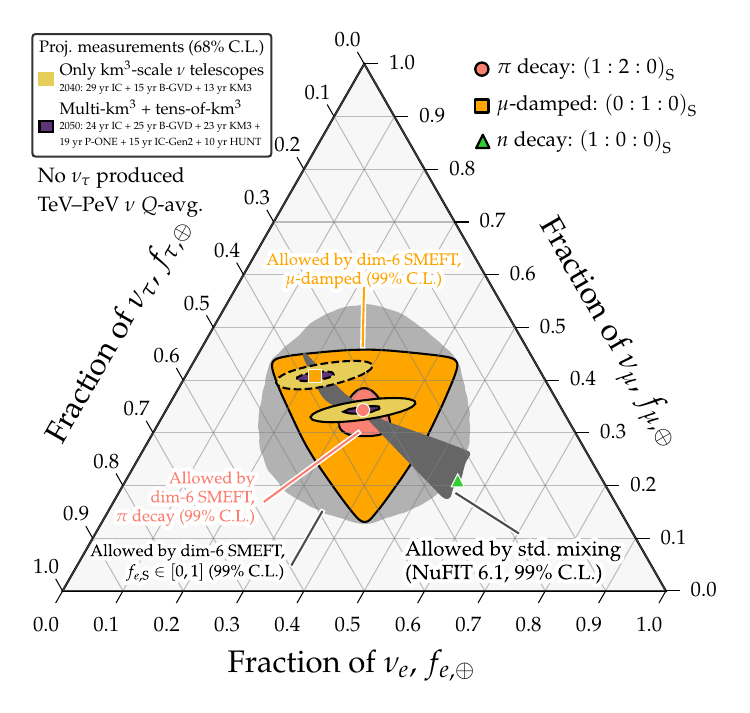}
 \caption{\textbf{Effect of the flavor composition at the sources on the SMEFT RG running.}  Similar to \figu{ternary_theory_smeft}, but now showing three separate scenarios of the flavor composition at the sources (all assuming no $\nu_\tau$ production): varying $f_{e, {\rm S}} \in [0,1]$, fixing $f_{e, {\rm S}} = 1/3$ from neutrino production from full pion decay, and fixing $f_{e, {\rm S}} = 0$ from neutrino production from muon-damped pion decay.  In each case, the RG-regions of flavor composition are generated by varying simultaneously all of the $C_{ij}$ SMEFT coefficients in \equ{wilson_coeff_matrix_general} (with fixed $c_{\rm SMEFT} = 1$ and $\Lambda_{\rm SMEFT} = 1$~TeV). Projected 68\% C.L. contours for 2040 (combining IceCube, Baikal-GVD, and KM3NeT) and 2050 (adding IceCube-Gen2, P-ONE, and HUNT) are shown for benchmark pion-decay and muon-damped source scenarios.}
 \label{fig:ternary_smeft_pion_vs_muon}
\end{figure}

Figure~\ref{fig:ternary_theory_smeft} generalizes the effect of SMEFT RG running by exploring how changes to the texture of the $C_{H\ell}^{(3)}$ SMEFT coupling matrix affect the allowed regions of $Q$-averaged flavor composition at Earth. To generate these regions, we vary simultaneously all of the $C_{ij}$ SMEFT coefficients in \equ{wilson_coeff_matrix_general} (with fixed $c_{\rm SMEFT} = 1$ and $\Lambda_{\rm SMEFT} = 1$~TeV), evolve the RG equation of the mixing parameters [\equ{rg_equation_smeft}], and compute the $Q$-averaged flavor ratios via \equ{flavor_ratios_q_avg}.

The left panel of \figu{ternary_theory_smeft} shows that choosing between $Q$-averaging using the $\mathcal{P}(Q)$ distribution from TeV--PeV or UHE neutrinos has negligible effect on the allowed SMEFT flavor regions, since the distributions are so similar (see Figs.~\ref{fig:main_results}, \ref{fig:flavor_smeft_rg_vs_Q}, and \ref{fig:combined_pq}).  In spite of UHE neutrinos having energies orders-of-magnitude larger than TeV--PeV neutrinos, the energy spectra of both fall steeply with energy, which makes their $Q$-distributions dominated by momenta in the 10--100~GeV range (see Sec.~\ref{sec:rg_running-momentum_dist} and  Appendix~\ref{app:momentum_distribution}).  This reveals that \textbf{\textit{future measurements of the UHE flavor composition~\cite{Testagrossa:2023ukh, Coleman:2024scd} will offer only a small gain when testing RG running of the mixing parameters.}}  

Figure~\ref{fig:ternary_smeft_pion_vs_muon} shows the effect on the SMEFT flavor regions of choosing a specific flavor composition at the sources from among our two benchmark choices: full ($f_{e, {\rm S}} = 1/3$) or muon-damped pion decay ($f_{e, {\rm S}} = 0$).  The relative size of these regions confirms our earlier findings from Figs.~\ref{fig:flavor_smeft_rg_vs_Q} and \ref{fig:flavor_smeft_rg_vs_c_smeft}: that the effect of RG running on the flavor composition is limited for full pion decay---remaining close to its standard-mixing expectation---and larger for muon-damped pion decay, for the same reasons presented in Sec.~\ref{sec:flavor_ratios-impact_sources}.  In fact, \figu{ternary_smeft_pion_vs_muon} reveals that the region generated assuming muon-damped pion decay spans nearly the entirety of the region generated by varying $f_{e, {\rm S}} \in [0,1]$.  This illustrates why, in our SMEFT constraints below, we find null sensitivity when assuming a projected flavor measurement centered on the standard-mixing expectation from full pion decay and positive sensitivity when assuming one centered on muon-damped pion decay.


\subsection{SMEFT limitations and approximations}

The SMEFT RG running in our analysis, while sufficiently general to capture the predominant physical features, relies on a few well-motivated approximations:
\begin{itemize}
 \item
  \textbf{Frozen SM couplings:} The gauge couplings ($g_1, g_2$), top Yukawa ($y_t$), and Higgs self-coupling ($\lambda$) are treated as constant over our integration window (0.1--100~GeV). The running of these parameters induces only sub-leading, second-order corrections to the mixing angles (\eg, \Refe~\cite{Antusch:2003kp}), which are negligible compared to the substantial uncertainties of current astrophysical measurements.
 \item
  \textbf{Kinematic decoupling in the $\overline{\text{MS}}$ scheme:} Our numerical integration employs the mass-independent modified minimal subtraction ($\overline{\text{MS}}$) renormalization scheme. In it, heavy particles (\eg, the tau lepton) are instantaneously integrated out via a step function at their mass threshold ($Q = m_\tau$), producing an abrupt ``kink'' in the RGE slopes rather than a smooth kinematic transition. While a formally exact EFT treatment requires finite one-loop threshold matching to smooth these boundaries, the unmatched step-function approximation is standard practice and sufficient for demonstrating appreciable running across wide $Q$ ranges.
 \item
  \textbf{Ultraviolet (UV) completion:} The dimension-6 Wilson coefficients are inserted as static spurions. While a complete one-loop SMEFT calculation would couple the running of $C_{H\ell}^{(3)}$ to the overall integration, a frozen insertion provides a proxy for the full physical mechanism. Physically, the effective dimensionless matrix used in our RGEs is related to the bare Wilson coefficients of the underlying high-energy theory via the standard EFT expansion, $C_{H\ell}^{(3)} \simeq C_{\text{bare}} (v^2 / \Lambda_{\rm SMEFT}^2)$. Thus, generating effective couplings of $\mathcal{O}(0.1)$ at low new-physics scales ($\Lambda_{\rm SMEFT} \gg 1$~TeV), like we have tacitly done, implies bare Wilson coefficients that challenge standard perturbativity limits ($C_{\text{bare}} \gtrsim 4\pi$). Consequently, our chosen SMEFT matrix should be interpreted merely as a phenomenological proof of principle: it serves to demonstrate how the flavor-violating tensor structure of dimension-6 operators can bridge the low- and high-$Q$ scales, rather than representing a strictly perturbative UV-complete model (which would require, \eg, introducing new $Z^\prime$ bosons or leptoquarks).
 \item
  \textbf{Electroweak matching:} We assume continuous running across the electroweak scale ($M_W \approx 80.4$~GeV). A rigorous treatment would entail matching SMEFT to the low-energy effective field theory by integrating out the $W^\pm$ and $Z$ bosons and the top quark, generating finite boundary shifts in the effective mass matrix. However, neglecting these sub-percent precision threshold corrections is well-justified, as our primary objective is to establish the phenomenology of RG mixing in high-energy astrophysical neutrinos, whose associated experimental errors are significantly larger.
\end{itemize}


\section{Statistical procedure}
\label{sec:statistics}

\textit{We establish the frequentist profile-likelihood procedure we use to extract constraints on the generic high-$Q$ mixing parameters and the SMEFT coefficients. Because simultaneous variations of all parameters result in destructive interference and degeneracies, meaningful experimental bounds must be derived by varying a single parameter or coefficient at a time. Crucially, we incorporate the unknown initial electron neutrino fraction ($f_{e, {\rm S}}$) as a nuisance parameter, ensuring that our resulting limits robustly reflect our underlying astrophysical ignorance, rather than artificially benefiting from assuming a perfectly known flavor composition at the sources.}

\medskip

To assess the sensitivity to the high-$Q$ mixing parameters and the SMEFT coefficients, we assume that the measured flavor composition at Earth is centered on its standard-oscillation expectations from neutrino production via full or muon-damped pion decay (\figu{ternary_theory}, right panel).
We employ a frequentist profile-likelihood methodology, generalizing the standard procedure used to extract standard mixing parameters from high-energy flavor measurements introduced in \Refe~\cite{Bustamante:2026aur}.

As stated in Sec.~\ref{sec:flavor_comp-measuring}, we base our present-day parameter constraints on the IceCube 11.4-year MESE flavor composition~\cite{Abbasi:2025fjc}.  We base our projected parameter constraints on our simulated multi-detector measurements of flavor composition combining HESE and through-going muons.  Since these simulated measurements are produced assuming flavor-composition expectations from standard mixing (\ie, contours centered on the full or muon-damped pion-decay expectations), \textbf{\textit{our projected  constraints represent limits on the parameter values, rather than discovery prospects}}.


\subsection{Constraints on high-$Q$ mixing parameters}
\label{sec:statistics-mixing_params}

To assess the sensitivity to the standard and high-$Q$ mixing parameters, $\boldsymbol{\theta}$ and $\boldsymbol{\theta}^\prime$, we use the $\chi^2$ function
\begin{equation}
 \chi_{\rm total}^2 
 (\boldsymbol{\theta}, \boldsymbol{\theta}^\prime, f_{e, {\rm S}}) 
 = 
 \chi_{\rm data}^2 (\boldsymbol{\theta}, \boldsymbol{\theta}^\prime, f_{e, {\rm S}})
 +
 \chi_{\rm prior}^2 (\boldsymbol{\theta}) 
 \;,
\end{equation}
where the comparison of our flavor-composition predictions against measurements is performed via the term
\begin{equation}
 \chi_{\rm data}^2
 (\boldsymbol{\theta}, \boldsymbol{\theta}^\prime, f_{e, {\rm S}}) 
 =
 -2 \ln \mathcal{L} (f_{\alpha, \oplus}(\boldsymbol{\theta}, \boldsymbol{\theta}^\prime, f_{e, {\rm S}})) 
 \;.
\end{equation}
Here, $\mathcal{L}$ is the aforementioned experimental likelihood of flavor-composition measurements, present or future (shown in the Supp.~Mat.~of \Refe~\cite{Bustamante:2026aur}).  The term $\chi_{\rm prior}^2$ represents penalty terms (\textit{pulls}) on the standard mixing parameters coming from fits to conventional sub-TeV oscillation measurements.

For present results, the pull terms are $\chi^2$ functions from the NuFIT~6.1~\cite{Esteban:2024eli} global oscillation fit (assuming normal mass ordering with Super-Kamiokande data; other choices change results negligibly). We account for correlations ($\theta_{12}$ vs.~$\theta_{13}$ and $\theta_{23}$ vs.~$\delta_{\rm CP}$) by using pairwise $\chi^2$ NuFIT functions.  For projections, the pull terms are Gaussians centered on NuFIT~6.1 best fits with narrower widths from \Refe~\cite{Song:2020nfh}.  In present and projected results, we let $f_{e, {\rm S}}$ float unconstrained in $[0,1]$, reflecting the large uncertainty in neutrino production.  

A simultaneous, all-parameter fit lacks the sensitivity required to break the severe degeneracies between the high-$Q$ parameters and between them and $f_{e, {\rm S}}$. As detailed in Appendix~\ref{app:smeft_regions_single_parameter}, when all parameters vary simultaneously, destructive interference washes out the extreme flavor topologies, rendering the allowed regions indistinguishable from standard mixing. 

To overcome this and establish meaningful bounds, we make the deliberate methodological decision to exclusively \textbf{\textit{constrain a single high-$Q$ parameter at a time}}, $\eta$ (one of $\theta_{12}^\prime$, $\theta_{23}^\prime$, $\theta_{13}^\prime$, or $\delta_{\rm CP}^\prime$). We achieve this by pinning the remaining high-$Q$ mixing parameters to their standard, low-$Q$ values, and profiling over all nuisance parameters, $\nu$, \ie, by computing
\begin{equation}
 \Delta \chi^2(\eta)
 =
 \min_{\nu} [\chi_{\rm total}^2(\eta, \nu)] 
 - 
 \chi^2_\text{global~min} \;.
\end{equation}
For instance, when constraining $\theta_{12}^\prime$, we pin $\theta_{23}^\prime = \theta_{23}$ and $\theta_{13}^\prime = \theta_{13}$, profile the likelihood over $\theta_{12}$, $\theta_{23}$, $\theta_{13}$, $\delta_{\rm CP}$, and $f_{e, {\rm S}}$, and report results on $\theta_{12}^\prime$. When reporting our results, we use Wilks' theorem~\cite{Wilks:1938dza} (valid asymptotically given the large event statistics of our projections, with inferred  constraints yielding conservative over-coverage) to find $1\sigma$, $2\sigma$, and $3\sigma$ confidence intervals of the parameters by demanding $\Delta\chi^2 \leq 1$, 4, and 9, respectively.

Counterintuitively, these targeted single-parameter contours can extend further out than the all-parameter flavor contours in \figu{ternary_theory} (for a fixed C.L.). This is because the all-parameter scans heavily weight the multidimensional probability density of the flavor ratios toward the highly-mixed center of the flavor triangle, statistically starving the boundaries of the allowed flavor region. This centralization occurs because broadly sampling the high-$Q$ mixing parameters effectively randomizes the high-$Q$ mixing matrix $U^\prime$, causing its squared elements to naturally average to $\langle|U^\prime_{\alpha i}|^2\rangle \approx 1/3$. Consequently, the flavor-transition probabilities $P_{\alpha\beta} = \sum_i |U_{\alpha i}|^2 |U^\prime_{\beta i}|^2$ [\equ{prob_modified}] factorize into $\frac{1}{3} \sum_i |U_{\alpha i}|^2$. At this point, the values of the low-$Q$ mixing matrix drop out entirely due to its own unitarity ($\sum_i |U_{\alpha i}|^2 = 1$), forcing the flavor composition at Earth closer to the equal-flavor mixture, $\left( \frac{1}{3}, \frac{1}{3}, \frac{1}{3} \right)_\oplus$, regardless of the flavor composition at the sources.  (However, rather than collapsing to a single point, the finite dimensionality of the parameter space ensures that statistical fluctuations maintain a non-zero variance around this isotropic mean, spreading the 99\%~C.L. probability mass over the finite area shown in \figu{ternary_theory}.)

In Appendix~\ref{app:pairwise_constraints}, we generalize the above statistical procedure to report two-dimensional profiled likelihoods, in order to highlight any underlying experimental correlation between measured high-$Q$ parameters.  In Appendix~\ref{app:results_with_nu_tau_production}, we generalize our one-dimensional procedure  to allow for $\nu_\tau$ production at the sources, effectively adding $f_{\mu, {\rm S}}$ to the list of nuisance parameters (after which the $\nu_\tau$ fraction becomes $f_{\tau, {\rm S}} = 1-f_{e, {\rm S}}-f_{\mu, {\rm S}}$).  


\subsection{Constraints on SMEFT couplings}
\label{sec:statistics-smeft}

In addition to constraining the high-$Q$ mixing parameters, we assess directly the sensitivity to the dimension-6 SMEFT coefficients,
\begin{eqnarray}
 \boldsymbol{\rho} 
 &=&
 \left\{ C_{11}, \text{Re}(C_{12}), \text{Im}(C_{12}), \text{Re}(C_{13}), \text{Im}(C_{13}), \right. 
 \nonumber \\
 && \left. C_{22}, \text{Re}(C_{23}), \text{Im}(C_{23}), C_{33} \right\} \;.
\end{eqnarray}
In analogy to Sec.~\ref{sec:statistics-mixing_params}, we use the $\chi^2$ function
\begin{equation}
 \chi_{\rm total}^2 
 (\boldsymbol{\theta}, \boldsymbol{\rho}, f_{e, {\rm S}}) 
 = 
 \chi_{\rm data}^2 (\boldsymbol{\theta}, \boldsymbol{\rho}, f_{e, {\rm S}})
 +
 \chi_{\rm prior}^2 (\boldsymbol{\theta}) 
 \;,
\end{equation}
where the comparison of our flavor-composition predictions against measurements is performed via the term
\begin{equation}
 \chi_{\rm data}^2
 (\boldsymbol{\theta}, \boldsymbol{\rho}, f_{e, {\rm S}}) 
 =
 -2 \ln \mathcal{L} (\tilde{f}_{\alpha, \oplus}(\boldsymbol{\theta}, \boldsymbol{\rho}, f_{e, {\rm S}})) 
 \;,
\end{equation}
and $\tilde{f}_{\alpha, \oplus}$ is the $Q$-averaged flavor composition at Earth from \equ{flavor_ratios_q_avg}.  The likelihood of flavor-composition measurements, $\mathcal{L}$, and the pull term on the standard mixing parameters, $\chi_{\rm prior}^2$, are the same as in Sec.~\ref{sec:statistics-mixing_params}.

As with the high-$Q$ mixing parameters in Sec.~\ref{sec:statistics-mixing_params}, we find no sensitivity to constrain all SMEFT coefficients simultaneously.  Thus, we instead constrain a single coefficient at a time (one of $C_{11}$, $\text{Re}(C_{12})$, $\text{Im}(C_{12})$, $\text{Re}(C_{13})$, $\text{Im}(C_{13})$, $C_{22}$, $\text{Re}(C_{23})$, $\text{Im}(C_{23})$, or $C_{33}$), by setting all other coefficients to zero and profiling over all remaining nuisance parameters (standard mixing parameters and possibly $f_{e, {\rm S}}$).  Using Wilks' theorem, we report one-dimensional allowed confidence intervals for each coefficient, and two-dimensional profiled likelihoods of each coefficient vs.~$f_{e, {\rm S}}$.

As in the case of the high-$Q$ mixing parameters in Sec.~\ref{sec:statistics-mixing_params}, this single-parameter profiling of the SMEFT coefficients yields allowed flavor regions that extend significantly further out than those obtained by varying all SMEFT coefficients simultaneously, which are detailed in Appendix~\ref{app:smeft_regions_single_parameter}. Notably, this contrast between the single-parameter and all-parameter contours is much more pronounced for the SMEFT coefficients than for the generic high-$Q$ mixing parameters. 

This amplification stems from two compounding effects. First, the higher dimensionality of the SMEFT parameter space (nine coefficients versus four high-$Q$ mixing parameters) exacerbates the statistical concentration of measure toward the center of the flavor triangle, rendering the fine-tuned parameter combinations required to reach extreme flavor configurations exceedingly rare in a global scan. Second, unlike the purely trigonometric mapping of the generic high-$Q$ scenario, the SMEFT flavor evolution is governed by non-linear RG equations. Simultaneously activating all SMEFT operators introduces competing effective potentials that destructively interfere, actively scrambling the eigenvectors of the effective neutrino mass matrix during its high-$Q$ evolution and driving the system toward isotropic flavor mixing. By isolating a single SMEFT coefficient, we eliminate both the multidimensional probability dilution and the dynamical interference, allowing the resulting contours to accurately trace the reach of each individual operator, from which our constraining power originates.


\section{Results}
\label{sec:results}

\begin{figure}[t!]
 \centering
 \includegraphics[width=\columnwidth]{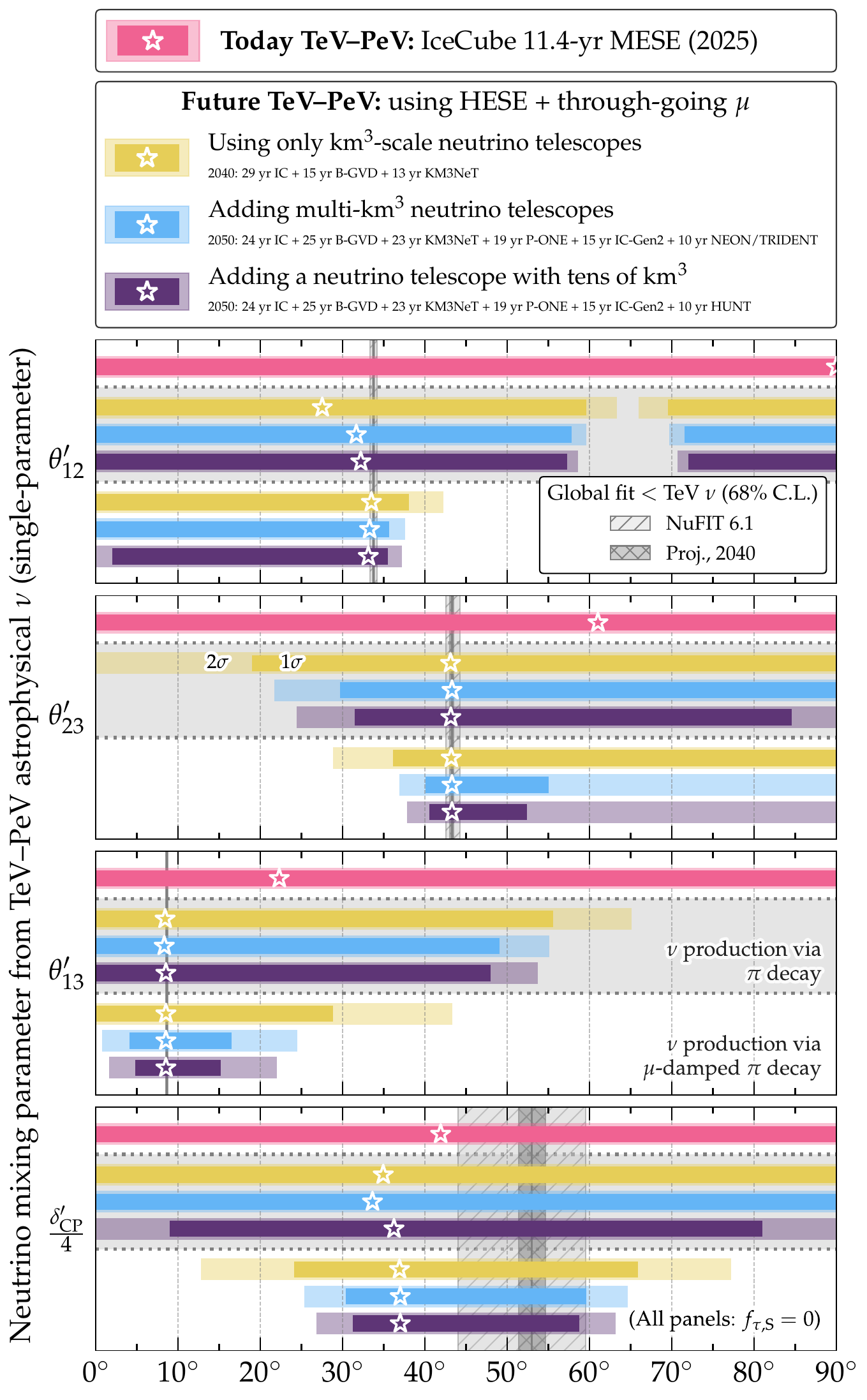}
 \vspace*{-0.5cm}
 \caption{\textbf{Constraints on the high-$Q$ neutrino mixing parameters from TeV--PeV astrophysical neutrinos.}  Present constraints from the 11.4-yr IceCube MESE sample~\cite{Abbasi:2025fjc} are consistent with three-flavor mixing but cannot constrain individual parameters. Projected measurements from multi-telescope observations show  sensitivity to $\theta_{23}^\prime$ and $\theta_{13}^\prime$ (and $\delta_{\rm CP}^\prime$ under neutrino production via muon-damped pion decay). For comparison, we show current~\cite{Esteban:2024eli} and projected~\cite{Song:2020nfh} sub-TeV global-fit ranges. \textit{These results represent the first rigorous assessment of mixing parameter sensitivity at a representative $Q \approx 20$~GeV (see \figu{main_results}).}}
 \label{fig:results_tev_pev}
\end{figure}

\begin{figure}[t!]
 \centering
 \includegraphics[width=\columnwidth]{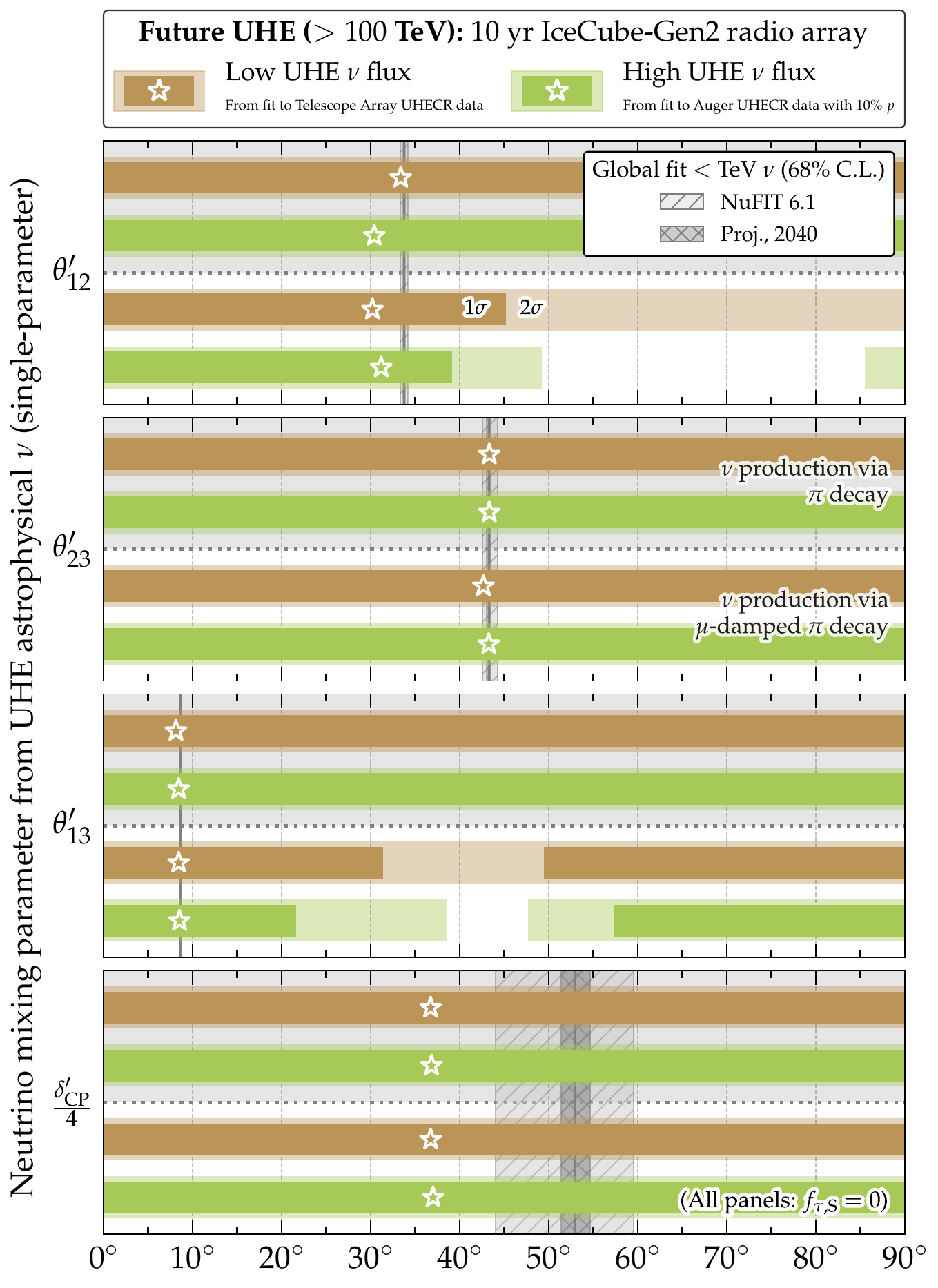}
 \vspace*{-0.5cm}
 \caption{\textbf{Constraints on the high-$Q$ neutrino mixing parameters from UHE astrophysical neutrinos.}  Similar to \figu{results_tev_pev}, but for UHE ($> 100$~PeV) neutrinos.  All constraints are projected, inferred from 10-year flavor measurements in the radio array of IceCube-Gen2 from \Refe~\cite{Coleman:2024scd}, assuming two possible neutrino flux predictions, low and high. For comparison, we show current~\cite{Esteban:2024eli} and projected~\cite{Song:2020nfh} sub-TeV global-fit ranges. \textit{These results represent the first rigorous assessment of mixing parameter sensitivity at a representative $Q \approx 40$~GeV (see \figu{main_results}).}}
 \label{fig:results_uhe}
\end{figure}

\begin{table*}[t!]
 \begin{ruledtabular}
    \centering
    \caption{\label{tab:results_mix_params_tev_pev}\textbf{\textit{Constraints on the high-$Q$ neutrino mixing parameters from TeV--PeV astrophysical neutrinos.}} All results shown are obtained by constraining one parameter at a time while keeping the others controlled via pull terms informed by present (NuFIT 6.1~\cite{Esteban:2024eli}) and future~\cite{Song:2020nfh} sub-TeV global oscillation fits.  We assume no $\nu_\tau$ production (\ie, $f_{\tau, {\rm S}} = 0$). Figures~\ref{fig:2d_profiled_likelihood_pion_2040} and \ref{fig:2d_profiled_likelihood_pion_2050} show joint parameter distributions. The values are stacked: for each parameter, the top line shows the best fit $\pm$ $1\sigma$, and the middle and bottom lines show (in brackets) the $2\sigma$ and $3\sigma$ range, respectively, profiled over all the other parameters. See Table~\ref{tab:results_mix_params_uhe} for results inferred from UHE ($>100$~PeV) neutrinos and Sec.~\ref{sec:results-proj_mix_params} for details.}
    \renewcommand{\arraystretch}{1.5}
    \setlength{\tabcolsep}{2pt} 
    \footnotesize
    \begin{tabular}{l c c c c c c c c}
        \multirow{3}{*}{Parameter} & \multirow{3}{*}{\makecell{Present \\ (IC MESE \\ 11.4 yr)}} & \multicolumn{6}{c}{Future from TeV--PeV $\nu$ (multi-detector projections using HESE plus through-going muons)} & \multirow{3}{*}{\makecell{Global \\ ($<$ TeV) \\ NuFIT 6.1}} \\
        \cline{3-8}
        & & \multicolumn{2}{c}{Only km$^3$-scale telescopes\footnotemark[1]} & \multicolumn{2}{c}{Plus multi-km$^3$ telescopes\footnotemark[2]} & \multicolumn{2}{c}{Plus tens-of-km$^3$ telescope\footnotemark[3]} & \\
        \cline{3-4} \cline{5-6} \cline{7-8}
        & & $\pi$ decay & $\mu$-damped & $\pi$ decay & $\mu$-damped & $\pi$ decay & $\mu$-damped & \\
        \hline \noalign{\vspace{3pt}}
        \multirow{3}{*}{ $\sin^2 \theta'_{12}$ } & $1\sigma$: $[0, 1]$ & \makecell{$0.214_{-0.214}^{+0.531}$ \\ $\cup \, [0.877, 1.000]$} & $0.305_{-0.305}^{+0.076}$ & \makecell{$0.276_{-0.276}^{+0.440}$ \\ $\cup \, [0.900, 1.000]$} & $0.301_{-0.301}^{+0.039}$ & \makecell{$0.284_{-0.284}^{+0.424}$ \\ $\cup \, [0.905, 1.000]$} & $0.299_{-0.297}^{+0.038}$ & $0.303_{-0.012}^{+0.012}$ \\
        & $2\sigma$: $[0, 1]$ & \makecell{$[0.000, 0.799]$ \\ $\cup \, [0.834, 1.000]$} & $[0.000, 0.452]$ & \makecell{$[0.000, 0.744]$ \\ $\cup \, [0.880, 1.000]$} & $[0.000, 0.372]$ & \makecell{$[0.000, 0.729]$ \\ $\cup \, [0.890, 1.000]$} & $[0.000, 0.366]$ & \makecell{$\cdots$} \\
        & $3\sigma$: $[0, 1]$ & $[0.000, 1.000]$ & $[0.000, 0.522]$ & \makecell{$[0.000, 0.771]$ \\ $\cup \, [0.860, 1.000]$} & $[0.000, 0.404]$ & \makecell{$[0.000, 0.749]$ \\ $\cup \, [0.876, 1.000]$} & $[0.000, 0.394]$ & $[0.270, 0.341]$ \\
        \cline{2-9} \noalign{\vspace{3pt}}
        \multirow{3}{*}{ $\theta'_{12}~[^\circ]$ } & $[0, 90]$ & \makecell{$27.53_{-27.53}^{+32.09}$ \\ $\cup \, [69.51, 90.00]$} & $33.50_{-33.50}^{+4.59}$ & \makecell{$31.67_{-31.67}^{+26.12}$ \\ $\cup \, [71.52, 90.00]$} & $33.28_{-33.28}^{+2.41}$ & \makecell{$32.21_{-32.21}^{+25.08}$ \\ $\cup \, [72.01, 90.00]$} & $33.13_{-31.11}^{+2.37}$ & $33.40_{-0.75}^{+0.74}$ \\
        & $[0, 90]$ & \makecell{$[0.00, 63.36]$ \\ $\cup \, [65.99, 90.00]$} & $[0.00, 42.26]$ & \makecell{$[0.00, 59.58]$ \\ $\cup \, [69.72, 90.00]$} & $[0.00, 37.60]$ & \makecell{$[0.00, 58.60]$ \\ $\cup \, [70.66, 90.00]$} & $[0.00, 37.21]$ & \makecell{$\cdots$} \\
        & $[0, 90]$ & $[0.00, 90.00]$ & $[0.00, 46.27]$ & \makecell{$[0.00, 61.40]$ \\ $\cup \, [68.00, 90.00]$} & $[0.00, 39.44]$ & \makecell{$[0.00, 59.91]$ \\ $\cup \, [69.40, 90.00]$} & $[0.00, 38.87]$ & $[31.31, 35.73]$ \\
        \hline \noalign{\vspace{0pt}}
        \multirow{3}{*}{ $\sin^2 \theta'_{23}$ } & $[0, 1]$ & $0.467_{-0.361}^{+0.533}$ & $0.469_{-0.122}^{+0.531}$ & $0.470_{-0.225}^{+0.530}$ & $0.470_{-0.056}^{+0.201}$ & $0.468_{-0.195}^{+0.523}$ & $0.470_{-0.047}^{+0.157}$ & $0.451_{-0.016}^{+0.019}$ \\
        & $[0, 1]$ & $[0.000, 1.000]$ & $[0.233, 1.000]$ & $[0.137, 1.000]$ & $[0.360, 1.000]$ & $[0.171, 1.000]$ & $[0.377, 1.000]$ & \makecell{$\cdots$} \\
        & $[0, 1]$ & $[0.000, 1.000]$ & $[0.127, 1.000]$ & $[0.000, 1.000]$ & $[0.308, 1.000]$ & \makecell{$[0.000, 0.008]$ \\ $\cup \, [0.095, 1.000]$} & $[0.331, 1.000]$ & $[0.408, 0.603]$ \\
        \cline{2-9} \noalign{\vspace{0pt}}
        \multirow{3}{*}{ $\theta'_{23}~[^\circ]$ } & $[0, 90]$ & $43.12_{-24.11}^{+46.88}$ & $43.21_{-7.11}^{+46.79}$ & $43.29_{-13.63}^{+46.71}$ & $43.29_{-3.21}^{+11.75}$ & $43.15_{-11.65}^{+41.44}$ & $43.29_{-2.73}^{+9.08}$ & $42.19_{-0.92}^{+1.09}$ \\
        & $[0, 90]$ & $[0.00, 90.00]$ & $[28.83, 90.00]$ & $[21.72, 90.00]$ & $[36.90, 90.00]$ & $[24.43, 90.00]$ & $[37.85, 90.00]$ & \makecell{$\cdots$} \\
        & $[0, 90]$ & $[0.00, 90.00]$ & $[20.91, 90.00]$ & $[0.00, 90.00]$ & $[33.70, 90.00]$ & \makecell{$[0.00, 5.18]$ \\ $\cup \, [17.96, 90.00]$} & $[35.15, 90.00]$ & $[39.70, 50.94]$ \\
        \hline \noalign{\vspace{0pt}}
        \multirow{3}{*}{ $\sin^2 \theta'_{13}$ } & $[0, 1]$ & $0.022_{-0.022}^{+0.659}$ & $0.022_{-0.022}^{+0.211}$ & $0.021_{-0.021}^{+0.550}$ & $0.022_{-0.017}^{+0.059}$ & $0.022_{-0.022}^{+0.530}$ & $0.022_{-0.015}^{+0.047}$ & $0.02248_{-0.00059}^{+0.00055}$ \\
        & $[0, 1]$ & $[0.000, 0.822]$ & $[0.000, 0.470]$ & $[0.000, 0.673]$ & $[0.000, 0.172]$ & $[0.000, 0.650]$ & $[0.001, 0.140]$ & \makecell{$\cdots$} \\
        & $[0, 1]$ & $[0.000, 1.000]$ & \makecell{$[0.000, 0.646]$ \\ $\cup \, [0.939, 1.000]$} & $[0.000, 0.751]$ & $[0.000, 0.275]$ & $[0.000, 0.721]$ & $[0.000, 0.228]$ & $[0.02064, 0.02418]$ \\
        \cline{2-9} \noalign{\vspace{0pt}}
        \multirow{3}{*}{ $\theta'_{13}~[^\circ]$ } & $[0, 90]$ & $8.43_{-8.43}^{+47.17}$ & $8.53_{-8.53}^{+20.34}$ & $8.33_{-8.33}^{+40.76}$ & $8.53_{-4.43}^{+8.00}$ & $8.53_{-8.53}^{+39.47}$ & $8.53_{-3.74}^{+6.67}$ & $8.54_{-0.12}^{+0.11}$ \\
        & $[0, 90]$ & $[0.00, 65.08]$ & $[0.00, 43.31]$ & $[0.00, 55.14]$ & $[0.81, 24.47]$ & $[0.00, 53.74]$ & $[1.64, 22.00]$ & \makecell{$\cdots$} \\
        & $[0, 90]$ & $[0.00, 90.00]$ & \makecell{$[0.00, 53.46]$ \\ $\cup \, [75.75, 90.00]$} & $[0.00, 60.08]$ & $[0.00, 31.62]$ & $[0.00, 58.13]$ & $[0.00, 28.49]$ & $[8.19, 8.90]$ \\
        \hline \noalign{\vspace{0pt}}
        \multirow{3}{*}{ $\delta'_{\rm CP}~[^\circ]$ } & $[0, 360]$ & $[0, 360]$ & $148_{-51}^{+116}$ & $[0, 360]$ & $148_{-26}^{+90}$ & $145_{-109}^{+179}$ & $148_{-23}^{+87}$ & $232_{-26}^{+36}$ \\
        & $[0, 360]$ & $[0, 360]$ & $[51, 309]$ & $[0, 360]$ & $[101, 259]$ & $[0, 360]$ & $[107, 253]$ & \makecell{$\cdots$} \\
        & $[0, 360]$ & $[0, 360]$ & $[0, 360]$ & $[0, 360]$ & $[83, 277]$ & $[0, 360]$ & $[91, 269]$ & $[133, 368]$ \\
    \end{tabular}
    \footnotetext[1]{Projections for 2040 using 29~yr of IceCube + 15~yr of Baikal-GVD + 13~yr of KM3NeT.}
    \footnotetext[2]{Projections for 2050 using 24~yr of IceCube + 25~yr of Baikal-GVD + 23~yr of KM3NeT + 19~yr of P-ONE + 15~yr of IceCube-Gen2 + 10~yr of NEON or TRIDENT.}
    \footnotetext[3]{Projections for 2050 using 24~yr of IceCube + 25~yr of Baikal-GVD + 23~yr of KM3NeT + 19~yr of P-ONE + 15~yr of IceCube-Gen2 + 10~yr of HUNT.}
 \end{ruledtabular}
\end{table*}

\begin{table*}[t!]
 \begin{ruledtabular}
    \centering
    \caption{\label{tab:results_mix_params_uhe}\textbf{\textit{Projected constraints to the high-$Q$ neutrino mixing parameters from ultra-high-energy ($> 100$~PeV) astrophysical neutrinos.}} Similar to Table~\ref{tab:results_mix_params_tev_pev}, but for UHE neutrinos.  All results shown are obtained by constraining one parameter at a time while keeping the others controlled via pull terms informed by future~\cite{Song:2020nfh} sub-TeV global oscillation fits.  Flavor-measurement projections are for 10 years of the radio array of IceCube-Gen2, as presented in \Refe~\cite{Coleman:2024scd}.  We assume no $\nu_\tau$ production (\ie, $f_{\tau, {\rm S}} = 0$). Figure~\ref{fig:2d_profiled_likelihood_uhe_muon_flux_0} shows joint parameter distributions. The values are stacked: for each parameter, the top line shows the best fit $\pm$ $1\sigma$, and the middle and bottom lines show (in brackets) the $2\sigma$ and $3\sigma$ range, respectively, profiled over all the other parameters. See Table~\ref{tab:results_mix_params_tev_pev} for results inferred from TeV--PeV neutrinos and Sec.~\ref{sec:results-proj_mix_params} for details.}
    \renewcommand{\arraystretch}{1.3}
    \setlength{\tabcolsep}{4pt} 
    \footnotesize
    \begin{tabular}{l c c c c c}
        \multirow{3}{*}{Parameter} & \multicolumn{4}{c}{Future from $> 100$~PeV $\nu$ (radio array of IceCube-Gen2)} & \multirow{3}{*}{\makecell{Global \\ ($<$ TeV) \\ NuFIT 6.1}} \\
        \cline{2-5}
        & \multicolumn{2}{c}{Low UHE $\nu$ flux\footnotemark[1]} & \multicolumn{2}{c}{High UHE $\nu$ flux\footnotemark[2]} & \\
        \cline{2-3} \cline{4-5}
        & $\pi$ decay & $\mu$-damped & $\pi$ decay & $\mu$-damped & \\
        \hline \noalign{\vspace{2pt}}
        \multirow{3}{*}{ $\sin^2 \theta'_{12}$ } & $1\sigma$: $[0, 1]$ & $0.253_{-0.253}^{+0.250}$ & $[0, 1]$ & $0.268_{-0.268}^{+0.130}$ & $0.303_{-0.012}^{+0.012}$ \\
        & $2\sigma$: $[0, 1]$ & $[0, 1]$ & $[0, 1]$ & \makecell{$[0.000, 0.574]$ \\ $\cup \, [0.994, 1.000]$} & \makecell{$\cdots$} \\
        & $3\sigma$: $[0, 1]$ & $[0, 1]$ & $[0, 1]$ & $[0, 1]$ & $[0.270, 0.341]$ \\
        \cline{2-6} \noalign{\vspace{2pt}}
        \multirow{3}{*}{ $\theta'_{12}~[^\circ]$ } & $[0, 90]$ & $30.21_{-30.21}^{+14.99}$ & $[0, 90]$ & $31.19_{-31.19}^{+7.96}$ & $33.40_{-0.75}^{+0.74}$ \\
        & $[0, 90]$ & $[0, 90]$ & $[0, 90]$ & \makecell{$[0.00, 49.24]$ \\ $\cup \, [85.53, 90.00]$} & \makecell{$\cdots$} \\
        & $[0, 90]$ & $[0, 90]$ & $[0, 90]$ & $[0, 90]$ & $[31.31, 35.73]$ \\
        \hline \noalign{\vspace{2pt}}
        \multirow{3}{*}{ $\sin^2 \theta'_{23}$ } & $[0, 1]$ & $[0, 1]$ & $[0, 1]$ & $[0, 1]$ & $0.451_{-0.016}^{+0.019}$ \\
        & $[0, 1]$ & $[0, 1]$ & $[0, 1]$ & $[0, 1]$ & \makecell{$\cdots$} \\
        & $[0, 1]$ & $[0, 1]$ & $[0, 1]$ & $[0, 1]$ & $[0.408, 0.603]$ \\
        \cline{2-6} \noalign{\vspace{2pt}}
        \multirow{3}{*}{ $\theta'_{23}~[^\circ]$ } & $[0, 90]$ & $[0, 90]$ & $[0, 90]$ & $[0, 90]$ & $42.19_{-0.92}^{+1.09}$ \\
        & $[0, 90]$ & $[0, 90]$ & $[0, 90]$ & $[0, 90]$ & \makecell{$\cdots$} \\
        & $[0, 90]$ & $[0, 90]$ & $[0, 90]$ & $[0, 90]$ & $[39.70, 50.94]$ \\
        \hline \noalign{\vspace{2pt}}
        \multirow{3}{*}{ $\sin^2 \theta'_{13}$ } & $[0, 1]$ & \makecell{$0.022_{-0.022}^{+0.250}$ \\ $\cup \, [0.578, 1.000]$} & $[0, 1]$ & \makecell{$0.022_{-0.022}^{+0.114}$ \\ $\cup \, [0.709, 1.000]$} & $0.02248_{-0.00059}^{+0.00055}$ \\
        & $[0, 1]$ & $[0, 1]$ & $[0, 1]$ & \makecell{$[0.000, 0.388]$ \\ $\cup \, [0.547, 1.000]$} & \makecell{$\cdots$} \\
        & $[0, 1]$ & $[0, 1]$ & $[0, 1]$ & $[0, 1]$ & $[0.02064, 0.02418]$ \\
        \cline{2-6} \noalign{\vspace{2pt}}
        \multirow{3}{*}{ $\theta'_{13}~[^\circ]$ } & $[0, 90]$ & \makecell{$8.43_{-8.43}^{+22.96}$ \\ $\cup \, [49.50, 90.00]$} & $[0, 90]$ & \makecell{$8.53_{-8.53}^{+13.07}$ \\ $\cup \, [57.32, 90.00]$} & $8.54_{-0.12}^{+0.11}$ \\
        & $[0, 90]$ & $[0, 90]$ & $[0, 90]$ & \makecell{$[0.00, 38.50]$ \\ $\cup \, [47.67, 90.00]$} & \makecell{$\cdots$} \\
        & $[0, 90]$ & $[0, 90]$ & $[0, 90]$ & $[0, 90]$ & $[8.19, 8.90]$ \\
        \hline \noalign{\vspace{2pt}}
        \multirow{3}{*}{ $\delta'_{\rm CP}~[^\circ]$ } & $[0, 360]$ & $[0, 360]$ & $[0, 360]$ & $[0, 360]$ & $232_{-26}^{+36}$ \\
        & $[0, 360]$ & $[0, 360]$ & $[0, 360]$ & $[0, 360]$ & \makecell{$\cdots$} \\
        & $[0, 360]$ & $[0, 360]$ & $[0, 360]$ & $[0, 360]$ & $[133, 368]$ \\
    \end{tabular}
    \footnotetext[1]{Low UHE neutrino flux derived from a fit to Auger UHECR data assuming 10\% of the cosmic rays are protons~\cite{vanVliet:2019nse}.  We add the IceCube flux derived from 9.5 years of through-going tracks~\cite{Abbasi:2021qfz}, extrapolated to ultra-high energies.}
    \footnotetext[2]{High UHE neutrino flux derived from a fit to Telescope Array UHECR data~\cite{Bergman:2021djm}.  We add the IceCube flux derived from 9.5 years of through-going tracks~\cite{Abbasi:2021qfz}, extrapolated to ultra-high energies.}
 \end{ruledtabular}
\end{table*}

\textit{We present the constraints derived from our analysis. While present-day IceCube MESE data lack the precision to constrain high-$Q$ mixing or SMEFT coefficients, future TeV--PeV multi-detector networks (IceCube-Gen2, KM3NeT, Baikal-GVD, P-ONE, TRIDENT, NEON, HUNT) will be capable of placing meaningful bounds by 2040 and 2050, whereas UHE measurements yield drastically weaker bounds due to the larger uncertainties of radio-based flavor tagging. These TeV--PeV limits are highly dependent on the astrophysical source model: we find no sensitivity if neutrinos are produced via full pion decay due to intrinsic probability averaging, but we place $\mathcal{O}(1)$ limits on SMEFT coefficients (at $\Lambda_{\rm SMEFT} = 1$~TeV) if production occurs via muon-damped pion decay. Finally, we highlight that these constraints exhibit a strong parameter hierarchy---tightest for $\theta_{23}^\prime$ and $C_{11}$---and offer a unique, complementary astrophysical probe capable of testing new physics and potentially distinguishing between Dirac and Majorana neutrinos.}


\subsection{Present constraints}
\label{sec:results-present}

We find no sensitivity to either the high-$Q$ mixing parameters or the dimension-6 SMEFT coefficients in the present TeV--PeV flavor-composition measurements based on the IceCube 11.4-year MESE sample.  This is due to the relatively large uncertainties associated with this measurement (\figu{ternary_theory}, right panel) and, in the case of the SMEFT coefficients, to the small-to-moderate size of the expected RG deviations induced by them (Figs.~\ref{fig:flavor_smeft_rg_vs_Q}, \ref{fig:flavor_smeft_rg_vs_c_smeft}).  Our projections based on multi-detector combinations, however, reveal upcoming meaningful sensitivity.


\subsection{Projected constraints on high-$Q$ mixing parameters}
\label{sec:results-proj_mix_params}

Figure~\ref{fig:results_tev_pev} and Table~\ref{tab:results_mix_params_tev_pev} show that measurements of the TeV--PeV neutrino flavor composition have the potential to constrain the high-$Q$ mixing parameters.  Figure~\ref{fig:results_uhe} and Table~\ref{tab:results_mix_params_uhe} show that, in contrast, measurements of the UHE ($> 100$~PeV) neutrino flavor composition have limited constraining power.  These results confirm our theory expectations from Sec.~\ref{sec:flavor_comp-regions_earth}, which we summarize in the salient points below.
Overall, our results show that \textbf{\textit{the TeV--PeV neutrino flavor composition is suitable to place constraints only on large deviations of the high-$Q$ mixing parameters relative to their standard values, while UHE flavor measurements currently lack the resolution to do so.}}  Later (Sec.~\ref{sec:results-proj_smeft}), we show how this translates to the specific case of constraining SMEFT coefficients.

\medskip

\textbf{\textit{Flavor composition does not provide precision constraints.---}}Although our TeV--PeV projections show meaningful constraints on the high-$Q$ mixing parameters, these constraints are wide, with $1\sigma$ relative uncertainties of 100--200\% being commonplace, the exact value depending on the parameter, the assumed neutrino production mechanism, and the multi-detector combination used for the projection.  In contrast, the present-day precision on the standard mixing parameters, from NuFIT~6.1, is $\lesssim 3\%$ for the mixing angles and 16\% for $\delta_{\rm CP}$ (at $1\sigma$), with expected improvements to come~\cite{Song:2020nfh}.  

The reason behind our wide constraints is two-fold.  The first reason is experimental, \ie, the significant uncertainty with which neutrino telescopes infer the flavor composition---as illustrated in \figu{ternary_theory}, right panel---reflecting the inherent challenge in separating signals from neutrinos of different flavor (Sec.~\ref{sec:flavor_comp-measuring}).  However, while this uncertainty remains overwhelmingly dominant for UHE neutrinos, it quickly becomes sub-dominant in our TeV--PeV multi-detector projections.

The second reason is theoretical; it exists today and survives into our projections, where it becomes the leading cause of our wide TeV--PeV constraints: our ignorance of the flavor composition at the sources.  As detailed in Sec.~\ref{sec:statistics-mixing_params}, this ignorance forces us to profile over the $\nu_e$ fraction at the sources, $f_{e, {\rm S}}$, when computing constraints on the high-$Q$ mixing parameters, and therefore weakens our sensitivity to them.  \textbf{\textit{This profiling is inevitable,}} since the flavor composition at the sources is inferred~\cite{Bustamante:2019sdb, Song:2020nfh} from the same observations that are used to constrain the high-$Q$ mixing parameters.  The co-dependence with $f_{e, {\rm S}}$ is illustrated by the two-dimensional profiled likelihood of the high-$Q$ mixing parameters vs.~$f_{e, {\rm S}}$ in Figs.~\ref{fig:2d_profiled_likelihood_pion_2040} and \ref{fig:2d_profiled_likelihood_pion_2050}, which show significant correlation.

Assuming instead the flavor composition to be known, as in \Refes~\cite{Bustamante:2010bf, Babu:2021cxe, Mir:2025fae}, yields high-$Q$ deviations in the flavor composition that appear deceivingly large enough to be detectable.  However, these prospects are overly optimistic, and must be tempered by the inescapable uncertainty on $f_{e, {\rm S}}$ to be made realistic, upon which they should become comparable to our TeV--PeV results.

\medskip

\textbf{\textit{UHE vs.~TeV--PeV constraints.---}}Despite UHE neutrinos probing an energy regime roughly three orders of magnitude higher than their TeV--PeV counterparts, their ability to constrain the high-$Q$ mixing parameters is drastically inferior. 

As \figu{results_uhe} and Table~\ref{tab:results_mix_params_uhe} demonstrate, 10 years of observation with the projected IceCube-Gen2 radio array leaves the high-$Q$ mixing parameters almost entirely unconstrained. The projected $1\sigma$ and $2\sigma$ allowed regions span nearly the full physical parameter space, regardless of the assumed UHE neutrino flux model (stemming either from optimistic fits to Telescope Array cosmic-ray~\cite{Bergman:2021djm} or pessimistic fits to Auger cosmic-ray data~\cite{vanVliet:2019nse}) or the neutrino production mechanism (full or muon-damped pion decay). 

This severe lack of sensitivity stems directly from the formidable experimental challenges at EeV energies: the inherently low expected event statistics~\cite{ Valera:2022ylt, Valera:2022wmu} and the extreme difficulty of performing precise flavor-tagging via radio-detection techniques~\cite{Coleman:2024scd} yield uncertainties in the measured flavor composition that are vastly larger than those at TeV--PeV energies. Consequently, while UHE neutrinos offer a slightly extended lever arm to constrain the high-$Q$ mixing parameters, the severe degradation in flavor resolution completely overrides this theoretical advantage when attempting to constrain them directly in a model-independent way.

\medskip

\textbf{\textit{Differentiated parameter sensitivity.---}}The different high-$Q$ mixing parameters are constrained to different degrees.  As detailed in Appendix~\ref{app:allowed_flavor_regions}, deviations of $\theta_{12}^\prime$ and $\theta_{23}^\prime$ from their standard values shift the region of allowed flavor composition at Earth away from its standard-oscillation expectation more than deviations of $\theta_{13}^\prime$ and $\delta_{\rm CP}^\prime$. We can understand this hierarchy through the analytical approximations derived in Appendix~\ref{app:analytical_approximation_smeft}. 

Specifically, the leading-order expansion of the shifts in the flavor fractions at Earth induced by the high-$Q$ mixing parameters [Eqs.~(\ref{equ:flavor_shift_electron_generic_feS})--(\ref{equ:flavor_shift_tau_generic_feS})] exposes a strong, direct dependence on $\theta_{12}^\prime$ and $\theta_{23}^\prime$, while remaining largely insensitive to small variations in $\theta_{13}^\prime$ and $\delta_{\rm CP}^\prime$. Figures~\ref{fig:flavor_ratios_pion} and \ref{fig:flavor_ratios_muon} show the same differentiated behavior.  This hierarchy translates imperfectly into the differences in the relative precision with which these parameters are constrained in \figu{results_tev_pev} and Table~\ref{tab:results_mix_params_tev_pev}.  In the UHE regime, however, this theoretical hierarchy is entirely washed out by the dominant experimental uncertainties (see \figu{results_uhe}).

The angle $\theta_{23}^\prime$ is the most tightly constrained high-$Q$ parameter, reaching a $1\sigma$ precision of about 33\% in our most ambitious 2050 TeV--PeV projections (assuming neutrino production via muon-damped pion decay).  The reason, as illuminated by Eqs.~(\ref{equ:flavor_shift_electron_generic_feS})--(\ref{equ:flavor_shift_tau_generic_feS}) in Appendix~\ref{app:analytical_approximation_smeft}, is that $\theta_{23}^\prime$ directly governs the breaking of the $\nu_\mu$--$\nu_\tau$ symmetry. Consequently, variations in $\theta_{23}^\prime$, together with $f_{e, {\rm S}}$, result in deviations in the flavor composition that move nearly orthogonally away from the center of our assumed flavor-composition measurement, as shown in \figu{ternary_individual}.  

In contrast, variations in $\theta_{12}^\prime$ predominantly scale the $\nu_e$ fraction at Earth via $P_{ee}$ (see \figu{probabilities}). While these variations are also prominent, they are aligned with the major axis of our assumed ellipse-like flavor-composition measurement, as shown also in \figu{ternary_individual}, rendering them relatively harder to constrain, especially when assuming flavor measurements centered on the standard-mixing full-pion-decay expectation (more on this below). Further, Figs.~\ref{fig:2d_profiled_likelihood_pion_2040} and \ref{fig:2d_profiled_likelihood_pion_2050} reveal stronger correlation between $\theta_{12}^\prime$ and $f_{e, {\rm S}}$ than between $\theta_{23}^\prime$ and $f_{e, {\rm S}}$, since the  former governs the $\nu_e$ content of the flux more directly.  As a result, the profiling over $f_{e, {\rm S}}$ weakens the constraints on $\theta_{12}^\prime$ more than it does the constraints on $\theta_{23}^\prime$.

\medskip

\textbf{\textit{Neutrino production via full pion decay yields weaker constraints.---}}The constraints on the high-$Q$ mixing parameters are weaker when the projected flavor-composition measurement is assumed to be entered on the expectation from full pion decay than when centered on the expectation from muon-damped pion decay.  This is due to the latter case being more sensitive to changes in individual flavor-transition channels, as explained in detail in Sec.~\ref{sec:flavor_ratios-impact_sources} and Appendix~\ref{app:analytical_approximation_smeft}. \textit{Cf.}~the sensitivity of the flavor ratios to the high-$Q$ mixing parameters under each production channel in Figs.~\ref{fig:flavor_ratios_pion} and \ref{fig:flavor_ratios_muon}.  The underlying reason, as presented in Appendix~\ref{app:analytical_approximation_smeft-flavor_distance_pion_decay}, is that for full pion decay ($f_{e, {\rm S}} = 1/3$) the $\nu_1$ and $\nu_2$ populations at the sources are too similar to one another, which dampens the effect of the high-$Q$ mixing parameters---notably, of $\theta_{12}^\prime$---on the flavor composition at Earth.

Concretely, Table~\ref{tab:results_mix_params_tev_pev} shows that under muon-damped pion decay the allowed $1\sigma$ intervals on the high-$Q$ mixing angles are 2--4 times narrower than under full pion decay, and the $3\sigma$ intervals do not span the full allowed physical range, unlike full pion decay. (For UHE neutrinos, as noted above, the constraints remain weak regardless of the production mechanism; see Table~\ref{tab:results_mix_params_uhe}.) Further, under the TeV--PeV full-pion-decay projections, there is little improvement in the relative precision with which the mixing parameters can be constrained over time.

Since, from theory, we expect full pion decay to be the nominal neutrino production mechanism, the above results have the unfortunate consequence of weakening our future prospects of constraining the high-$Q$ mixing parameters.  Nevertheless, muon-damped pion production represents a viable possibility, especially towards higher neutrino energies~\cite{Song:2020nfh}, where muon synchrotron losses may become significant~\cite{Winter:2013cla, Bustamante:2020bxp}, depending on the intensity of the magnetic fields harbored by the sources.

\medskip

\textbf{\textit{Existing TeV--PeV neutrino telescopes can provide meaningful constraints.---}}Our projections show that even combining observations exclusively by existing km$^3$-scale neutrino telescopes IceCube, Baikal-GVD, and KM3NeT, it may be possible to constrain the high-$Q$ mixing parameters by 2040.  The relative uncertainties on the high-$Q$ mixing parameters in these projected constraints are large: 100--200\% (at $1\sigma$) on $\theta_{12}^\prime$ and $\theta_{23}^\prime$, factor-9-to-30 uncertainties on $\theta_{13}^\prime$, and 78\% or null uncertainty on $\delta_{\rm CP}^\prime$ assuming neutrino production via muon-damped or full pion decay, respectively.  These uncertainties represent the baseline benchmarks for the predicted size that the high-$Q$ mixing parameters must have in order to be experimentally distinguishable from their standard-mixing counterparts.  (Reference~\cite{Bustamante:2026aur} presents a similar argument geared at the measurement of the standard mixing parameters.)

\medskip

\textbf{\textit{The limits of precision.---}}While our most ambitious TeV--PeV projections---combining multi-km$^3$ telescopes by 2040 and adding a tens-of-km$^3$ telescope by 2050---show improvement in the parameter constraints, this improvement is moderate.  Broadly stated, the $1\sigma$ constraints remain roughly at the same level they would reach by 2040 using only existing telescopes.  In parallel, $3\sigma$ constraints become viable for parameters for which they previously were not.  The reason for the slow improvement is the same as before: the need to profile over $f_{e, {\rm S}}$, which remains throughout even in our farthest projections.  The one exception occurs for $\theta_{23}^\prime$---our best constrained parameter (see above)---in our projections assuming neutrino production via muon-damped pion decay.  In this case, there is a more appreciable improvement in the $1\sigma$ precision, evolving from more than 100\% using only existing telescopes by 2040 to 33\% combining all detectors by 2050. 

\medskip

Figure~\ref{fig:main_results} summarizes our results on high-$Q$ mixing-parameter constraints. It contrasts the allowed regions dictated by current low-energy global fits with our projected measurements for 2040 and 2050. The figure illustrates the hierarchy of parameter sensitivity: projected TeV--PeV multi-detector measurements place meaningful bounds on $\theta_{23}^\prime$ and $\theta_{13}^\prime$, whereas UHE radio-detection projections remain nearly entirely washed out due to the resolution limits of EeV flavor tagging. This divergence in sensitivity occurs despite both regimes probing remarkably similar kinematics, sharing largely overlapping 68\% and 99\% $Q$-range containment regions with representative momentum transfers of $Q \approx 20$~GeV (TeV--PeV) and $Q \approx 40$~GeV (UHE), as explained above.

\medskip

In summary, while high precision eludes us, the capacity of TeV--PeV flavor measurements to bound high-$Q$ mixing provides a robust, model-independent foundation for testing new-physics scenarios, to which we now turn.


\subsection{Projected constraints on SMEFT coefficients}
\label{sec:results-proj_smeft}

\begin{figure*}[t!]
 \centering
 \includegraphics[width=\textwidth]{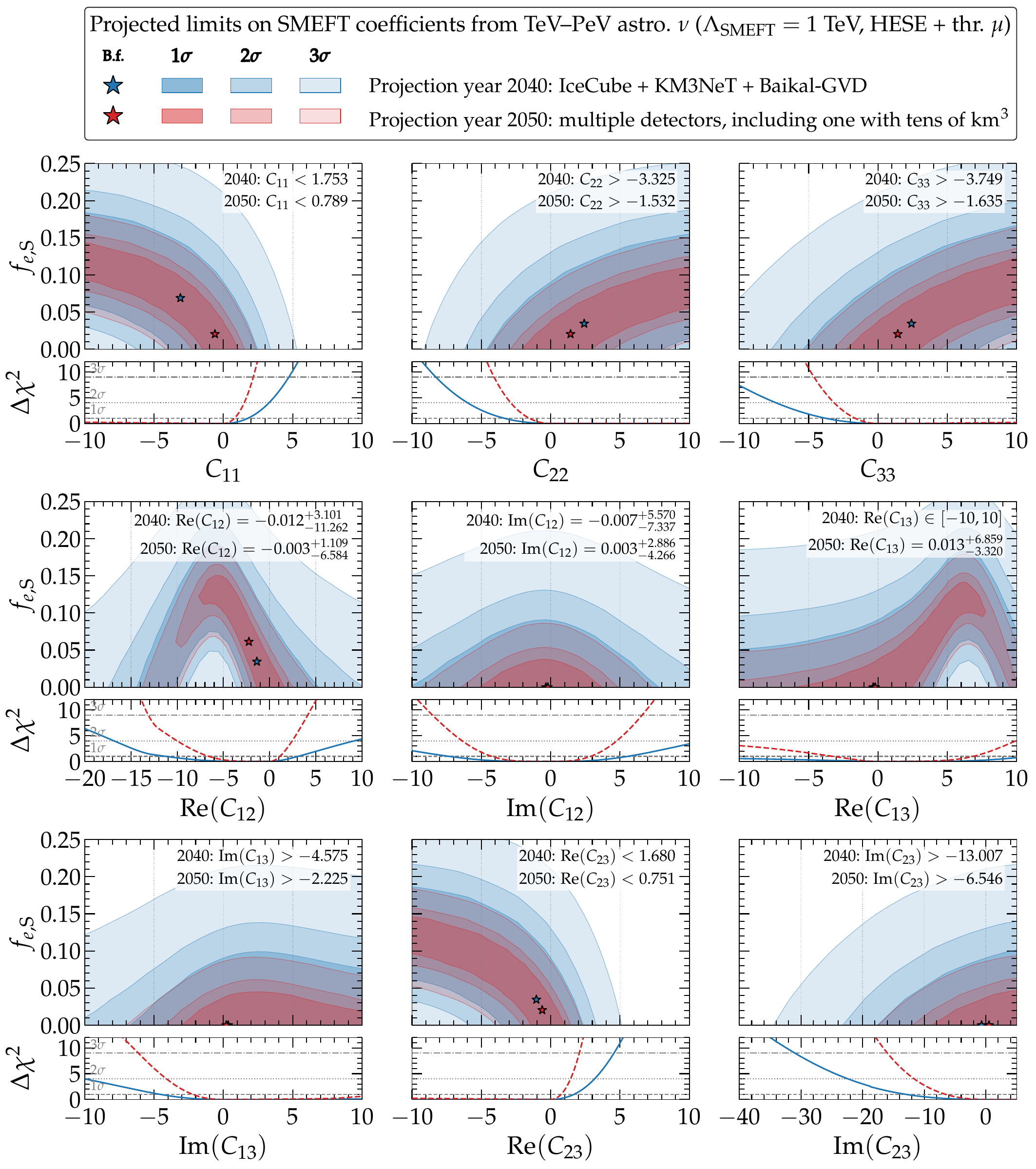}
 \caption{\textbf{Projected constraints on dimension-6 SMEFT coefficients using TeV--PeV astrophysical neutrinos.}  Projections are for 2040, combining only existing neutrino telescopes IceCube, KM3NeT, and Baikal-GVD, and for 2050, combining them with future telescopes P-ONE, IceCube-Gen2, TRIDENT, NEON, and HUNT. Neutrino production is assumed to be via muon-damped pion decay.  Constraints use the neutrino flavor composition inferred from HESE plus through-going muons.   For each parameter, the lower panels show the 2D allowed regions at the 1$\sigma$, 2$\sigma$, and 3$\sigma$ C.L.~in the plane of the coupling versus the source electron flavor fraction, $f_{e, {\rm S}}$. The corresponding 1D profiled $\Delta\chi^2$ distributions are displayed in the upper panels. The numerical bounds shown in each panel are at 1$\sigma$ C.L. The limits assume a reference new physics scale of $\Lambda_{\text{SMEFT}} = 1 \text{ TeV}$ (see the main text to transform to other scales).  Limits are derived under the single-operator assumption, where only one SMEFT coefficient is non-zero at a time.   When computing constraints, we assume no $\nu_\tau$ production (\ie, $f_{\tau, {\rm S}} = 0$).  See Sec.~\ref{sec:results-proj_smeft} for details, Table~\ref{tab:results_smeft_full} for one-dimensional parameter constraints, Appendix~\ref{app:pairwise_constraints} for pairwise constraints, and \figu{results_2d_smeft_uhe} for constraints from UHE neutrinos.  \textbf{\textit{Meaningful, robust constraints can be placed even by 2040 using the network of existing neutrino telescopes.}}}
 \label{fig:results_2d_smeft_muon}
\end{figure*}

\begin{figure*}[t!]
 \centering
 \includegraphics[width=\textwidth]{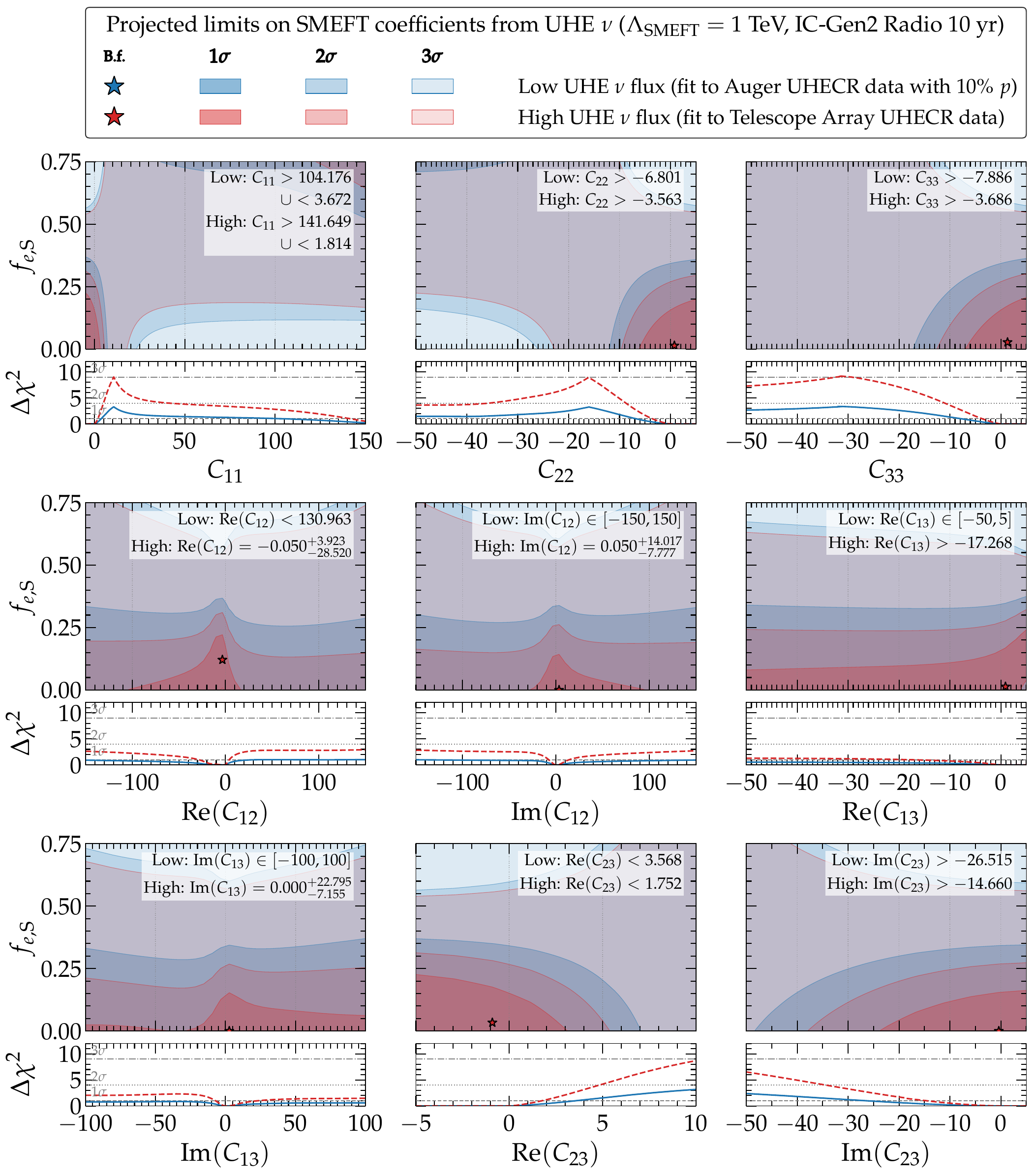}
 \caption{\textbf{Projected constraints on dimension-6 SMEFT coefficients using ultra-high-energy astrophysical neutrinos.}  Similar to \figu{results_2d_smeft_muon}, but for neutrinos of $> 100$~PeV.  Limits are derived under the single-operator assumption, where only one SMEFT coefficient is non-zero at a time.  Neutrino production is assumed to be via muon-damped pion decay.  Flavor-measurement projections are for 10 years of the radio array of IceCube-Gen2, as presented in \Refe~\cite{Coleman:2024scd}.  When computing constraints, we assume no $\nu_\tau$ production (\ie, $f_{\tau, {\rm S}} = 0$).  See Sec.~\ref{sec:results-proj_smeft} for details, Table~\ref{tab:results_smeft_uhe} for one-dimensional parameter constraints, Appendix~\ref{app:pairwise_constraints} for pairwise constraints, and \figu{results_2d_smeft_muon} for constraints from TeV--PeV neutrinos. \textbf{\textit{Constraints from UHE neutrinos are weaker than from TeV--PeV neutrinos because UHE flavor measurements are more uncertain.}}}
 \label{fig:results_2d_smeft_uhe}
\end{figure*}

\begin{table*}[t!]
 \begin{ruledtabular}
    \centering
    \caption{\label{tab:results_smeft_full}\textbf{\textit{Projected constraints on dimension-6 SMEFT coefficients from TeV--PeV astrophysical neutrinos.}} Constraints on individual SMEFT coefficients $C_{ij} \equiv [C_{Hl}^{(3)}]_{ij}$ at $\Lambda = 1$~TeV. Results are obtained by constraining one parameter at a time while keeping the others at zero. Because the physical effects of these dimension-6 operators scale as $\Lambda^{-2}$, the limits for an arbitrary new-physics scale $\Lambda'$ can be obtained by multiplying the tabulated values by $(\Lambda' / 1~\text{TeV})^2$. Figures~\ref{fig:2d_profiled_likelihood_pion_2040} and \ref{fig:2d_profiled_likelihood_pion_2050} show joint parameter distributions. The values are stacked: for each parameter, the top line shows the best fit $\pm$ $1\sigma$, the middle line shows the $2\sigma$ range, and the bottom line shows the $3\sigma$ range. Whenever an interval reaches the limits of our numerical scan window, it is printed as an upper ($<$) or lower ($>$) limit.  See Sec.~\ref{sec:results-proj_smeft} for details, Table~\ref{tab:results_smeft_uhe} for results from UHE ($> 100$~PeV) neutrinos, and Appendix~\ref{app:pairwise_constraints} for pairwise parameter constraints.}
    \renewcommand{\arraystretch}{1.3}
    \setlength{\tabcolsep}{4pt} 
    \footnotesize
    \begin{tabular}{l c c c c c c c c}
        \multirow{3}{*}{Parameter} & \multirow{3}{*}{\makecell{Present \\ (IC MESE \\ 11.4 yr)}} & \multicolumn{6}{c}{Future from TeV--PeV $\nu$ (multi-detector projections using HESE plus through-going muons)} & \multirow{3}{*}{\makecell{Approximate \\ existing \\ sub-TeV limits}} \\
        \cline{3-8}
         & & \multicolumn{2}{c}{Only km$^3$-scale telescopes\footnotemark[1]} & \multicolumn{2}{c}{Plus multi-km$^3$ telescopes\footnotemark[2]} & \multicolumn{2}{c}{Plus tens-of-km$^3$ telescope\footnotemark[3]} & \\
        \cline{3-4} \cline{5-6} \cline{7-8}
         & & $\pi$ decay & $\mu$-damped & $\pi$ decay & $\mu$-damped & $\pi$ decay & $\mu$-damped & \\
        \hline
        \multirow{3}{*}{ $C_{11}$ } & \multirow{3}{*}{\makecell{$\cdots$}} & \multirow{3}{*}{\makecell{$\cdots$}} & {\scriptsize 1$\sigma$:} $< 1.75$ & \multirow{3}{*}{\makecell{$\cdots$}} & $< 0.86$ & \multirow{3}{*}{\makecell{$\cdots$}} & $< 0.79$ & \multirow{3}{*}{$[-0.01, 0.01]$\footnotemark[4]} \\
         &  &  & {\scriptsize 2$\sigma$:} $< 3.28$ &  & $< 1.63$ &  & $< 1.50$ &  \\
         &  &  & {\scriptsize 3$\sigma$:} $< 4.72$ &  & $< 2.34$ &  & $< 2.15$ &  \\
        \cline{2-9}
        \multirow{3}{*}{ $C_{22}$ } & \multirow{3}{*}{\makecell{$\cdots$}} & \multirow{3}{*}{\makecell{$\cdots$}} & $> -3.33$ & \multirow{3}{*}{\makecell{$\cdots$}} & $> -1.67$ & \multirow{3}{*}{\makecell{$\cdots$}} & $> -1.53$ & \multirow{3}{*}{$[-0.01, 0.01]$\footnotemark[4]} \\
         &  &  & $> -5.99$ &  & $> -3.12$ &  & $> -2.88$ &  \\
         &  &  & $> -8.28$ &  & $> -4.42$ &  & $> -4.09$ &  \\
        \cline{2-9}
        \multirow{3}{*}{ $C_{33}$ } & \multirow{3}{*}{\makecell{$\cdots$}} & \multirow{3}{*}{\makecell{$\cdots$}} & $> -3.75$ & \multirow{3}{*}{\makecell{$\cdots$}} & $> -1.78$ & \multirow{3}{*}{\makecell{$\cdots$}} & $> -1.63$ & \multirow{3}{*}{$[-0.01, 0.01]$\footnotemark[4]} \\
         &  &  & $> -7.34$ &  & $> -3.44$ &  & $> -3.16$ &  \\
         &  &  & \makecell{$\cdots$} &  & $> -5.04$ &  & $> -4.62$ &  \\
        \hline
        \multirow{3}{*}{ ${\rm Re}(C_{12})$ } & \multirow{3}{*}{\makecell{$\cdots$}} & \multirow{3}{*}{\makecell{$\cdots$}} & $-0.01_{-11.26}^{+3.10}$ & \multirow{3}{*}{\makecell{$\cdots$}} & $-0.01_{-7.31}^{+1.23}$ & \multirow{3}{*}{\makecell{$\cdots$}} & $-0.00_{-6.58}^{+1.11}$ & \multirow{3}{*}{$[-10^{-5}, 10^{-5}]$\footnotemark[5]} \\
         &  &  & $[-16.84, 9.14]$ &  & $[-11.65, 2.76]$ &  & $[-10.01, 2.44]$ &  \\
         &  &  & \makecell{$\cdots$} &  & $[-14.24, 4.75]$ &  & $[-13.13, 4.12]$ &  \\
        \cline{2-9}
        \multirow{3}{*}{ ${\rm Im}(C_{12})$ } & \multirow{3}{*}{\makecell{$\cdots$}} & \multirow{3}{*}{\makecell{$\cdots$}} & $-0.01_{-7.34}^{+5.57}$ & \multirow{3}{*}{\makecell{$\cdots$}} & $0.00_{-4.47}^{+3.17}$ & \multirow{3}{*}{\makecell{$\cdots$}} & $0.00_{-4.27}^{+2.89}$ & \multirow{3}{*}{$[-10^{-5}, 10^{-5}]$\footnotemark[5]} \\
         &  &  & \makecell{$\cdots$} &  & $[-6.82, 5.26]$ &  & $[-6.38, 4.76]$ &  \\
         &  &  & \makecell{$\cdots$} &  & $[-9.32, 7.38]$ &  & $[-8.53, 6.60]$ &  \\
        \hline
        \multirow{3}{*}{ ${\rm Re}(C_{23})$ } & \multirow{3}{*}{\makecell{$\cdots$}} & \multirow{3}{*}{\makecell{$\cdots$}} & $< 1.68$ & \multirow{3}{*}{\makecell{$\cdots$}} & $< 0.82$ & \multirow{3}{*}{\makecell{$\cdots$}} & $< 0.75$ & \multirow{3}{*}{$[-10^{-3}, 10^{-3}]$\footnotemark[6]} \\
         &  &  & $< 3.16$ &  & $< 1.56$ &  & $< 1.43$ &  \\
         &  &  & $< 4.57$ &  & $< 2.25$ &  & $< 2.06$ &  \\
        \cline{2-9}
        \multirow{3}{*}{ ${\rm Im}(C_{23})$ } & \multirow{3}{*}{\makecell{$\cdots$}} & \multirow{3}{*}{\makecell{$\cdots$}} & $> -13.01$ & \multirow{3}{*}{\makecell{$\cdots$}} & $> -7.09$ & \multirow{3}{*}{\makecell{$\cdots$}} & $> -6.55$ & \multirow{3}{*}{$[-10^{-3}, 10^{-3}]$\footnotemark[6]} \\
         &  &  & $> -22.38$ &  & $> -12.52$ &  & $> -11.63$ &  \\
         &  &  & $> -31.01$ &  & $> -17.18$ &  & $> -15.98$ &  \\
        \hline
        \multirow{3}{*}{ ${\rm Re}(C_{13})$ } & \multirow{3}{*}{\makecell{$\cdots$}} & \multirow{3}{*}{\makecell{$\cdots$}} & \makecell{$\cdots$} & \multirow{3}{*}{\makecell{$\cdots$}} & $0.01_{-3.76}^{+7.50}$ & \multirow{3}{*}{\makecell{$\cdots$}} & $0.01_{-3.32}^{+6.86}$ & \multirow{3}{*}{$[-10^{-3}, 10^{-3}]$\footnotemark[6]} \\
         &  &  & \makecell{$\cdots$} &  & \makecell{$\cdots$} &  & $< 9.92$ &  \\
         &  &  & \makecell{$\cdots$} &  & \makecell{$\cdots$} &  & \makecell{$\cdots$} &  \\
        \cline{2-9}
        \multirow{3}{*}{ ${\rm Im}(C_{13})$ } & \multirow{3}{*}{\makecell{$\cdots$}} & \multirow{3}{*}{\makecell{$\cdots$}} & $> -4.57$ & \multirow{3}{*}{\makecell{$\cdots$}} & $> -2.43$ & \multirow{3}{*}{\makecell{$\cdots$}} & $> -2.23$ & \multirow{3}{*}{$[-10^{-3}, 10^{-3}]$\footnotemark[6]} \\
         &  &  & \makecell{$\cdots$} &  & $> -4.49$ &  & $> -4.11$ &  \\
         &  &  & \makecell{$\cdots$} &  & $> -6.82$ &  & $> -6.16$ &  \\
    \end{tabular}
    \footnotetext[1]{Projections for 2040 using 29~yr of IceCube + 15~yr of Baikal-GVD + 13~yr of KM3NeT.}
    \footnotetext[2]{Projections for 2050 using 24~yr of IceCube + 25~yr of Baikal-GVD + 23~yr of KM3NeT + 19~yr of P-ONE + 15~yr of IceCube-Gen2 + 10~yr of NEON or TRIDENT.}
    \footnotetext[3]{Projections for 2050 using 24~yr of IceCube + 25~yr of Baikal-GVD + 23~yr of KM3NeT + 19~yr of P-ONE + 15~yr of IceCube-Gen2 + 10~yr of HUNT.}
    \footnotetext[4]{Limits on flavor-diagonal components are dominated by electroweak precision data (EWPD) and $Z$-pole measurements. Values represent typical approximate 95\% CL bounds on $C_{Hl}^{(3)} / \Lambda^2$ (in TeV$^{-2}$) from global SMEFT fits, \eg, \Refe~\cite{Ellis:2020unq}.}
    \footnotetext[5]{Limits on $e-\mu$ flavor-violating components are severely constrained by $\mu \to e$ transitions (such as $\mu \to 3e$ and $\mu-e$ conversion in nuclei), which tightly bound flavor-violating $Z$ couplings. Values represent typical approximate 90\% CL bounds on $C_{Hl}^{(3)} / \Lambda^2$ (in TeV$^{-2}$), as derived in SMEFT cLFV analyses, \eg, \Refe~\cite{Crivellin:2013hpa}.}
    \footnotetext[6]{Limits on components involving the $\tau$ lepton are derived from $\tau \to 3e$ and $\tau \to 3\mu$ cLFV decays. Values represent typical approximate 90\% CL bounds on $C_{Hl}^{(3)} / \Lambda^2$ (in TeV$^{-2}$), \eg, \Refe~\cite{Crivellin:2013hpa}.}
 \end{ruledtabular}
\end{table*}

\begin{table*}[t!]
 \begin{ruledtabular}
    \centering
    \caption{\label{tab:results_smeft_uhe}\textbf{\textit{Constraints on dimension-6 SMEFT coefficients from ultra-high-energy astrophysical neutrinos.}} Similar to Table~\ref{tab:results_smeft_full}, but for neutrinos of $> 100$~PeV.  Results are obtained by constraining one parameter at a time while keeping the others controlled via pull terms informed by future~\cite{Song:2020nfh} sub-TeV global oscillation fits.  Flavor-measurement projections are for 10 years of the radio array of IceCube-Gen2 for a low and a high UHE neutrino flux model, as presented in \Refe~\cite{Coleman:2024scd}.  We assume no $\nu_\tau$ production (\ie, $f_{\tau, {\rm S}} = 0$). See Sec.~\ref{sec:results-proj_smeft} for details, Table~\ref{tab:results_smeft_full} for results from TeV--PeV neutrinos, and Appendix~\ref{app:pairwise_constraints} for pairwise parameter constraints.}
    \renewcommand{\arraystretch}{1.3}
    \setlength{\tabcolsep}{4pt} 
    \footnotesize
    \begin{tabular}{l c c c c c}
        \multirow{3}{*}{Parameter} & \multicolumn{4}{c}{Future from UHE ($> 100$~TeV): 10~yr IceCube-Gen2 radio array} & \multirow{3}{*}{\makecell{Approximate \\ existing \\ sub-TeV limits\footnotemark[3]}} \\
        \cline{2-5}
         & \multicolumn{2}{c}{Low UHE neutrino flux\footnotemark[1]} & \multicolumn{2}{c}{High UHE neutrino flux\footnotemark[2]} & \\
        \cline{2-3} \cline{4-5}
         & $\pi$ decay & $\mu$-damped & $\pi$ decay & $\mu$-damped & \\
        \hline
        \multirow{3}{*}{ $C_{11}$ } & \multirow{3}{*}{\makecell{$\cdots$}} & {\scriptsize 1$\sigma$:} \makecell{$< 3.67$ \\ $\cup \, > 104.18$} & \multirow{3}{*}{\makecell{$\cdots$}} & \makecell{$< 1.81$ \\ $\cup \, > 141.65$} & \multirow{3}{*}{$[-0.01, 0.01]$} \\
         &  & {\scriptsize 2$\sigma$:} \makecell{$\cdots$} &  & \makecell{$< 4.93$ \\ $\cup \, > 40.98$} &  \\
         &  & {\scriptsize 3$\sigma$:} \makecell{$\cdots$} &  & \makecell{$\cdots$} &  \\
        \cline{2-6}
        \multirow{3}{*}{ $C_{22}$ } & \multirow{3}{*}{\makecell{$\cdots$}} & $> -6.80$ & \multirow{3}{*}{\makecell{$\cdots$}} & $> -3.56$ & \multirow{3}{*}{$[-0.01, 0.01]$} \\
         &  & \makecell{$\cdots$} &  & \makecell{$< -35.61$ \\ $\cup \, > -8.78$} &  \\
         &  & \makecell{$\cdots$} &  & \makecell{$\cdots$} &  \\
        \cline{2-6}
        \multirow{3}{*}{ $C_{33}$ } & \multirow{3}{*}{\makecell{$\cdots$}} & $> -7.89$ & \multirow{3}{*}{\makecell{$\cdots$}} & $> -3.69$ & \multirow{3}{*}{$[-0.01, 0.01]$} \\
         &  & \makecell{$\cdots$} &  & $> -11.01$ &  \\
         &  & \makecell{$\cdots$} &  & \makecell{$< -32.31$ \\ $\cup \, > -29.33$} &  \\
        \hline
        \multirow{3}{*}{ ${\rm Re}(C_{12})$ } & \multirow{3}{*}{\makecell{$\cdots$}} & $< 130.96$ & \multirow{3}{*}{\makecell{$\cdots$}} & $-0.05_{-28.52}^{+3.92}$ & \multirow{3}{*}{$[-10^{-5}, 10^{-5}]$} \\
         &  & \makecell{$\cdots$} &  & \makecell{$\cdots$} &  \\
         &  & \makecell{$\cdots$} &  & \makecell{$\cdots$} &  \\
        \cline{2-6}
        \multirow{3}{*}{ ${\rm Im}(C_{12})$ } & \multirow{3}{*}{\makecell{$\cdots$}} & \makecell{$\cdots$} & \multirow{3}{*}{\makecell{$\cdots$}} & $0.05_{-7.78}^{+14.02}$ & \multirow{3}{*}{$[-10^{-5}, 10^{-5}]$} \\
         &  & \makecell{$\cdots$} &  & \makecell{$\cdots$} &  \\
         &  & \makecell{$\cdots$} &  & \makecell{$\cdots$} &  \\
        \hline
        \multirow{3}{*}{ ${\rm Re}(C_{23})$ } & \multirow{3}{*}{\makecell{$\cdots$}} & $< 3.57$ & \multirow{3}{*}{\makecell{$\cdots$}} & $< 1.75$ & \multirow{3}{*}{$[-10^{-3}, 10^{-3}]$} \\
         &  & \makecell{$\cdots$} &  & $< 4.80$ &  \\
         &  & \makecell{$\cdots$} &  & \makecell{$\cdots$} &  \\
        \cline{2-6}
        \multirow{3}{*}{ ${\rm Im}(C_{23})$ } & \multirow{3}{*}{\makecell{$\cdots$}} & $> -26.51$ & \multirow{3}{*}{\makecell{$\cdots$}} & $> -14.66$ & \multirow{3}{*}{$[-10^{-3}, 10^{-3}]$} \\
         &  & \makecell{$\cdots$} &  & $> -34.32$ &  \\
         &  & \makecell{$\cdots$} &  & \makecell{$\cdots$} &  \\
        \hline
        \multirow{3}{*}{ ${\rm Re}(C_{13})$ } & \multirow{3}{*}{\makecell{$\cdots$}} & \makecell{$\cdots$} & \multirow{3}{*}{\makecell{$\cdots$}} & $> -17.27$ & \multirow{3}{*}{$[-10^{-3}, 10^{-3}]$} \\
         &  & \makecell{$\cdots$} &  & \makecell{$\cdots$} &  \\
         &  & \makecell{$\cdots$} &  & \makecell{$\cdots$} &  \\
        \cline{2-6}
        \multirow{3}{*}{ ${\rm Im}(C_{13})$ } & \multirow{3}{*}{\makecell{$\cdots$}} & \makecell{$\cdots$} & \multirow{3}{*}{\makecell{$\cdots$}} & $0.00_{-7.16}^{+22.79}$ & \multirow{3}{*}{$[-10^{-3}, 10^{-3}]$} \\
         &  & \makecell{$\cdots$} &  & \makecell{$\cdots$} &  \\
         &  & \makecell{$\cdots$} &  & \makecell{$\cdots$} &  \\
    \end{tabular}
    \footnotetext[1]{Low UHE neutrino flux derived from a fit to Auger UHECR data assuming 10\% of the cosmic rays are protons~\cite{vanVliet:2019nse}.  We add the IceCube flux derived from 9.5 years of through-going tracks~\cite{Abbasi:2021qfz}, extrapolated to ultra-high energies.}
    \footnotetext[2]{High UHE neutrino flux derived from a fit to Telescope Array UHECR data~\cite{Bergman:2021djm}.  We add the IceCube flux derived from 9.5 years of through-going tracks~\cite{Abbasi:2021qfz}, extrapolated to ultra-high energies.}
    \footnotetext[3]{See footnotes in Table~\ref{tab:results_smeft_full} for details.}
 \end{ruledtabular}
\end{table*}

Figure~\ref{fig:results_2d_smeft_muon} and Table~\ref{tab:results_smeft_full} show the projected constraints on the SMEFT coefficients, based on TeV--PeV astrophysical neutrinos, while \figu{results_2d_smeft_uhe} and Table~\ref{tab:results_smeft_uhe} show the corresponding constraints from UHE neutrinos.  Our constraints are exclusively for the case of neutrino production via muon-damped pion decay; we find no sensitivity in the case of production via full pion decay even under our most optimistic TeV--PeV projections.  Our results show that, under muon-damped pion decay, combining exclusively existing TeV--PeV telescopes IceCube, KM3NeT, and Baikal-GVD it will be possible to meaningfully constrain most of the coefficients by 2040.  By 2050, even larger multi-detector combinations will deliver sensitivity to all the coefficients. 

The constraints in Figs.~\ref{fig:results_2d_smeft_muon} and \ref{fig:results_2d_smeft_uhe}, and in Tables~\ref{tab:results_smeft_full} and \ref{tab:results_smeft_uhe}, are derived assuming a fixed new-physics reference scale of $\Lambda_{\text{SMEFT}} = 1$ TeV in \equ{wilson_coeff_matrix}.  Because the effects of dimension-6 operators scale with the ratio $C_{ij} / \Lambda_{\text{SMEFT}}^2$, these constraints  can be easily recast for any arbitrary new-physics scale $\Lambda_{\text{SMEFT}}^\prime$. A bound on $C_{ij}$ at our reference scale translates to a bound on a new coupling $C_{ij}^\prime$ at the scale $\Lambda_{\text{SMEFT}}^\prime$ via  $C_{ij}^\prime = C_{ij} (\Lambda_{\text{SMEFT}}^\prime / \Lambda_{\text{SMEFT}})^2$.

\medskip

\textbf{\textit{UHE vs.~TeV--PeV constraints.---}}Because the effects of dimension-6 SMEFT operators scale as $E_\nu/\Lambda_{\rm SMEFT}^2$, we might intuitively expect the UHE regime to provide the most stringent bounds. However, our results demonstrate the opposite: UHE neutrinos yield drastically weaker constraints than TeV--PeV neutrinos. As shown in \figu{results_2d_smeft_uhe} and Table~\ref{tab:results_smeft_uhe}, 10 years of observation with the projected IceCube-Gen2 radio array leaves the SMEFT coefficients largely unconstrained, with the $1\sigma$ and $2\sigma$ allowed regions spanning essentially the entire massive numerical scan window, regardless of the assumed UHE flux model. 

This severe degradation in sensitivity is driven by two effects.  The first effect---as prefigured in Sec.~\ref{sec:rg_running-trajectory_flavor_space}---is that, despite the larger EeV energies of UHE neutrinos, their accessible $Q$-distribution is only slightly higher than that of TeV--PeV neutrinos, as shown in Figs.~\ref{fig:main_results} and \ref{fig:flavor_smeft_rg_vs_Q} (also \figu{ternary_theory_smeft}, left panel).  This means that effectively the RG deviations accessible by UHE neutrinos are comparable to those accessible by TeV--PeV neutrinos.  The second, dominant effect is the experimental limitation of EeV flavor measurements, as shown in \Refe~\cite{Coleman:2024scd}. Although the SMEFT-induced flavor shifts at UHE scales are theoretically slightly magnified, the massive uncertainties associated with radio-based flavor tagging completely swallow these deviations. Consequently, the competitive precision achievable with optical TeV--PeV neutrino telescopes (\figu{results_2d_smeft_muon}) cannot be replicated in the UHE regime, cementing the TeV--PeV band as the optimal window for these model-independent flavor studies.  

Combining the UHE flavor measurements performed in the radio array of IceCube-Gen2 with $\nu_\tau$-dedicated measurements in next-generation detectors GRAND~\cite{GRAND:2018iaj}, TAMBO~\cite{TAMBO:2025jio}, or Trinity~\cite{Otte:2018uxj}, could boost the precision of UHE flavor measurements.  This combination would tighten the flavor measurements along the $\nu_\tau$ direction, which is roughly orthogonal to the orientation of the measurement contours accessible by IceCube-Gen2, as shown explicitly for GRAND in \Refe~\cite{Testagrossa:2023ukh}.

\medskip

\textbf{\textit{No all-coefficient sensitivity.---}}Appendix~\ref{app:smeft_regions_single_parameter} shows that, while the flavor regions generated by varying all SMEFT coefficients simultaneously  (\figu{ternary_theory_smeft}) remains too close to standard expectations for any parameter to be bounded by projected TeV--PeV flavor measurements, the regions generated by varying a single SMEFT coefficient at a time reach far enough across the flavor triangle to be bounded.  The relatively featureless all-coefficient region in \figu{ternary_theory_smeft} is the result of destructive interference between multiple SMEFT coefficients, which does not occur in the single-coefficient scenario, yielding more extreme flavor predictions.  This explains why we are able to place robust constraints on the SMEFT coefficients using TeV--PeV neutrinos when they are varied individually (Table~\ref{tab:results_smeft_full}), but cannot do so when all coefficients are allowed to float simultaneously.  

\medskip

\textbf{\textit{No sensitivity to full pion decay.---}}We find no sensitivity to the SMEFT coefficients in the case of neutrino production via full pion decay. Appendix~\ref{app:analytical_approximation_smeft} explains why in detail: the RG-induced deviations of the flavor composition are too small compared to even our 2050 TeV--PeV flavor-measurement projections to detect (\cf~Figs.~\ref{fig:flavor_smeft_rg_vs_Q}, \ref{fig:flavor_smeft_rg_vs_c_smeft} vs.~\figu{ternary_theory}).  The underlying reason stems from the weak sensitivity to the high-$Q$ mixing parameters that we found in Sec.~\ref{sec:results-proj_mix_params}: the value of $f_{e, {\rm S}} = 1/3$ implies only a small difference in the $\nu_1$ and $\nu_2$ populations produced by the sources, which dampens any RG-induced modification of the mixing parameters (see Appendix~\ref{app:analytical_approximation_smeft-flavor_distance_pion_decay}). In contrast, the difference between the $\nu_1$ and $\nu_2$ populations is larger in the case of neutrino production via muon-damped pion decay, leading to our ability to constrain the SMEFT coefficients in this case using TeV--PeV neutrinos.

\medskip

\textbf{\textit{Parameter sensitivity.---}}To understand the differences in the constraints on the different SMEFT coefficients seen in \figu{results_2d_smeft_muon} and Table~\ref{tab:results_smeft_full}, we must map how each coefficient alters specific mixing angles, how those angles subsequently affect the predicted flavor composition at Earth, and whether their effect can be identified in view of the experimental sensitivity to the flavor composition.  Appendix~\ref{app:analytical_approximation_smeft} explains in detail the origin of the sensitivity to the different SMEFT coefficients.  Below, we present only the salient features.

As shown in \equ{th12_evol_gen} and, more clearly, \equ{shift_smeft_th12}, the SMEFT RG evolution affects most prominently the angle $\theta_{12}$, due to its RG equation being enhanced by the larger value of $1/\Delta m_{21}^2$ (vs.~the smaller $1/\Delta m_{31}^2$ that drives the RG evolution of the other mixing parameters).  The combination of values of the standard, low-$Q$ mixing parameters (adopted from NuFIT 6.1) assigns the highest impact on the evolution to the specific SMEFT parameters $C_{11}$ and ${\rm Re}(C_{23})$, via large multiplicative factors; see Eqs.~(\ref{equ:shift_smeft_th12}) and (\ref{equ:distance_smeft_muon_damped}).  

This makes our analysis particularly sensitive to these two parameters, as evidenced by the fact that they are the most tightly constrained ones in the TeV--PeV regime (in \figu{results_2d_smeft_muon} and Table~\ref{tab:results_smeft_full}).  Increasing the values of these two coefficients increases the electron content at Earth ($f_{e, \oplus}$) and decreases the muon and tau flavor content ($f_{\mu, \oplus}$, $f_{\tau, \oplus}$), visibly shifting the flavor composition away from our projected flavor measurement centered on the standard-mixing expectation from muon-damped pion decay, making the shift experimentally detectable in the TeV--PeV band, and allowing us to place upper limits on the coefficients.  In contrast, the other SMEFT coefficients carry comparatively smaller weights in the RG evolution of the mixing parameters and induce smaller flavor shifts [see \equ{distance_smeft_muon_damped}], explaining the weaker constraints on them.

\medskip

\textbf{\textit{Comparison to existing bounds.---}}Table~\ref{tab:results_smeft_full} shows that our projected constraints on the dimension-6 SMEFT coefficients complement existing bounds from terrestrial global fits, which synthesize data from LEP, the LHC, and low-energy flavor experiments. While current terrestrial limits are formidable---restricting flavor-diagonal couplings to $10^{-3}$--$10^{-1}$ and lepton-flavor-violating (LFV) couplings to $10^{-6}$ for $\Lambda_{\rm SMEFT} = 1$ TeV---they operate in different kinematic regimes. 

Bounds on the Higgs, diboson, and top sectors rely on hard-scattering events from the LHC (such as those analyzed by the SMEFiT Collaboration~\cite{Ethier:2021bye}) with momentum transfers $Q \lesssim 3$~TeV.  Meanwhile, precision LFV bounds rely on high-intensity, low-energy (MeV--GeV) processes, such as rare charged-lepton decays. In contrast, our limits exploit astrophysical neutrinos interacting and propagating in the unprecedented TeV–PeV regime (a feat currently out of reach for radio-based UHE observatories). \textbf{\textit{Consequently, while terrestrial experiments yield tighter absolute bounds, high-energy astrophysical neutrinos provide an entirely independent dynamical test---one based exclusively on neutrinos---at scales inaccessible to other neutrino experiments.}} 

However, comparing these projections directly to terrestrial bounds requires care. Modern collider constraints are predominantly derived from global, marginalized fits where numerous operators vary simultaneously, whereas the limits derived here assume a single active operator at a time. Further, terrestrial limits are evaluated at the hard scattering scale (\eg, $\mu \sim M_Z$ or the TeV scale), while our astrophysical bounds inherently integrate over the RG evolution of the couplings from the production energy scale to the new-physics detection scale $\Lambda_{\rm SMEFT}$. 

\medskip

In summary, the capacity to bound dimension-6 SMEFT coefficients using high-energy astrophysical neutrinos complements terrestrial experiments. While the UHE regime lacks the experimental flavor resolution necessary to meaningfully restrict these parameters, the TeV--PeV regime will soon reach it. 


\subsection{Perspectives: distinguishing between Dirac and Majorana neutrinos}

The model-independent framework developed here has a natural extension to the question of the fundamental nature of neutrinos.
A Majorana heavy neutral lepton (HNL) with mass $M$ in the GeV range, if integrated
out at the scale $Q \sim M$, generates a genuine threshold correction to the
light neutrino mixing angles---a physical shift in $U'$ that is absent for a
Dirac state with identical mass and flavor couplings, since no Majorana mass
matrix exists to correct. Importantly, the pattern of this correction across the three mixing angles is
not arbitrary: it is determined by the flavor structure of the couplings of the HNL to the active neutrinos, so that different HNL flavor alignments
produce distinct and predictable relative shifts in $\theta_{12}$,
$\theta_{23}$, and $\theta_{13}$. This flavor fingerprint is in principle observable.

The SHiP experiment~\cite{SHiP:2015vad} is specifically designed to probe exactly this
mass range, with sensitivity to active-sterile mixing-matrix elements $|U_{\alpha N}|^2 \sim 10^{-5}$---the low-mass, small-coupling corner inaccessible to HNL searches at the LHC---and can measure the mass and flavor couplings $|U_{eN}|^2$, $|U_{\mu N}|^2$, $|U_{\tau N}|^2$.

Crucially, the $Q$ values corresponding to this mass range overlap with the transferred momentum distribution of TeV--PeV astrophysical neutrinos
(top panel of \figu{main_results}, and Figs.~\ref{fig:flavor_smeft_rg_vs_Q} and \ref{fig:combined_pq}). This makes the combination of SHiP measurements and high-energy astrophysical flavor measurements a powerful probe of HNLs. SHiP provides $M$ and the flavor couplings, from which the expected threshold corrections to $\theta_{12}$, $\theta_{23}$, $\theta_{13}$, and $\delta_{\rm CP}$ can be computed.  Next-generation neutrino telescopes then test whether $U'$ deviates from $U$ at the corresponding $Q$, by the predicted amount, and in the predicted flavor pattern. A Dirac state of identical mass and couplings would produce no such deviation.

This interpretation becomes progressively sharper in light of null results at the LHC: if no BSM states are discovered in collider searches, including indirect constraints from precision electroweak observables and lepton flavor violation, the only sector capable of generating a physical threshold correction in $U'(Q)$ at accessible scales while remaining consistent with all collider bounds is the feeble-coupling neutrino sector itself. In this sense, a comprehensive LHC null result combined with an observed deviation in $U'(Q)$ at a scale $Q \sim M$ would constitute model-independent evidence for the Majorana nature of neutrinos, with \ship~providing the direct corroboration through the identification of the responsible state and the measurement of its flavor couplings.

Since the two measurements carry entirely uncorrelated systematic uncertainties---one being a controlled laboratory experiment, the other integrating neutrino propagation over cosmological baselines---their
consistency under the Majorana hypothesis, and inconsistency under the Dirac one, would be a compelling and independent probe of the nature of neutrino mass in a regime inaccessible to neutrinoless double beta decay searches.

However, this program is only sensitive to threshold corrections above some minimum size,
set by the precision achievable in astrophysical flavor-ratio measurements. As our results show, this precision improves substantially with multi-telescope combinations, but remains a limiting factor for HNLs with very small mixing angles to active neutrinos. We defer a quantitative development of this argument---including the mapping of threshold corrections onto the \ship~sensitivity contours and a systematic
comparison with the reach of neutrinoless double beta decay searches---to a
forthcoming study.


\section{Summary and outlook}
\label{sec:summary}

High-energy astrophysical neutrinos offer a largely untapped regime to test the standard neutrino mixing paradigm. While the neutrino mixing parameters are measured with high precision in low-energy oscillation experiments (momentum transfer $Q \sim \text{GeV}$), new physics---such as dimension-6 operators in the Standard Model Effective Field Theory (SMEFT)---can induce renormalization group (RG) running of these parameters at the 10--100~GeV momentum scale accessible by TeV--PeV astrophysical neutrinos in optical detectors like IceCube, and by neutrinos in excess of 100~PeV in planned ultra-high-energy (UHE) radio detectors like the radio array of IceCube-Gen2.

We explored the sensitivity of the flavor composition of high-energy astrophysical neutrinos at Earth---the proportions of $\nu_e$, $\nu_\mu$, and $\nu_\tau$ in the flux---to such high-$Q$ modifications through a two-pronged approach (Sec.~\ref{sec:flavor_comp}). First, to establish broad, model-independent bounds, we treated the high-$Q$ mixing angles ($\theta_{12}^\prime, \theta_{23}^\prime, \theta_{13}^\prime$) and the CP-violation phase ($\delta_{\rm CP}^\prime$) as phenomenological parameters (Sec.~\ref{sec:flavor_comp-osc_mod}). Second, to ground these effects in theory, we evaluated the RG evolution of the mixing parameters induced by dimension-6 SMEFT operators (Sec.~\ref{sec:rg_running}). Across both frameworks, we computed the resulting deformations to the expected flavor fractions at Earth, $f_{\alpha, \oplus}$, and determined the capability of current (IceCube MESE) and future (multi-detector optical-Cherenkov and UHE radio arrays) neutrino telescopes to constrain them (Secs.~\ref{sec:statistics} and \ref{sec:results}).

The main takeaways of our analysis are as follows:
\begin{itemize}
 \item
  \textbf{High-$Q$ sensitivity is driven by single-parameter dominance:} Allowing a single high-$Q$ mixing parameter to deviate from its standard low-$Q$ value forces the flavor composition along narrow, extreme trajectories in the flavor triangle. This allows combinations of future TeV--PeV neutrino telescopes to place constraints on individual high-$Q$ parameters with a precision of about 10\% (while future UHE telescopes are expected to yield significantly weaker constraints). Conversely, if all parameters run simultaneously, the resulting cross-terms and operator interference wash out these extreme deviations. The accessible flavor space shrinks toward the standard-mixing expectation, precluding competitive bounds on the fully flexible high-$Q$ mixing matrix (see Sec.~\ref{sec:statistics-mixing_params} and \ref{sec:flavor_comp-regions_earth}).
 \item
  \textbf{Hierarchy of parameter sensitivity:} Assuming the experimental resolution of TeV--PeV neutrino telescopes, the flavor composition at Earth is most sensitive to modifications in $\theta_{23}^\prime$ and $\theta_{13}^\prime$. Deviations in $\theta_{23}^\prime$ break the $\mu$-$\tau$ symmetry dominant in standard mixing, resulting in a measurable asymmetry between $f_{\mu, \oplus}$ and $f_{\tau, \oplus}$. Modifications to $\theta_{13}^\prime$ exert a disproportionately large effect due to the smallness of the standard $\theta_{13}$ value. In contrast, sensitivity to $\theta_{12}^\prime$ is degraded by its degeneracy with uncertainties in the initial source composition, and the impact of $\delta_{\rm CP}^\prime$ remains negligible, as under standard mixing (see Sec.~\ref{sec:results-proj_mix_params} and Appendix~\ref{app:analytical_approximation_smeft}).
 \item
  \textbf{Purer astrophysical sources maximize sensitivity:} The ability to constrain high-$Q$ mixing is heavily dependent on the neutrino production mechanism. The canonical full pion-decay scenario, yielding $\left( \frac{1}{3}, \frac{2}{3}, 0 \right)_{\rm S}$, produces highly mixed transition probabilities that inherently dilute the observable impact of RG running. In contrast, initial states with higher flavor purity---such as muon-damped pion decay [$(0, 1, 0)_{\rm S}$] or neutron decay [$(1, 0, 0)_{\rm S}$]---transfer high-$Q$ mixing modifications directly to the flavor fractions at Earth, enhancing experimental sensitivity (see Sec.~\ref{sec:flavor_ratios-theory} and \ref{sec:flavor_ratios-impact_sources}).
 \item
  \textbf{Astrophysical unknowns dilute constraints:} Because the exact mechanism of high-energy neutrino production remains observationally unknown, robust parameter extraction requires our limit-setting procedure to profile over all possible flavor compositions at the sources. Adhering to the standard assumption that $\nu_\tau$ production at the sources is negligible ($f_{\tau, \rm S} = 0$), we must allow the initial electron fraction, $f_{e, \rm S}$, to vary freely. This broad profiling introduces an unavoidable degeneracy: an observed flavor composition at Earth could be the result of high-$Q$ modifications to the mixing, or it could simply stem from an atypical but standard astrophysical source mixture. Thus, accommodating our ignorance of the true flavor composition at the sources severely dilutes our limit-setting power on the high-$Q$ mixing parameters and the SMEFT coefficients (see Sec.~\ref{sec:statistics-mixing_params} and Appendix~\ref{app:results_with_nu_tau_production}).
 \item
  \textbf{Current limits versus the future precision:} Present flavor measurements, such as the IceCube 11.4-year MESE analysis, remain too broad to meaningfully distinguish between standard mixing and extreme RG-modified scenarios. However, by 2040 and 2050, the combined statistics of km$^3$-scale and tens-of-km$^3$-scale optical-Cherenkov networks (IceCube-Gen2, Baikal-GVD, KM3NeT, P-ONE, NEON, TRIDENT, HUNT) will shrink observational uncertainties to the precision required to test these high-$Q$ effects (see Sec.~\ref{sec:flavor_comp-measuring} and \ref{sec:results-proj_mix_params}).
 \item
  \textbf{UHE radio arrays as an independent diagnostic:} At energies above 100 PeV, planned large-scale UHE telescopes like the IceCube-Gen2 radio array will provide independent constraints on the UHE flavor composition. By leveraging the LPM effect to identify $\nu_e$-induced electromagnetic showers and catastrophic energy losses to identify $\nu_\mu$- and $\nu_\tau$-induced showers, these arrays will constrain the $f_{e, \oplus}$ and $f_{\mu, \oplus} + f_{\tau, \oplus}$ flavor fractions in the energy regime where beyond-the-Standard-Model effects are expected to be further enhanced, albeit with significantly lower precision than in TeV--PeV telescopes (see Sec.~\ref{sec:flavor_comp-measuring}).
 \item
  \textbf{Bounds only on isolated SMEFT coefficients:} When assuming neutrino mixing is modified only by a single dimension-6 SMEFT operator (setting all others to zero), the RG evolution of the mixing parameters forces the expected flavor composition at Earth into extreme, extended trajectories (``spikes'') on the flavor triangle. While current measurements (\ie, 11.4 years of IceCube MESE) are too broad to restrict these trajectories, the projected precision of 2040 and 2050 TeV--PeV multi-detector combinations will be capable of placing robust limits on individual SMEFT coefficients (see Sec.~\ref{sec:results-proj_smeft} and Appendix~\ref{app:smeft_regions_single_parameter}).
 \item
  \textbf{Operator interference limits global SMEFT constraints:} A critical limitation arises when all SMEFT coefficients are allowed to float simultaneously. In this generalized scenario, RG evolution induces extensive operator mixing and cross-terms. This interference predominantly results in cancellations that wash out the extreme topologies of the single-parameter cases, causing the accessible flavor footprint to shrink back toward the standard-mixing expectation. Consequently, placing simultaneous constraints on the full suite of SMEFT coefficients using flavor composition alone is precluded by this inherent degeneracy, even under our most optimistic TeV--PeV projections (see Sec.~\ref{sec:statistics-smeft} and Appendix~\ref{app:smeft_regions_single_parameter}).
 \item
  \textbf{UHE constraints are severely limited by experimental resolution:} Although the linear energy scaling of dimension-6 SMEFT effects ($E_\nu/\Lambda_{\rm SMEFT}^2$) intuitively suggests UHE neutrinos should provide the most stringent bounds, the RG-induced mixing modifications scale only logarithmically with momentum transfer $Q$, yielding remarkably similar $Q$-distributions for both TeV--PeV and UHE neutrinos. This is driven both by the steeply falling astrophysical flux, which weights event samples toward lower energies, and by the weak-boson propagator, which strongly suppresses momentum transfers much above the electroweak scale ($Q \sim M_W$). Further, we have shown that constraints derived from UHE neutrinos are drastically weaker than from TeV--PeV neutrinos. The experimental challenges of EeV radio-detection---namely, low event statistics and larger uncertainties in flavor tagging compared to optical-Cherenkov telescopes---complicate finding any anomalous BSM flavor shifts. This cements the high-statistics TeV--PeV regime as the primary driver to constrain high-energy neutrino mixing (see Sec.~\ref{sec:results-proj_mix_params}, \ref{sec:results-proj_smeft}, and Appendix~\ref{app:momentum_distribution}).
\end{itemize}

How could future tests of the RG evolution of neutrino mixing with high-energy astrophysical neutrinos be improved?  Our forecasts have implicitly assumed that the experimental and analysis techniques for high-energy neutrinos will remain essentially unchanged from their present status.  However, two advances could boost tests of RG (and of new physics in general):
\begin{itemize}
 \item 
  \textbf{Improving flavor separation:} The use of late-time, dim Cherenkov light echoes from low-energy muons and neutrons following the primary light from a $\nu N$ DIS shower could help differentiate between $\nu_e$ and $\nu_\tau$, breaking the prevailing degeneracy between these two flavors.  This proposal, first presented in \Refe~\cite{Li:2016kra}, has seen promising preliminary results based on IceCube data~\cite{Farrag:2023jut, Dutta:2025qgk}. 
 \item 
  \textbf{Breaking the source flavor degeneracy:} Currently, the precision of our bounds is fundamentally limited by the need to marginalize over the unknown flavor composition at the sources. However, future coincident electromagnetic observations of transient neutrino sources---such as blazar flares, tidal disruption events, or gamma-ray bursts---could directly probe the physical environment of the neutrino-production region. If multi-messenger data confirm conditions like extreme magnetic fields or dense photon targets that guarantee catastrophic muon cooling, the source composition for those events can be set to muon-damped pion decay. By stacking the sparse neutrino detections from multiple identically-classed sources, we could build the statistical power necessary to measure their specific flavor composition at Earth.  Breaking this astrophysical degeneracy would drastically shrink the allowed flavor parameter space at Earth, strengthening our constraints. 
\end{itemize}

Ultimately, the standard, energy-independent formulation of three-flavor neutrino mixing projects a remarkably narrow and rigid region of allowed flavor compositions at Earth (see Sec.~\ref{sec:flavor_comp-regions_earth}). This rigidity constitutes an exceptionally pristine null hypothesis. As the global network of neutrino telescopes expands and matures over the next two decades, the measurement of high-energy astrophysical neutrino flavor composition will transition from a statistically limited estimation to a precision probe. If the high-statistics flavor measurements of the 2040s and 2050s converge anywhere outside this narrow standard corridor, it will provide a definitive signature of new dynamics governing neutrino mixing at high energies.


\acknowledgements

The authors are grateful for the contribution of Bernanda Telalovic in the early stages of this work.   MB is supported by {\sc Villum Fonden} under project no.~29388. This work used the Tycho supercomputer hosted at the SCIENCE High Performance Computing Center at the University of Copenhagen.  QL is supported by Canada First Research Excellence Fund and Natural Sciences and Engineering Research Council of Canada through the Arthur B. McDonald Canadian Astroparticle Physics Research Institute.  GB is supported by the Spanish grants CIPROM/2021/054 (Generalitat Valenciana), PID2023-151418NB-I00 funded by MCIU/AEI/10.13039/501100011033/, and by the European ITN project HIDDeN (H2020-MSCA-ITN-2019/860881-HIDDeN).



%


\newpage
\clearpage
\appendix

\begin{center}
\textbf{List of appendices}
\end{center}
\vspace{-4.5em}
\begin{itemize}[label=, itemsep=0pt, leftmargin=*]
 \item
  \textbf{Appendix \ref{app:rge_approximations}:} Analytical details of the RG evolution
 \item
  \textbf{Appendix \ref{app:momentum_distribution}:} Distribution of momentum transfer in neutrino-nucleon deep inelastic scattering
 \item
  \textbf{Appendix \ref{app:flavor_composition_muon_damped}:} Flavor composition at Earth with RG running under muon-damped pion decay
 \item
  \textbf{Appendix \ref{app:allowed_flavor_regions}:} Breakdown of allowed regions of flavor composition at Earth
 \item
  \textbf{Appendix \ref{app:pairwise_constraints}:} Detailed numerical results
 \item
  \textbf{Appendix \ref{app:results_with_nu_tau_production}:} Parameter constraints allowing $\nu_\tau$ production
 \item
  \textbf{Appendix \ref{app:analytical_approximation_smeft}:} Analytical approximation of high-$Q$ and SMEFT-induced flavor shifts
 \item
  \textbf{Appendix \ref{app:smeft_regions_single_parameter}:} Single-parameter SMEFT-induced flavor composition regions
\end{itemize}
\vspace{1em}


\section{Analytical details of the RG evolution}
\label{app:rge_approximations}

\renewcommand{\theequation}{A\arabic{equation}}
\renewcommand{\thefigure}{A\arabic{figure}}
\renewcommand{\thetable}{A\arabic{table}}
\setcounter{figure}{0} 
\setcounter{table}{0} 

In this appendix, we provide the supplementary derivations linking the flavor-basis SMEFT couplings to the exact PMNS generator expansions needed to fully understand and reproduce the analytical approximations of the RG evolution of the mixing parameters presented in Sec.~\ref{sec:rg_running-impact_smeft_coefficients} in the main text.


\subsection{Explicit mass-basis operator expansions}

To derive the coefficient-by-coefficient approximations presented in the main text (Sec.~\ref{sec:rg_running-impact_smeft_coefficients}), we must expand the mass-basis elements $\tilde{C}_{ij} = \sum_{\alpha, \beta} \tilde{U}_{\alpha i}^* C_{\alpha \beta} \tilde{U}_{\beta j}$. Because the standard reactor angle is small ($s_{13} \approx 0.15$), we evaluate these at leading order, \ie, taking $s_{13} \to 0$ and $\tilde{U}$ as real. Splitting $C_{\alpha\beta}$ into its real and imaginary parts reveals how the operators cleanly partition into angle-driving and phase-driving components:
\begin{align}
 \text{Re}(\tilde{C}_{12})
 &=
 \frac{1}{2}\sin 2\theta_{12} (C_{11} - c_{23}^2 C_{22} - s_{23}^2 C_{33}) \nonumber \\
 &\quad + c_{23} \cos 2\theta_{12} \text{Re}(C_{12})
 - s_{23} \cos 2\theta_{12} \text{Re}(C_{13}) \nonumber \\
 &\quad + \frac{1}{2} \sin 2\theta_{12} \sin 2\theta_{23} \text{Re}(C_{23}) \;, \\
 \text{Im}(\tilde{C}_{12})
 &=
 c_{23} \text{Im}(C_{12}) - s_{23} \text{Im}(C_{13}) \;, \\
 \text{Re}(\tilde{C}_{13})
 &=
 \frac{1}{2} s_{12} \sin 2\theta_{23} (C_{33} - C_{22}) \nonumber \\
 &\quad + c_{12} s_{23} \text{Re}(C_{12}) + c_{12} c_{23} \text{Re}(C_{13}) \nonumber \\
 &\quad - s_{12} \cos 2\theta_{23} \text{Re}(C_{23}) \;, \\
 \text{Im}(\tilde{C}_{13})
 &=
 c_{12} s_{23} \text{Im}(C_{12}) + c_{12} c_{23} \text{Im}(C_{13}) \nonumber \\
 &\quad - s_{12} \text{Im}(C_{23}) \;, \\
 \text{Re}(\tilde{C}_{23})
 &=
 \frac{1}{2} c_{12} \sin 2\theta_{23} (C_{22} - C_{33}) \nonumber \\
 &\quad + s_{12} s_{23} \text{Re}(C_{12}) + s_{12} c_{23} \text{Re}(C_{13}) \nonumber \\
 &\quad + c_{12} \cos 2\theta_{23} \text{Re}(C_{23}) \;, \\
 \text{Im}(\tilde{C}_{23}) &= s_{12} s_{23} \text{Im}(C_{12}) + s_{12} c_{23} \text{Im}(C_{13}) \nonumber \\
 &\quad + c_{12} \text{Im}(C_{23}) \;.
\end{align}


\subsection{Exact PMNS generator expansions}

To extract the derivatives of the individual PMNS parameters from $T_{ij}$ in \equ{Tij} in the main text, we evaluate the anti-Hermitian generator $T = \tilde{U}^\dagger \dot{\tilde{U}}$ using the standard parametrization of $\tilde{U}$. The exact expansions for the off-diagonal elements are:
\begin{align}
 T_{12}
 &=
 \dot{\theta}_{12} + s_{13} \cos\delta_{\text{CP}} \dot{\theta}_{23} \\
 &\quad + i \left( s_{12} c_{12} s_{13}^2 \dot{\delta}_{\text{CP}} - s_{13} \sin\delta_{\text{CP}} \cos 2\theta_{12} \dot{\theta}_{23} \right) \;, 
 \nonumber \\
 T_{13}
 &=
 c_{12} e^{-i\delta_{\text{CP}}} \dot{\theta}_{13} - s_{12} c_{13} \dot{\theta}_{23} \nonumber \\
 &\quad - i c_{12} s_{13} c_{13} e^{-i\delta_{\text{CP}}} \dot{\delta}_{\text{CP}} \;, 
 \\ 
 T_{23}
 &=
 s_{12} e^{-i\delta_{\text{CP}}} \dot{\theta}_{13} + c_{12} c_{13} \dot{\theta}_{23} \nonumber \\
 &\quad - i s_{12} s_{13} c_{13} e^{-i\delta_{\text{CP}}} \dot{\delta}_{\text{CP}} \;. 
\end{align}

By extracting the real and imaginary parts of these $T_{ij}$ expressions (applying the necessary phase rotation $e^{i\delta_{\text{CP}}}$ to $T_{13}$), equating them to the right-hand side of \equ{Tij}, and taking $s_{13} \to 0$ alongside the explicit $\tilde{C}_{ij}$ components derived above, we recover the leading-order coefficient-by-coefficient approximations presented in Sec.~\ref{sec:rg_running-impact_smeft_coefficients} in the main text.


\section{Distribution of momentum transfer in neutrino-nucleon deep inelastic scattering}
\label{app:momentum_distribution}

\renewcommand{\theequation}{B\arabic{equation}}
\renewcommand{\thefigure}{B\arabic{figure}}
\renewcommand{\thetable}{B\arabic{table}}
\setcounter{figure}{0} 
\setcounter{table}{0} 

\begin{figure*}[t!]
 \centering
 \includegraphics[width=\textwidth]{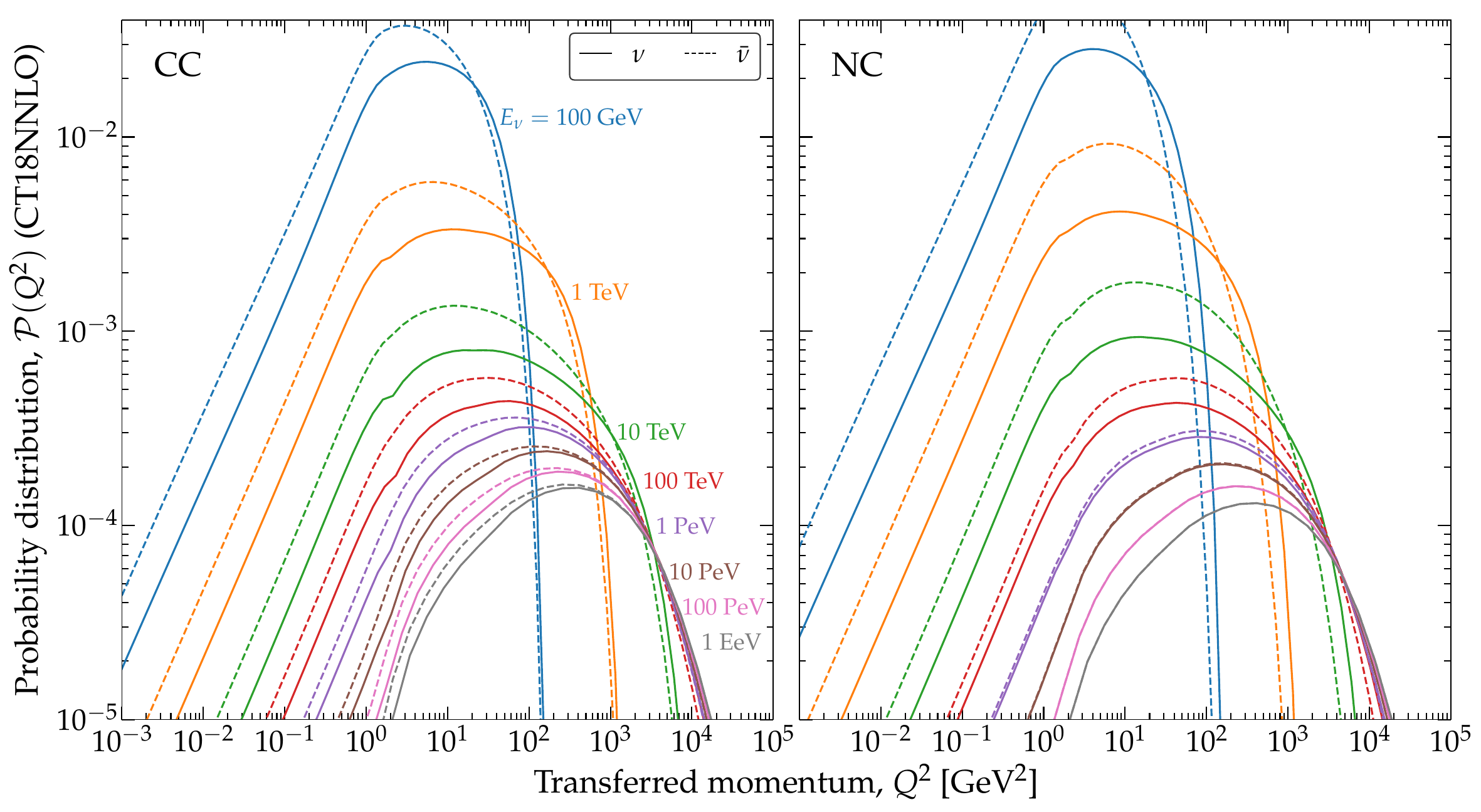}
 \caption{\textbf{Distribution of transferred momentum for fixed neutrino energies.} The distributions are of $Q$, the transferred momentum in neutrino-nucleon ($\nu N$) deep inelastic scattering (DIS).  They are computed for different representative choices of fixed neutrino energy via \equ{pq2_fixed_energy}, using the CT18NNL0 PDFs~\cite{Hou:2019efy} to calculate the $\nu N$ DIS cross section.  \textit{Left:} for charged-current (CC) interactions.  \textit{Right:} for neutral-current (NC) interactions.}
 \label{fig:fixed_energy_q2}
\end{figure*}

To evaluate the probability distribution of momentum transferred in high-energy astrophysical neutrino interactions, we compute the probability density function of the momentum transfer $Q = \sqrt{Q^2}$. This requires calculating the interaction cross sections at the parton level, transforming the kinematic variables, and subsequently folding these distributions over the expected astrophysical neutrino energy spectra. 


\subsection{Differential cross sections and fixed-energy momentum distributions}
\label{subsec:diff_cross_sections}

\begin{figure}[t!]
 \centering
 \includegraphics[width=\columnwidth]{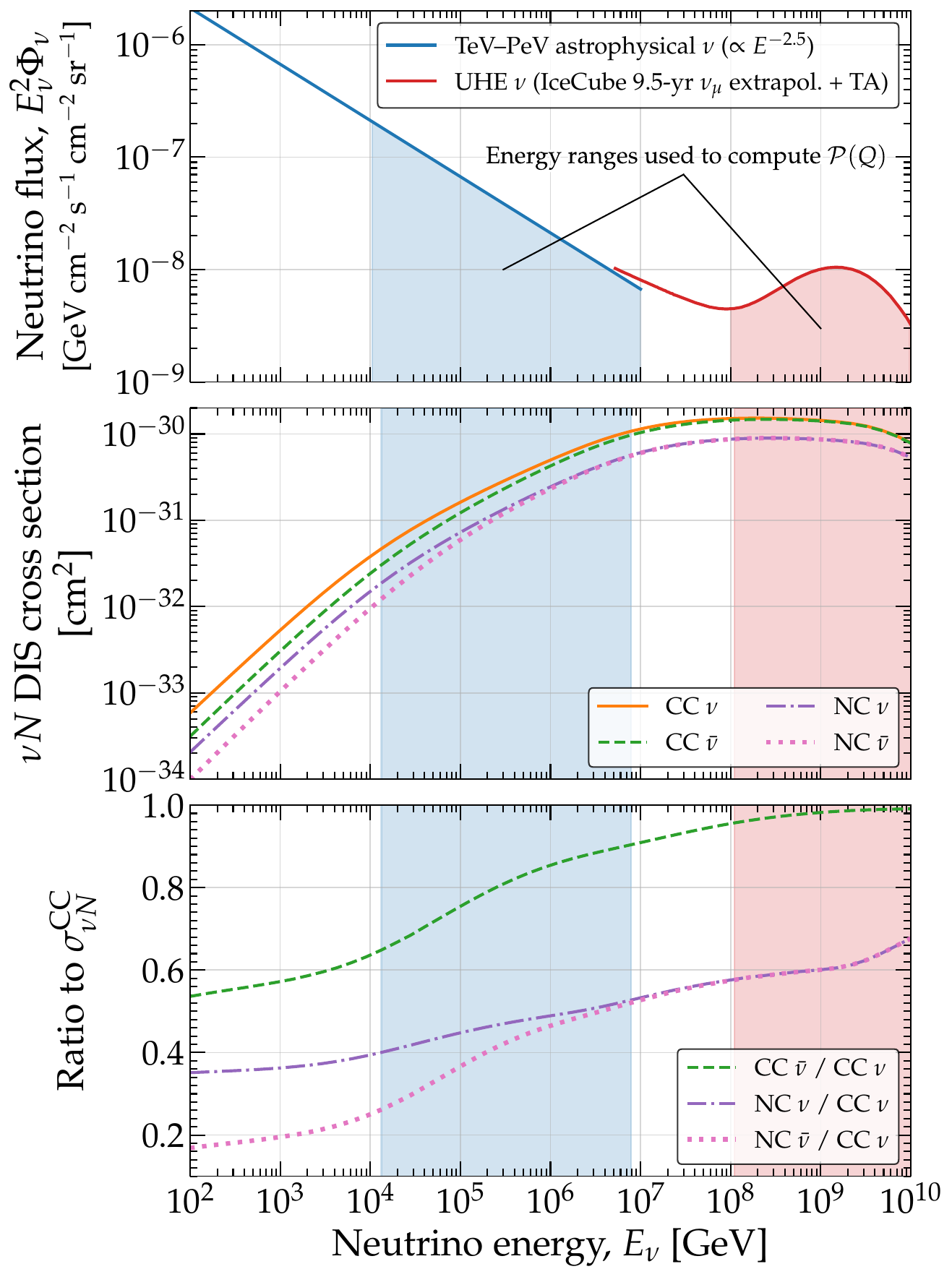}
 \caption{\textbf{Ingredients to compute the probability distribution of transferred momentum.}  \textit{Top:} High-energy astrophysical neutrino flux models to compute the energy weighing in TeV--PeV and UHE ranges, $\mathcal{P}(Q)$ in \equ{P_final}.  \textit{Middle:} Total cross section for neutrino and anti-neutrino charged-current (CC) and neutral-current (NC) deep inelastic scattering on nucleons, computed using the CT18NNLO parton distribution functions.  \textit{Bottom}: Ratios between the different interaction channels, used when computing the channel-combined $Q$-distribution, $\mathcal{P}_\text{comb}$ in \equ{P_comb}.}
 \label{fig:cross_sections_and_fluxes}
\end{figure}

We consider deep inelastic scattering (DIS) of neutrinos ($\nu$) and anti-neutrinos ($\bar{\nu}$) on nucleons ($N$) via charged-current (CC, mediated by the $W^\pm$ boson) and neutral-current (NC, mediated by the $Z$ boson) interactions. The inclusive double-differential cross section with respect to the Bjorken scaling variable $x$ and the inelasticity $y$ is given by $\frac{d^2\sigma_i}{dx dy}$, where $i \in \{\text{CC}\nu, \text{CC}\bar{\nu}, \text{NC}\nu, \text{NC}\bar{\nu}\}$ denotes the interaction channel.  In our results, these cross sections are evaluated using the CT18NNLO parton distribution functions (PDFs).

At the parton level, the double-differential cross section is evaluated by convolving the point-like neutrino-parton scattering cross sections with the PDFs, which describe the probability of finding a specific parton flavor carrying a momentum fraction $x$ of the nucleon at a resolution scale $Q^2$. For example, the leading-order charged-current cross section for neutrino-nucleon scattering explicitly combines the quark ($q$) and anti-quark ($\bar{q}$) distributions as
\begin{widetext}
\begin{equation}
    \frac{d^2\sigma_{{\rm CC}\nu}}{dx dy} = \frac{G_F^2 M_N E_\nu}{\pi} \left( \frac{M_W^2}{Q^2 + M_W^2} \right)^2 x \left[ \sum_{q=d,s,b} q(x,Q^2) + (1-y)^2 \sum_{\bar{q}=\bar{u},\bar{c}} \bar{q}(x,Q^2) \right] \;,
\end{equation}
\end{widetext}
where $G_F$ is the Fermi coupling constant and $M_W$ is the $W$-boson mass. In this expression, the $(1-y)^2$ kinematic factor reflects the helicity suppression inherent to the vector-minus-axial-vector ($V-A$) structure of the weak interaction, which suppresses the scattering of left-handed neutrinos on right-handed anti-quarks at large inelasticities. Analogous combinations are constructed for anti-neutrino interactions, which probe the $u$, $c$, $\bar{d}$, $\bar{s}$, and $\bar{b}$ parton flavors, and for neutral-current interactions, which sum over all quark flavors weighted by their left- and right-handed chiral couplings to the $Z$ boson.

The squared four-momentum transfer, $Q^2$, is related to $x$, $y$, and the incoming neutrino energy, $E_\nu$, through the kinematic relation
\begin{equation}
    Q^2 = 2 M_N E_\nu x y \;,
\end{equation}
where $M_N$ is the nucleon mass. To obtain the single-differential cross section with respect to $Q^2$, we change variables from $x$ to $Q^2$. The Jacobian of this transformation is $\left| \frac{\partial x}{\partial Q^2} \right| = (2 M_N E_\nu y)^{-1}$. Integrating over the kinematically allowed range of $y \in [y_{\min}, 1]$, we obtain
\begin{equation}
\begin{aligned}
      &  \frac{d\sigma_i}{dQ^2}(E_\nu) =  \\ & \int_{y_{\min}}^{1} dy \, \frac{1}{2 M_N E_\nu y} \frac{d^2\sigma_i}{dx dy}\left(x = \frac{Q^2}{2 M_N E_\nu y}, y, E_\nu\right) \;,
\end{aligned}
\end{equation}
where the integrand evaluates to zero if $x > 1$. The normalized probability distribution of $Q^2$ for a fixed neutrino energy and a specific interaction channel $i$ is then
\begin{equation}
    \mathcal{P}_i(Q^2 | E_\nu) = \frac{1}{\sigma_i(E_\nu)} \frac{d\sigma_i}{dQ^2}(E_\nu) \;,
    \label{equ:pq2_fixed_energy}
\end{equation}
where $\sigma_i(E_\nu) = \int \int dx dy \frac{d^2\sigma_i}{dxdy}$ is the total cross section for channel $i$. 

To express this distribution in terms of the momentum transfer $Q = \sqrt{Q^2}$, we apply the transformation $\mathcal{P}(Q)dQ = \mathcal{P}(Q^2)dQ^2$, yielding
\begin{equation}
    \mathcal{P}_i(Q | E_\nu) = 2Q \, \mathcal{P}_i(Q^2 | E_\nu) \;.
\end{equation}
Figure~\ref{fig:fixed_energy_q2} illustrates the behavior of $\mathcal{P}(Q^2 | E_\nu)$ for various discrete neutrino energies, demonstrating how the distributions shift towards higher momentum transfers as $E_\nu$ increases.


\subsection{Channel combination and spectrum weighting}
\label{subsec:spectrum_weighting}

High-energy neutrino telescopes detect a flux astrophysical neutrinos composed of a mixture of neutrinos and anti-neutrinos interacting via the CC and NC channels. To construct a physically representative $Q$ distribution, we must combine the $\mathcal{P}_i(Q|E_\nu)$ distributions weighted by their respective interaction probabilities. 

At a given energy $E_\nu$, the probability that an interacting neutrino underwent channel $i$ is proportional to its cross section, $\sigma_i(E_\nu)$. The combined, energy-conditional probability distribution is
\begin{equation}
 \mathcal{P}_{\rm comb}(Q | E_\nu) = \frac{\sum_i \sigma_i(E_\nu) \mathcal{P}_i(Q | E_\nu)}{\sum_i \sigma_i(E_\nu)} \;.
 \label{equ:P_comb}
\end{equation}
Figure~\ref{fig:cross_sections_and_fluxes} shows the total cross sections, $\sigma_i(E_\nu)$, and their relative ratios (\eg, $\sigma_{\text{CC}\bar{\nu}}/\sigma_{\text{CC}\nu}$).

Next, we fold this combined distribution over the incoming astrophysical neutrino energy spectrum, $\Phi(E_\nu)$. We define a normalized energy probability density function over a relevant energy range $[E_{\min}, E_{\max}]$:
\begin{equation}
    \mathcal{P}(E_\nu) = \frac{\Phi(E_\nu)}{\int_{E_{\min}}^{E_{\max}} \Phi(E^\prime_\nu) dE^\prime_\nu} \;.
\end{equation}
The final, spectrum-weighted probability distribution of $Q$ is obtained by marginalizing over the neutrino energy:
\begin{equation}
    \mathcal{P}(Q) = \int_{E_{\min}}^{E_{\max}} dE_\nu \, \mathcal{P}_{\rm comb}(Q | E_\nu) \mathcal{P}(E_\nu) \;.
    \label{equ:P_final}
\end{equation}
Numerically, $\mathcal{P}_{\rm comb}(Q|E_\nu)$ is computed on a logarithmically spaced grid of $E_\nu$ and interpolated to perform the continuous integration over the spectrum.


\subsection{Astrophysical flux models and containment intervals}
\label{subsec:flux_models}

\begin{figure}[b!]
 \centering
 \includegraphics[width=\columnwidth]{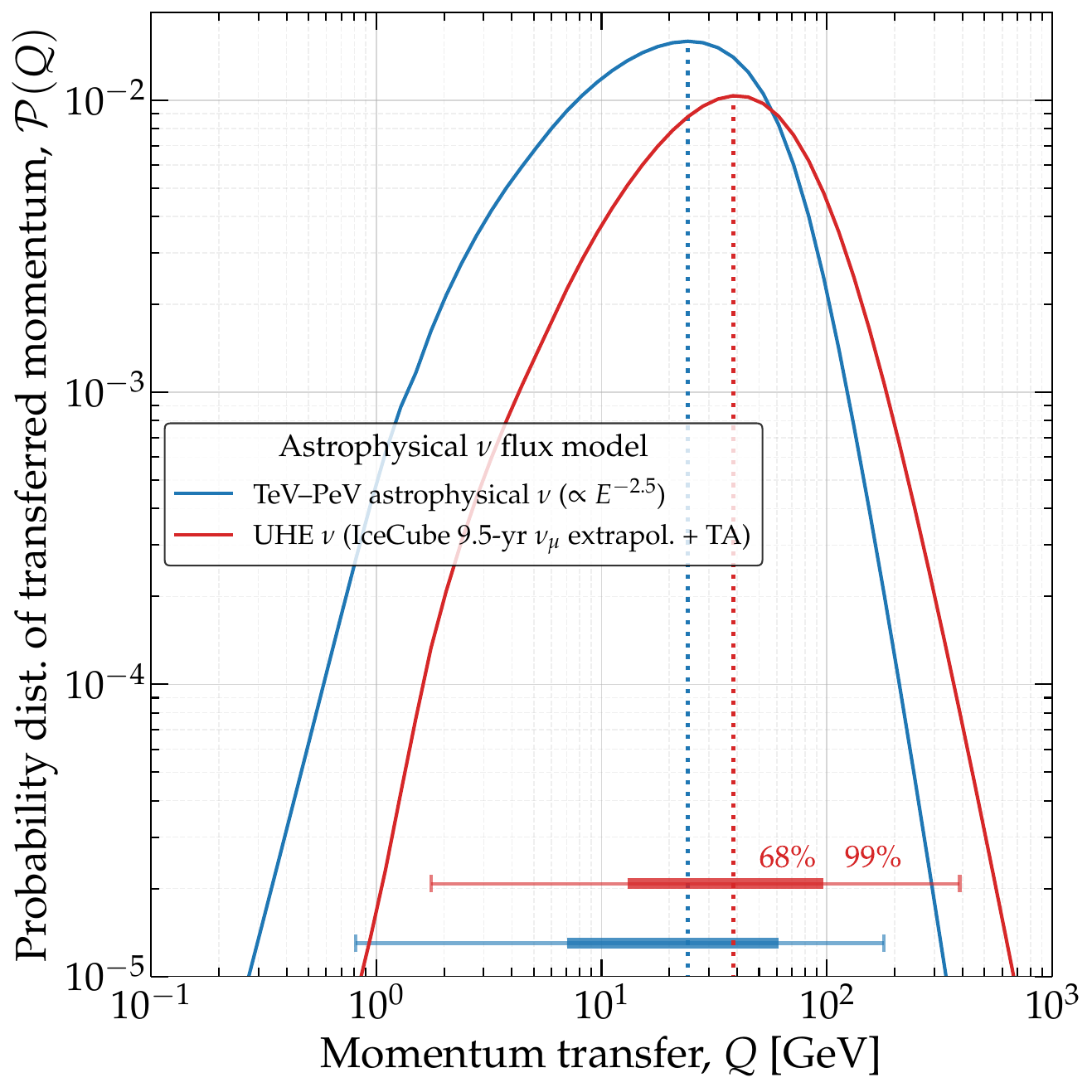}
 \caption{\textbf{Probability distribution of transferred momentum in $\nu N$ deep inelastic scattering.}  The $Q$-distribution shown, $\mathcal{P}(Q)$ in \equ{P_final}, is computed by weighing the distributions corresponding to individual neutrino energies by the energy spectrum of high-energy astrophysical neutrinos, separately for the TeV--PeV and UHE ($> 100$ PeV) ranges, as shown in \figu{cross_sections_and_fluxes}.}
 \label{fig:combined_pq}
\end{figure}

We apply this formalism to two distinct astrophysical neutrino flux models, shown in the top panel of Fig.~\ref{fig:cross_sections_and_fluxes}:
\begin{enumerate}
    \item \textbf{TeV--PeV astrophysical flux}: Modeled as an unbroken power law $\Phi(E_\nu) \propto E_\nu^{-2.5}$, integrated over the range $E_\nu \in [10^4, 10^7]$~GeV. 
    \item \textbf{Ultra-high-energy (UHE) flux}: Modeled as the cosmogenic neutrino flux inferred in \Refe~\cite{Bergman:2021djm} from fitting a predicted UHE cosmic-ray flux to Telescope Array data, and added to the IceCube 9.5-year $\nu_\mu$ extrapolation to the UHE range~\cite{Abbasi:2021qfz}.  We integrate over the range $E_\nu \in [10^8, 10^{10}]$~GeV.
\end{enumerate}

Figure~\ref{fig:combined_pq} shows the resulting spectrum-weighted distributions, $\mathcal{P}(Q)$, for both flux models. To quantify the typical momentum transfers, we compute the highest posterior density containment intervals. These intervals are determined numerically by identifying the peak of the distribution (mode) and integrating the probability mass $d\mathcal{P} = \mathcal{P}(Q) dQ$ downwards in probability density until 68\% and 99\% of the total probability is contained. These boundaries are represented as horizontal span bars in Fig.~\ref{fig:combined_pq}, providing a robust measure of the expected $Q$ range for each astrophysical flux scenario.


\section{Flavor composition at Earth with RG running under muon-damped pion decay}
\label{app:flavor_composition_muon_damped}

\renewcommand{\theequation}{C\arabic{equation}}
\renewcommand{\thefigure}{C\arabic{figure}}
\renewcommand{\thetable}{C\arabic{table}}
\setcounter{figure}{0} 
\setcounter{table}{0} 

\begin{figure*}[htpb]
 \centering
 \includegraphics[width=\textwidth]{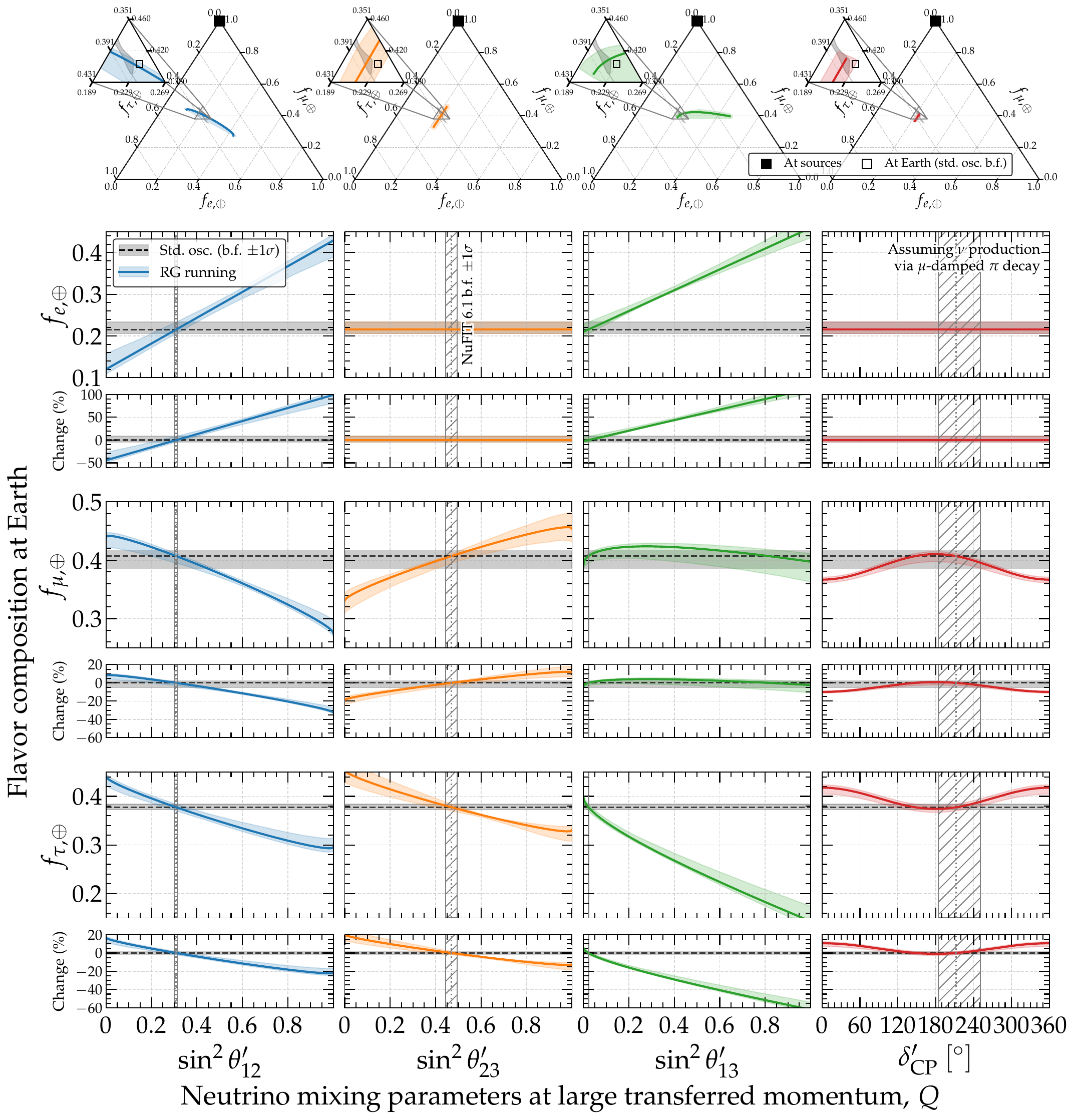}
 \caption{\textbf{Flavor composition of high-energy astrophysical neutrinos at Earth as a function of the modified high-$Q$ mixing parameters.}  Same as \figu{flavor_ratios_pion} in the main text, but assuming neutrino production via muon-damped pion decay instead, \ie, the flavor composition at the sources is $(0, 1, 0)_{\rm S}$.  See Appendix~\ref{app:flavor_composition_muon_damped} and Sec.~\ref{sec:flavor_ratios-theory} in the main text for details.}
 \label{fig:flavor_ratios_muon}
\end{figure*}

Figure~\ref{fig:flavor_ratios_muon} shows the variation of the flavor ratios at Earth with the high-$Q$ mixing parameters assuming neutrino production via muon-damped pion decay.  This is to be compared with \figu{flavor_ratios_pion} in the main text, which assumes neutrino production via full pion decay.  Compared to that figure, the impact of changing the high-$Q$ mixing parameters in \figu{flavor_ratios_muon} is more significant.  Section~\ref{sec:flavor_ratios-impact_sources} in the main text explains why.


\section{Breakdown of allowed regions of flavor composition at Earth}
\label{app:allowed_flavor_regions}

\renewcommand{\theequation}{D\arabic{equation}}
\renewcommand{\thefigure}{D\arabic{figure}}
\renewcommand{\thetable}{D\arabic{table}}
\setcounter{figure}{0} 
\setcounter{table}{0} 

To disentangle the drivers behind the expansion of the RG-allowed flavor space presented in Sec.~\ref{sec:flavor_comp-regions_earth} in the main text, we can isolate the effects of the high-$Q$ mixing parameters. In this appendix, we construct the theoretically allowed regions of flavor composition at Earth by varying a single high-$Q$ mixing parameter ($\theta_{12}^\prime, \theta_{23}^\prime, \theta_{13}^\prime$, or $\delta_{\mathrm{CP}}^\prime$) at a time across its full physically allowed range. 

Like in the main text, we vary the source composition freely across $f_{e,S} \in [0,1]$, with $f_{\tau,S} = 0$. We sample the standard mixing parameters from the NuFIT 6.1 pairwise $\Delta\chi^2$ profiles for $(\delta_{\mathrm{CP}}, \sin^2\theta_{23})$ and for $(\sin^2\theta_{12}, \sin^2\theta_{13})$. For each evaluation, the high-$Q$ mixing parameters not actively being varied are pinned to their respective standard (``low-$Q$'') values. 

\begin{figure*}[t!]
 \centering
 \includegraphics[width=0.495\textwidth]{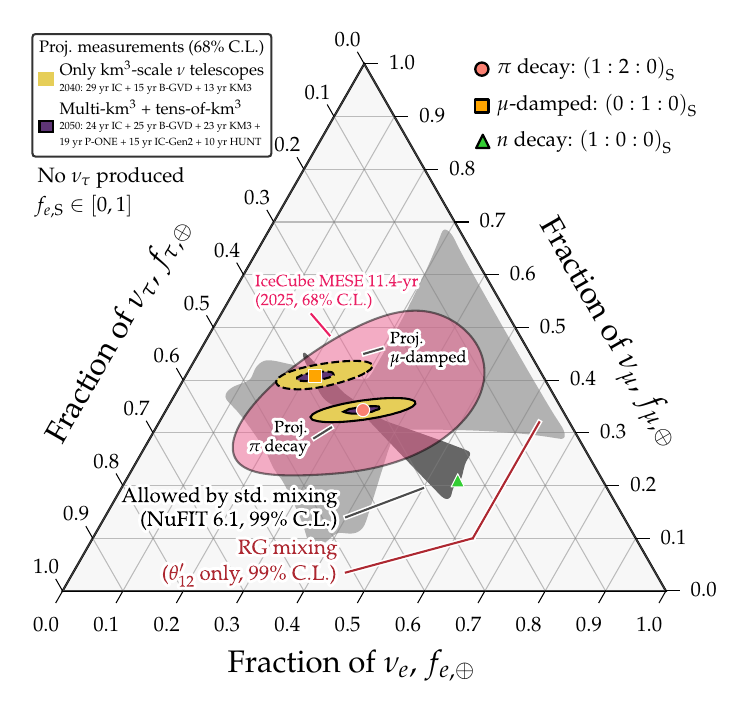}
 \hfill
 \includegraphics[width=0.495\textwidth]{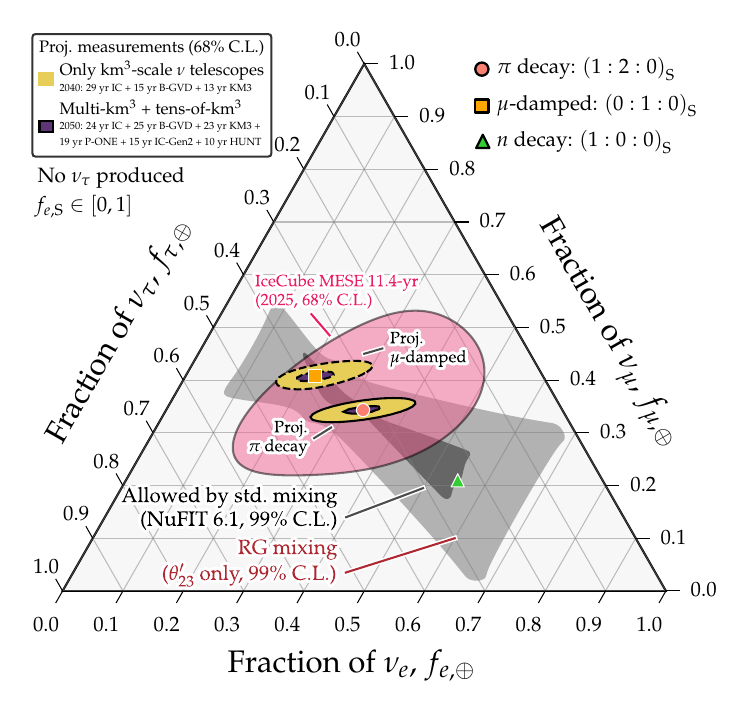}\\
 \vspace{0.4cm}
 \includegraphics[width=0.495\textwidth]{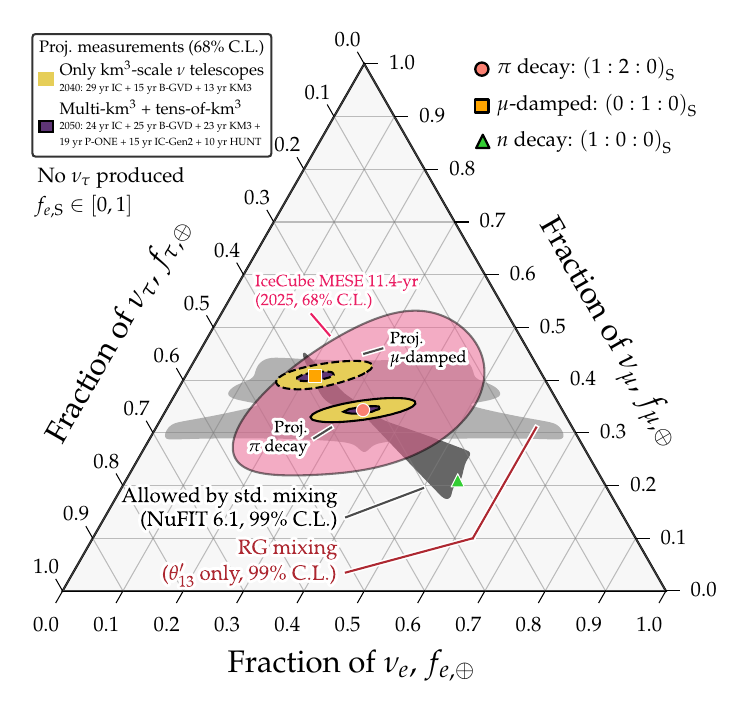}
 \hfill
 \includegraphics[width=0.495\textwidth]{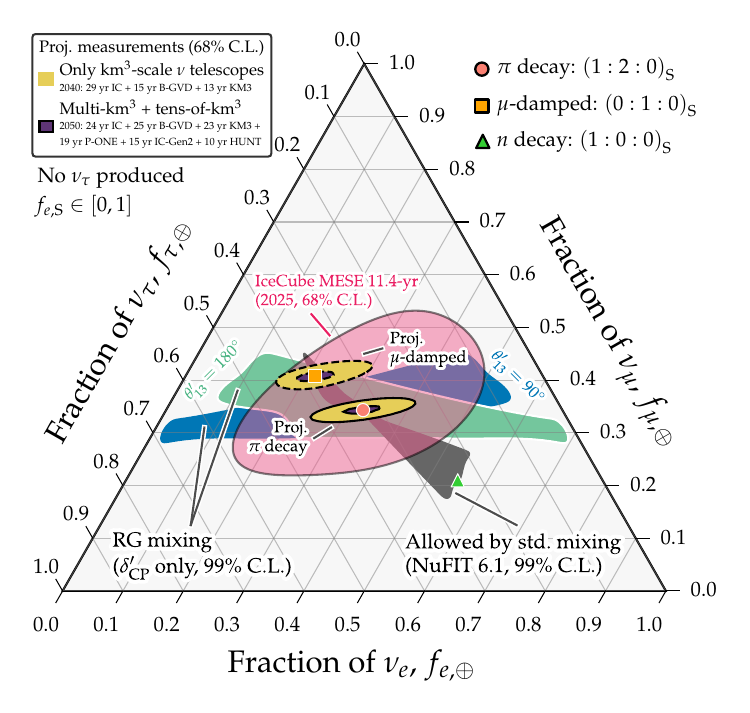}
 \caption{\textbf{Impact of individual high-$Q$ mixing parameters on the expected flavor composition at Earth.} Each panel displays the 99\% C.L. allowed region when only one high-$Q$ parameter is varied (top left: $\theta_{12}^\prime$, top right: $\theta_{23}^\prime$, bottom left: $\theta_{13}^\prime$, bottom right: $\delta_{\mathrm{CP}}^\prime$), while the others remain fixed to their standard values. The initial source composition is generalized as $f_{e, {\rm S}} \in [0,1]$ and $f_{\tau, {\rm S}} = 0$. The projections of future km$^3$-scale and tens-of-km$^3$-scale detector sensitivities are overlaid to demonstrate diagnostic potential.}
 \label{fig:ternary_individual}
\end{figure*}

Figure~\ref{fig:ternary_individual} illustrates the resulting 99\% C.L.~allowed flavor regions for these isolated variations, overlaid with current and projected experimental sensitivities. This decomposition reveals how different high-$Q$ mixing parameters uniquely deform the allowed flavor space, establishing the physical basis for our parameter sensitivities.

\smallskip

\textbf{\textit{Impact of $\theta_{12}^\prime$ and $\theta_{23}^\prime$.---}}The high-$Q$ solar and atmospheric mixing angles generate the largest expansions of the allowed flavor space. Varying $\theta_{12}^\prime$ breaks the standard correlation between $f_{e, \oplus}$ and the combined $f_{\mu, \oplus} + f_{\tau, \oplus}$, opening up regions of significant $f_{e, \oplus}$ suppression and enhancement. Conversely, varying $\theta_{23}^\prime$ (top right) breaks the standard $\mu$-$\tau$ symmetry, opening up regions of $f_{\mu, \oplus}$ and $f_{\tau, \oplus}$ well beyond standard expectations. For a muon-damped source, these robust deviations allow future detectors to effectively constrain both parameters. For full pion decay, the strong $\mu$-$\tau$ breaking preserves high sensitivity to $\theta_{23}^\prime$, though the sensitivity to $\theta_{12}^\prime$ is partially diminished by probability averaging across the different possibilities of flavor composition at the sources.

\smallskip

\textbf{\textit{Impact of $\theta_{13}^\prime$ and $\delta_{\mathrm{CP}}^\prime$.---}}Varying the high-$Q$ reactor angle $\theta_{13}^\prime$ yields an asymmetric expansion, bowing the allowed region toward the pure or null $\nu_e$ vertices, depending on the value of $\delta_{\rm CP}$. Although the total area gained is smaller than for $\theta_{12}^\prime$ or $\theta_{23}^\prime$, the induced shifts are phenomenologically critical. For both full pion decay and muon-damped sources, variations in $\theta_{13}^\prime$ push the expected flavor ratios far enough from their standard baseline to be constrained by future precision measurements. Finally, varying the phase $\delta_{\mathrm{CP}}^\prime$ produces variations similar to those of $\theta_{13}^\prime$. While the impact of $\delta_{\mathrm{CP}}^\prime$ remains largely obscured under the averaging effects of full pion decay, the unbuffered nature of a muon-damped source (see Sec.~\ref{sec:flavor_ratios-impact_sources}) amplifies these variations, making it possible for future combinations of multi-km$^3$ neutrino telescopes to constrain all four high-$Q$ parameters.


\section{Pairwise parameter constraints}
\label{app:pairwise_constraints}

\renewcommand{\theequation}{E\arabic{equation}}
\renewcommand{\thefigure}{E\arabic{figure}}
\renewcommand{\thetable}{E\arabic{table}}
\setcounter{figure}{0} 
\setcounter{table}{0} 

\begin{figure*}[t!]
 \centering
 \includegraphics[width=\textwidth]{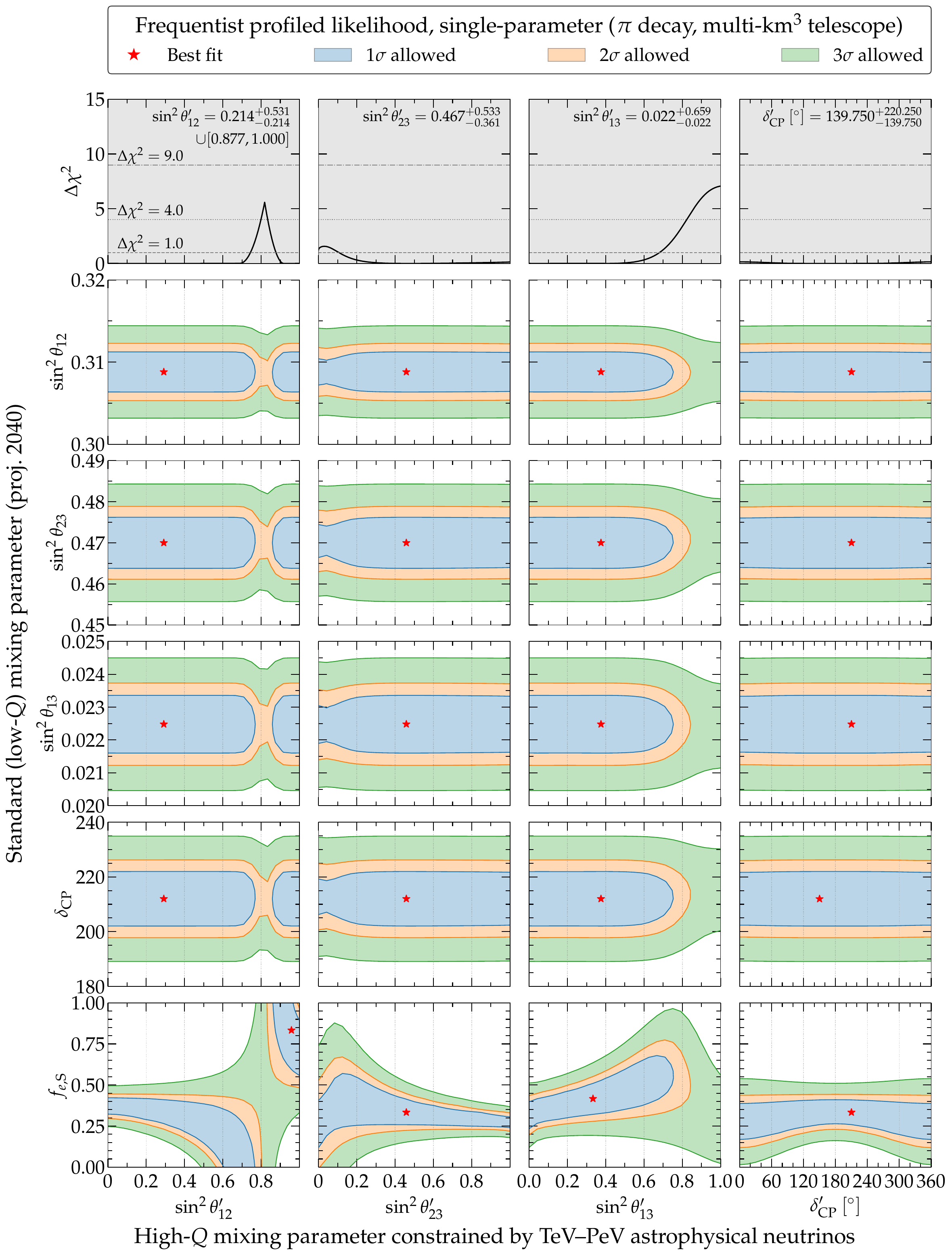}
 \caption{\textbf{\textit{Future TeV--PeV pairwise profiled likelihood using multi-km$^3$ detectors.}} The flavor-measurement likelihood uses the combined exposure of IceCube, Baikal-GVD, and KM3NeT to the year 2040, assuming neutrino production via full pion decay and $f_{\tau, {\rm S}} = 0$. See Appendix~\ref{app:pairwise_constraints} for  details.}
 \label{fig:2d_profiled_likelihood_pion_2040}
\end{figure*}

\begin{figure*}[t!]
 \centering
 \includegraphics[width=\textwidth]{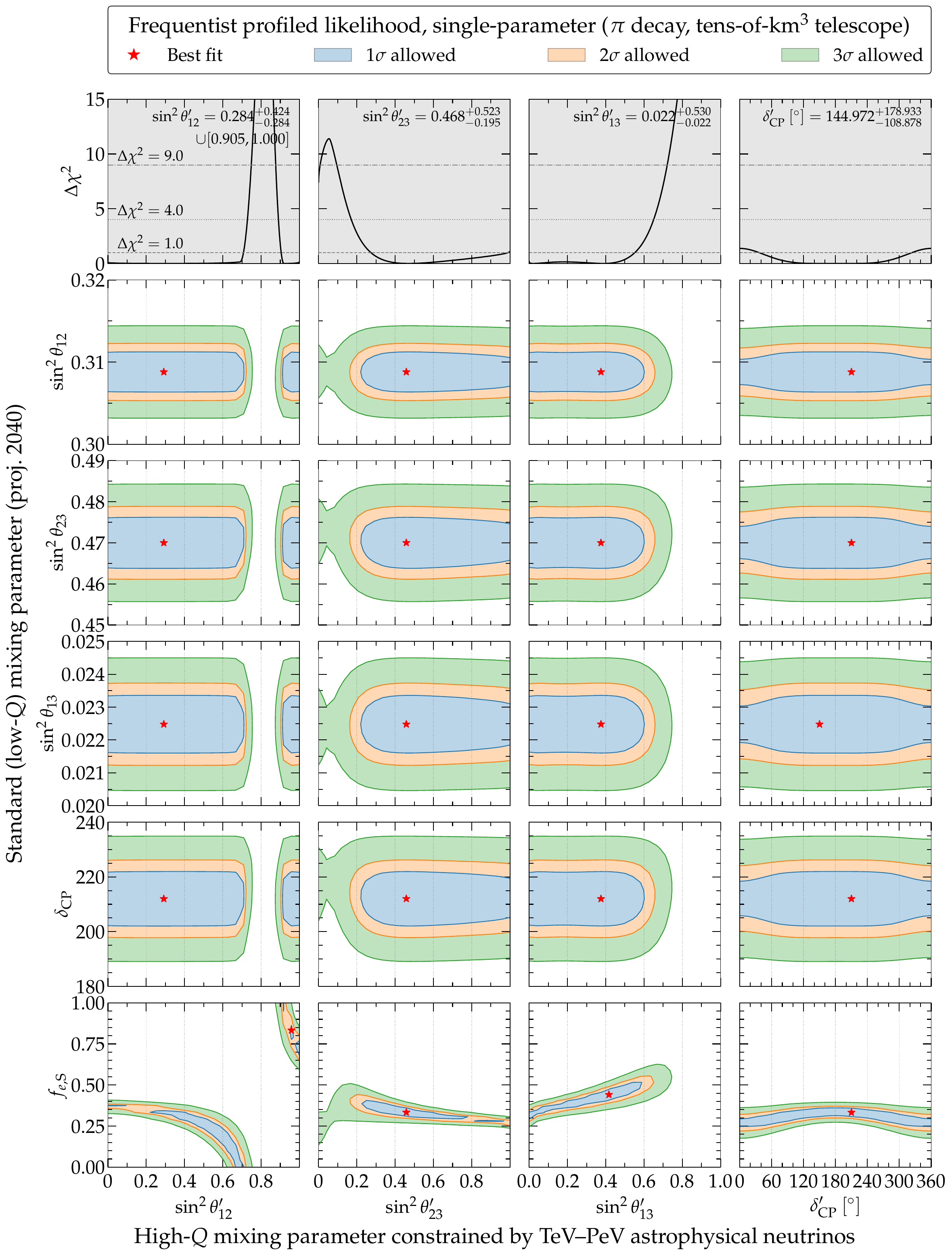}
 \caption{\textbf{\textit{Future TeV--PeV pairwise profiled likelihood using multi-km$^3$ detectors.}} Same as \figu{2d_profiled_likelihood_pion_2040}, but the flavor-measurement likelihood uses the combined exposure of IceCube, Baikal-GVD, KM3NeT, P-ONE, IceCube-Gen2, and HUNT to the year 2050, assuming neutrino production via full pion decay  and $f_{\tau, {\rm S}} = 0$.  See Appendix~\ref{app:pairwise_constraints} for details.}
 \label{fig:2d_profiled_likelihood_pion_2050}
\end{figure*}

\begin{figure*}[t!]
 \centering
 \includegraphics[width=\textwidth]{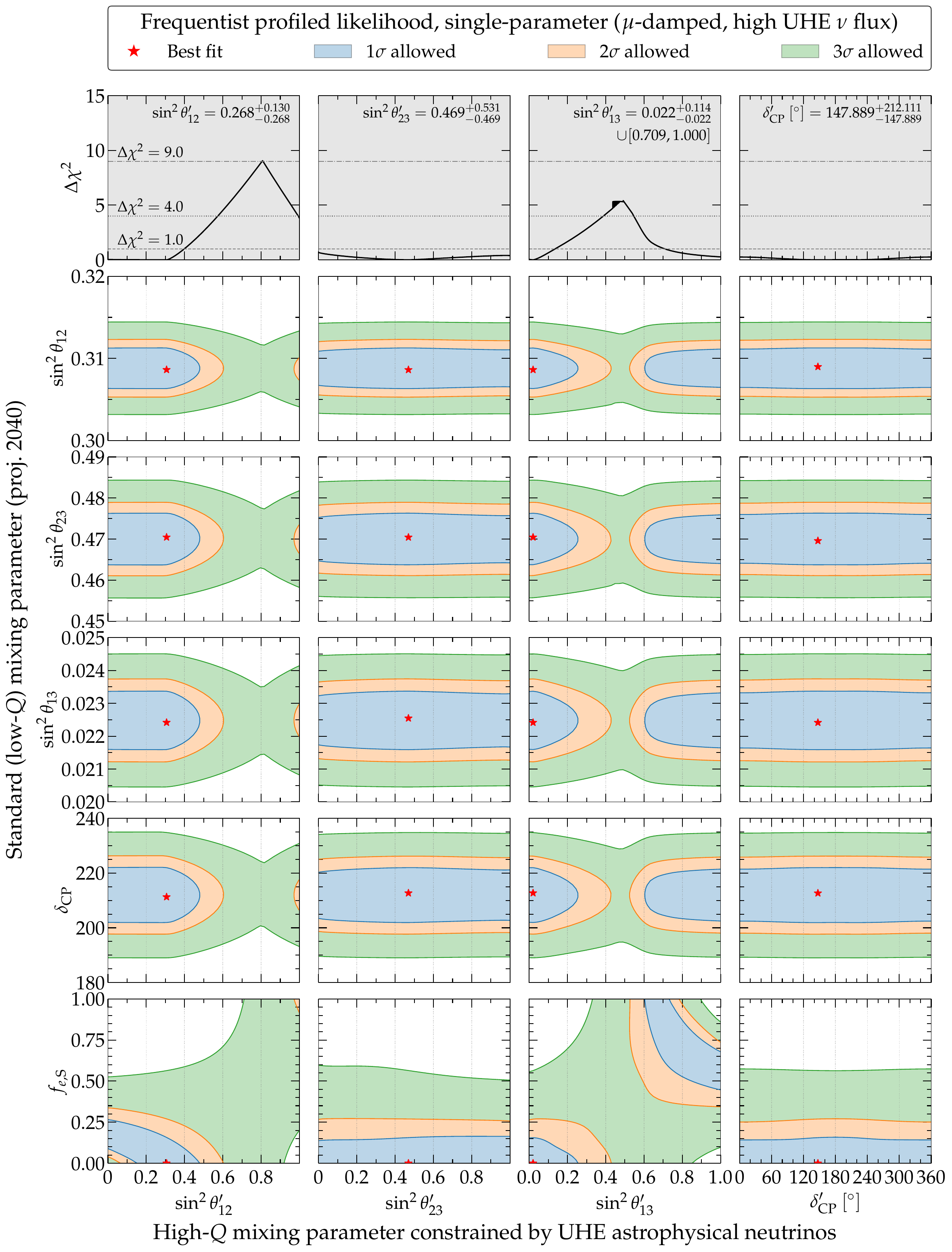}
 \caption{\textbf{\textit{Future UHE pairwise profiled likelihood.}} Similar to Figs.~\ref{fig:2d_profiled_likelihood_pion_2040} and \ref{fig:2d_profiled_likelihood_pion_2050}, but using UHE flavor measurements in the radio array of IceCube-Gen2 instead, assuming a high UHE neutrino flux, neutrino production via muon-damped pion decay, and $f_{\tau, {\rm S}} = 0$.  See Appendix~\ref{app:pairwise_constraints} for details.}
 \label{fig:2d_profiled_likelihood_uhe_muon_flux_0}
\end{figure*}

Figures~\ref{fig:2d_profiled_likelihood_pion_2040} and \ref{fig:2d_profiled_likelihood_pion_2050} show projected pairwise profiled likelihood functions, under our single-parameter approach while keeping the others constrained with sub-TeV pull terms, using multi-km$^3$ neutrino telescopes and adding a tens-of-km$^3$ telescope, respectively.  The pairwise profiled likelihood is a generalization of the one-dimensional profiled likelihood defined in the main text (Sec.~\ref{sec:statistics-mixing_params}), for two degrees of freedom instead of one.  The constraints in Figs.~\ref{fig:2d_profiled_likelihood_pion_2040} and \ref{fig:2d_profiled_likelihood_pion_2050} complement the one-dimensional constraints in Table~\ref{tab:results_mix_params_tev_pev} in the main text.

Figure~\ref{fig:2d_profiled_likelihood_uhe_muon_flux_0} similarly shows projected pairwise profiled likelihood functions, but for UHE neutrino flavor measurements in the radio array of IceCube-Gen2.  We show this exclusively for the assumptions of neutrino production via muon-damped pion decay and high UHE neutrino flux because no constraints can be placed assuming production via full pion decay or a low neutrino flux.   The constraints in \figu{2d_profiled_likelihood_uhe_muon_flux_0} complement the one-dimensional constraints in Table~\ref{tab:results_mix_params_uhe} in the main text.


\section{Parameter constraints allowing $\nu_\tau$ production}
\label{app:results_with_nu_tau_production}

\renewcommand{\theequation}{F\arabic{equation}}
\renewcommand{\thefigure}{F\arabic{figure}}
\renewcommand{\thetable}{F\arabic{table}}
\setcounter{figure}{0} 
\setcounter{table}{0} 

\begin{figure}[t!]
 \centering
 \includegraphics[width=\columnwidth]{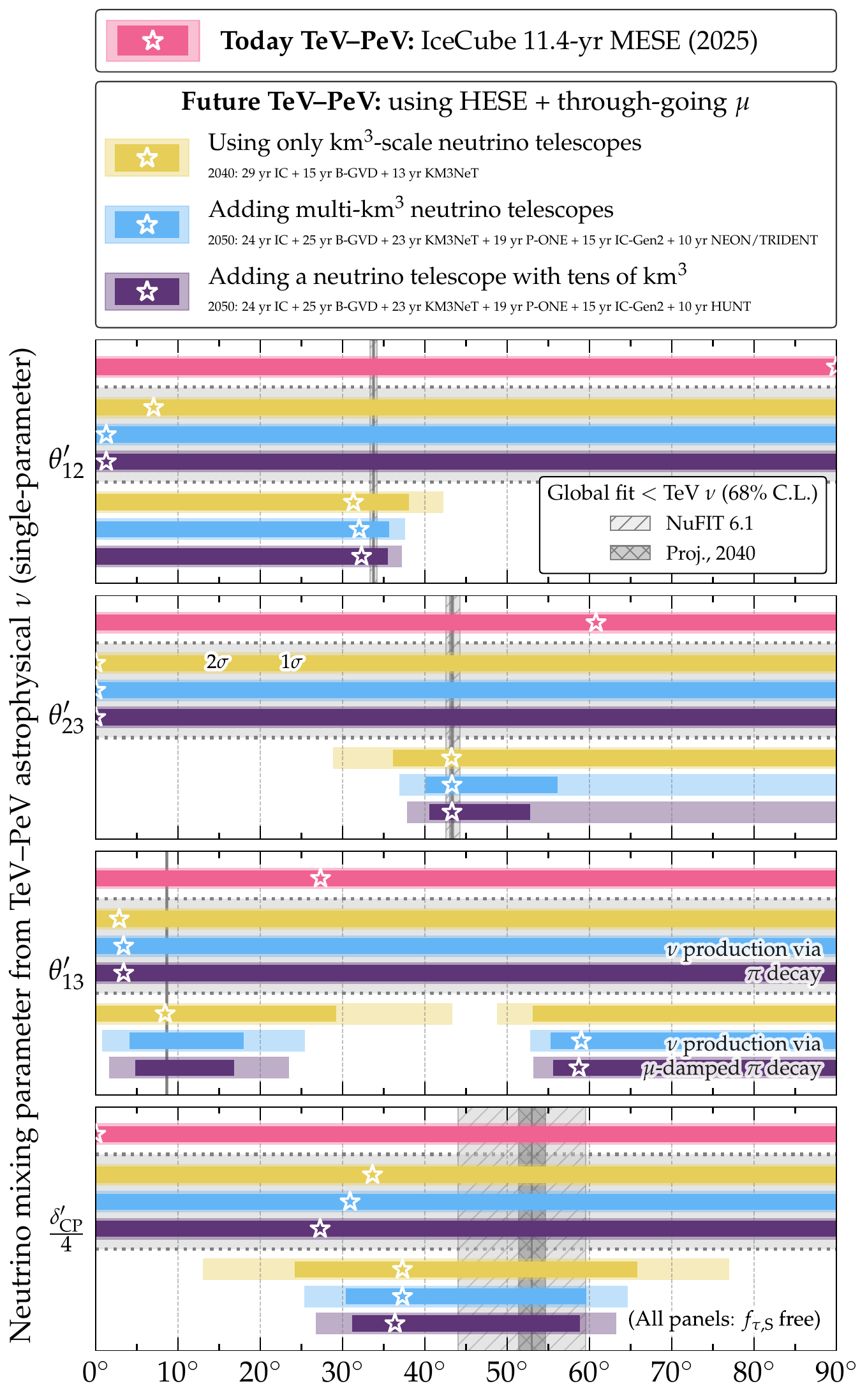}
 \vspace*{-0.5cm}
 \caption{\textbf{\textit{Constraints on the high-$Q$ neutrino mixing parameters with TeV--PeV astrophysical neutrinos, with $f_{\tau, {\rm S}}$ free.}} Same as \figu{results_tev_pev} in the main text, but allowing for $\nu_\tau$ production at the astrophysical sources, \ie, allowing $f_{\tau, {\rm S}} \in [0, 1]$ to float when constraining the mixing parameters.}
 \label{fig:results_tev_pev_ftS_free}
\end{figure}

In the main text, we reported as our main results constraints on the high-$Q$ mixing parameters obtained assuming no $\nu_\tau$ production in the astrophysical sources, \ie, fixing $f_{\tau, {\rm S}} = 0$.  Here we lift this restriction and allow $f_{\tau, {\rm S}}$ to float between 0 and 1, together with $f_{e, {\rm S}}$, when constraining the mixing parameters.  We illustrate the impact of this on the parameter measurements using TeV--PeV astrophysical neutrinos only.

Figure~\ref{fig:results_tev_pev_ftS_free} shows that, like in our main results (\cf, \figu{results_tev_pev} in the main text), the mixing parameters remain unconstrained by present-day IceCube MESE data when $f_{\tau, {\rm S}}$ is allowed to float.  This is to be expected, given that adding an additional free parameter dilutes the already inexistent constraining power in present data.

The largest change compared to our main results is in the projections under neutrino production via full pion decay: by allowing $f_{\tau, {\rm S}}$ to float, the constraints on $\theta_{23}^\prime$ and $\theta_{13}^\prime$ disappear.  The reason for this is that these two parameters regulate the appearance of $\nu_\tau$ via flavor transitions, the effects of which become harder to isolate when $\nu_\tau$ emission directly from the sources is allowed. 

In contrast, in the projections under neutrino production via muon-damped pion decay, the parameter constraints survive a floating $f_{\tau, {\rm S}}$, with constraints on $\theta_{12}^\prime$, $\theta_{23}^\prime$ and $\delta_{\rm CP}^\prime$ virtually unscathed. The constraint on $\theta_{13}^\prime$ is weakened relative to our main results by acquiring a second allowed interval at high values.  The relative robustness of the muon-damped case compared to full pion decay stems from the higher initial flavor purity of the former, as explained in Sec.~\ref{sec:flavor_ratios-impact_sources} in the main text.


\section{Analytical approximation of high-$Q$ and SMEFT-induced flavor shifts}
\label{app:analytical_approximation_smeft}

\renewcommand{\theequation}{G\arabic{equation}}
\renewcommand{\thefigure}{G\arabic{figure}}
\renewcommand{\thetable}{G\arabic{table}}
\setcounter{figure}{0} 
\setcounter{table}{0} 

To understand the origins of the bounds placed on the generic high-$Q$ mixing parameters and the individual SMEFT coefficients in Sec.~\ref{sec:results} in the main text, we can construct an analytical approximation of the expected flavor composition at Earth. By treating the shift in the high-$Q$ mixing parameters---whether floated freely or driven by SMEFT RG running---as a first-order perturbation on the standard-mixing expectation, we obtain explicit expressions that isolate the interplay between the astrophysical source composition, the standard mixing, and the high-$Q$ new physics.


\subsection{Flavor-transition probabilities and flavor shifts}

We consider the flavor-transition probability $P_{\alpha\beta} = \sum_{i=1}^3 |U_{\alpha i}|^2 |U'_{\beta i}|^2$ of high-energy astrophysical neutrinos from $\nu_\alpha$ at the source to $\nu_\beta$ at the detector, defined in \equ{prob_modified} in the main text. The $U_{\alpha i}$ are the elements of the standard, low-$Q$ PMNS mixing matrix at production and $U_{\beta i}^\prime$ are the elements of the high-$Q$ mixing matrix at detection. The latter contains the mixing parameters ($\theta_{12}^\prime, \theta_{13}^\prime, \theta_{23}^\prime, \delta_{\rm CP}^\prime$) that may deviate from their standard low-$Q$ values due to either generic new physics or specific SMEFT RG evolution. The predicted fraction of $\nu_\beta$ at Earth, $f_{\beta, \oplus}$, is the convolution of these transition probabilities with the flavor fractions at the sources, $f_{\alpha, {\rm S}}$ [\equ{flavor_ratios} in the main text],
\begin{equation}
 f_{\beta, \oplus} = \sum_{\alpha} f_{\alpha, {\rm S}} P_{\alpha\beta} = \sum_{\alpha} \sum_{i=1}^3 f_{\alpha, {\rm S}} |U_{\alpha i}|^2 |U'_{\beta i}|^2 \;.
\end{equation}

Defining the deviation of the high-$Q$ parameters from their standard values as $\Delta x = x' - x$, where $x \in \{\theta_{12}, \theta_{13}, \theta_{23}, \delta_{\rm CP}\}$, we expand $f_{\beta, \oplus}$ to first order around the standard baseline:
\begin{equation}
 f_{\beta, \oplus} \approx f_{\beta, \oplus}^{\text{std}} + \Delta f_{\beta, \oplus} \;,
\end{equation}
where $f_{\beta, \oplus}^{\text{std}}$ is the unperturbed composition (evaluated at $U' = U$), and the induced shift is
\begin{equation}
 \Delta f_{\beta, \oplus} \approx \sum_{x} \mathcal{S}_{\beta}^x \, \Delta x \;.
 \label{equ:master_shift}
\end{equation}
Here, $\mathcal{S}_{\beta}^x$ represents the geometric sensitivity of the flavor fraction to a perturbation in the parameter $x$, given by
\begin{equation}
 \mathcal{S}_{\beta}^x \equiv \frac{\partial f_{\beta, \oplus}}{\partial x} = \sum_{\alpha} f_{\alpha, {\rm S}} \sum_{i=1}^3 |U_{\alpha i}|^2 \frac{\partial |U_{\beta i}|^2}{\partial x} \;.
 \label{equ:geometric_sensitivity}
\end{equation}

For the generic analysis, the shifts $\Delta x$ are treated as free parameters. For the SMEFT analysis, to leading order in the logarithmic momentum interval $\Delta t = \ln(Q/Q_0)$, the shift is constrained to $\Delta x \approx \dot{x}\Delta t$, where the SMEFT RG evolution dictates $\dot{x}$, as shown in Eqs.~(\ref{equ:th12_evol_gen})--(\ref{equ:dcp_rg_running_smeft}) in the main text. This factorization isolates the layers driving the constraints: the astrophysical production ($\sum_\alpha f_{\alpha, {\rm S}} |U_{\alpha i}|^2$), the sensitivity of the PMNS matrix ($\mathcal{S}_{\beta}^x$), and the high-$Q$ deviations ($\Delta x$ or $\dot{x}\Delta t$).


\subsection{PMNS sensitivity derivatives}
\label{app:analytical_approximation_smeft-derivatives}

To evaluate \equ{geometric_sensitivity}, we compute the partial derivatives of the squared PMNS matrix elements $|U_{\beta i}|^2$ with respect to each mixing parameter. We employ the standard shorthand $c_{ij} \equiv \cos\theta_{ij}$ and $s_{ij} \equiv \sin\theta_{ij}$. Because unitarity enforces $\sum_{i=1}^3 |U_{\beta i}|^2 = 1$, it follows that $\sum_{i=1}^3 \frac{\partial |U_{\beta i}|^2}{\partial x} = 0$ for any parameter $x$.


\subsubsection{Sensitivity to the solar angle ($\theta_{12}$)}
\label{app:analytical_approximation_smeft-derivatives_th12}

Because $\nu_3$ does not mix with $\theta_{12}$ under standard mixing, $\frac{\partial |U_{\beta 3}|^2}{\partial \theta_{12}} = 0$ for all flavors, enforcing $\frac{\partial |U_{\beta 1}|^2}{\partial \theta_{12}} = - \frac{\partial |U_{\beta 2}|^2}{\partial \theta_{12}}$. The non-zero derivatives are:
\begin{align}
 \frac{\partial |U_{e2}|^2}{\partial \theta_{12}} 
 &= \sin 2\theta_{12} c_{13}^2 \;, \\
 \frac{\partial |U_{\mu 2}|^2}{\partial \theta_{12}} 
 &= -\sin 2\theta_{12} (c_{23}^2 - s_{23}^2 s_{13}^2) 
 \nonumber \\
 &\quad - 2 \cos 2\theta_{12} s_{23} c_{23} s_{13} \cos\delta_{\rm CP} \;, \\
 \frac{\partial |U_{\tau 2}|^2}{\partial \theta_{12}} 
 &= -\sin 2\theta_{12} (s_{23}^2 - c_{23}^2 s_{13}^2) 
 \nonumber \\
 &\quad + 2 \cos 2\theta_{12} s_{23} c_{23} s_{13} \cos\delta_{\rm CP} \;.
\end{align}


\subsubsection{Sensitivity to the reactor angle ($\theta_{13}$)}

The derivatives with respect to the reactor angle are, for the electron flavor,
\begin{align}
 \frac{\partial |U_{e 1}|^2}{\partial \theta_{13}} &= -c_{12}^2 \sin 2\theta_{13} \;, \\
 \frac{\partial |U_{e 2}|^2}{\partial \theta_{13}} &= -s_{12}^2 \sin 2\theta_{13} \;, \\
 \frac{\partial |U_{e 3}|^2}{\partial \theta_{13}} &= \sin 2\theta_{13} \;.
\end{align}
For the muon flavor, they are:
\begin{align}
 \frac{\partial |U_{\mu 1}|^2}{\partial \theta_{13}} 
 &= c_{12}^2 s_{23}^2 \sin 2\theta_{13} 
 \nonumber \\
 &\quad + \sin 2\theta_{12} s_{23} c_{23} c_{13} \cos\delta_{\rm CP} \;, \\
 \frac{\partial |U_{\mu 2}|^2}{\partial \theta_{13}} 
 &= s_{12}^2 s_{23}^2 \sin 2\theta_{13} 
 \nonumber \\
 &\quad - \sin 2\theta_{12} s_{23} c_{23} c_{13} \cos\delta_{\rm CP} \;, \\
 \frac{\partial |U_{\mu 3}|^2}{\partial \theta_{13}} &= -s_{23}^2 \sin 2\theta_{13} \;.
\end{align}
For the tau flavor, they are:
\begin{align}
 \frac{\partial |U_{\tau 1}|^2}{\partial \theta_{13}}
 &= c_{12}^2 c_{23}^2 \sin 2\theta_{13} 
 \nonumber \\
 &\quad - \sin 2\theta_{12} s_{23} c_{23} c_{13} \cos\delta_{\rm CP} \;, \\
 \frac{\partial |U_{\tau 2}|^2}{\partial \theta_{13}}
 &= s_{12}^2 c_{23}^2 \sin 2\theta_{13} 
 \nonumber \\
 &\quad + \sin 2\theta_{12} s_{23} c_{23} c_{13} \cos\delta_{\rm CP} \;, \\
 \frac{\partial |U_{\tau 3}|^2}{\partial \theta_{13}} &= -c_{23}^2 \sin 2\theta_{13} \;.
\end{align}


\subsubsection{Sensitivity to the atmospheric angle ($\theta_{23}$)}

Because $\theta_{23}$ does not participate in the electron row of the PMNS matrix, $\frac{\partial |U_{ei}|^2}{\partial \theta_{23}} = 0$ for all $i$. For the muon and tau flavors, the derivatives are strictly equal and opposite $\left( \frac{\partial |U_{\mu i}|^2}{\partial \theta_{23}} = -\frac{\partial |U_{\tau i}|^2}{\partial \theta_{23}} \right)$. The muon-flavor derivatives are:
\begin{align}
 \frac{\partial |U_{\mu 1}|^2}{\partial \theta_{23}} 
 &= -\sin 2\theta_{23} (s_{12}^2 - c_{12}^2 s_{13}^2) 
 \nonumber \\
 &\quad + 2 s_{12} c_{12} s_{13} \cos 2\theta_{23} \cos\delta_{\rm CP} \;, \\
 \frac{\partial |U_{\mu 2}|^2}{\partial \theta_{23}} 
 &= -\sin 2\theta_{23} (c_{12}^2 - s_{12}^2 s_{13}^2) 
 \nonumber \\
 &\quad - 2 s_{12} c_{12} s_{13} \cos 2\theta_{23} \cos\delta_{\rm CP} \;, \\
 \frac{\partial |U_{\mu 3}|^2}{\partial \theta_{23}} &= c_{13}^2 \sin 2\theta_{23} \;.
\end{align}


\subsubsection{Sensitivity to the Dirac CP-violation phase ($\delta_{\rm CP}$)}

The CP-violation phase only appears in the interference terms of the first and second mass eigenstates ($\nu_1$ and $\nu_2$) for the muon and tau flavors. Consequently, $\frac{\partial |U_{ei}|^2}{\partial \delta_{\rm CP}} = 0$ and $\frac{\partial |U_{\beta 3}|^2}{\partial \delta_{\rm CP}} = 0$. The remaining non-zero derivatives are:
\begin{align}
 \frac{\partial |U_{\mu 1}|^2}{\partial \delta_{\rm CP}} 
 &= -\frac{\partial |U_{\mu 2}|^2}{\partial \delta_{\rm CP}} \\
 &= -\sin 2\theta_{12} s_{23} c_{23} s_{13} \sin\delta_{\rm CP} \;, \nonumber \\
 \frac{\partial |U_{\tau 1}|^2}{\partial \delta_{\rm CP}} &= -\frac{\partial |U_{\tau 2}|^2}{\partial \delta_{\rm CP}} \\
 &= \sin 2\theta_{12} s_{23} c_{23} s_{13} \sin\delta_{\rm CP} \nonumber \;.
\end{align}


\subsection{Approximate flavor shifts and SMEFT RG running}
\label{app:analytical_approximation_smeft-flavor_shifts}

To obtain the explicit shift in the flavor composition at Earth, we substitute the sensitivities derived above [\equ{geometric_sensitivity}] into the generic master equation, \equ{master_shift}. 

For the generic scenario, the flavor composition can be displaced by an arbitrary combination of $\Delta\theta_{12}, \Delta\theta_{13}, \Delta\theta_{23},$ and $\Delta\delta_{\rm CP}$. However, for the SMEFT scenario, these parameter shifts are strictly coupled. As shown in Sec.~\ref{sec:rg_running-impact_smeft_coefficients} in the main text, the SMEFT RG evolution is overwhelmingly dominated by the solar angle due to the kinematically enhanced $\Delta m_{21}^{-2}$ mass splitting (\ie, $\dot{\theta}_{12} \gg \dot{\theta}_{13}, \dot{\theta}_{23}, \dot{\delta}_{\rm CP}$). The leading-order behavior of the SMEFT-induced flavor shift is accurately captured by isolating the $\theta_{12}$ trajectory, \ie,
\begin{equation}
 \Delta f_{\beta, \oplus} \approx \mathcal{S}_{\beta}^{\theta_{12}} \, \Delta\theta_{12} \;.
\end{equation}
By factoring in the constraint $\frac{\partial |U_{\beta 1}|^2}{\partial \theta_{12}} = - \frac{\partial |U_{\beta 2}|^2}{\partial \theta_{12}}$ (Appendix~\ref{app:analytical_approximation_smeft-derivatives_th12}), we separate the geometric sensitivity, $\mathcal{S}_{\beta}^{\theta_{12}}$, into a source-dependent eigenstate asymmetry and a derivative on $\theta_{12}$:
\begin{align}
 \Delta f_{\beta, \oplus} 
 &\approx \left( \frac{\partial |U_{\beta 2}|^2}{\partial \theta_{12}} \right) \left[ \sum_{\alpha} f_{\alpha, {\rm S}} \left( |U_{\alpha 2}|^2 - |U_{\alpha 1}|^2 \right) \right] \Delta\theta_{12} \;.
 \label{equ:flavor_shift_approx}
\end{align}
This expression dictates that the massive flavor shifts driven by coefficients like $C_{11}$ (which dominate the SMEFT generation of $\Delta\theta_{12}$) are directly proportional to the population asymmetry between $\nu_1$ and $\nu_2$ at the astrophysical sources. If a specific source production mechanism happens to populate $\nu_1$ and $\nu_2$ equally (\ie, if $\sum_{\alpha} f_{\alpha, {\rm S}} |U_{\alpha 2}|^2 = \sum_{\alpha} f_{\alpha, {\rm S}} |U_{\alpha 1}|^2$), the bracketed term exactly vanishes, rendering the drastic $\theta_{12}$ modification completely invisible at Earth.

To construct the complete analytical model, which is necessary to evaluate the constraints on the remaining generic high-$Q$ parameters and the sub-dominant SMEFT coefficients, we base our forthcoming derivations on the full sum across all parameters:
\begin{align}
 \Delta f_{\beta, \oplus} 
 &\approx \mathcal{S}_{\beta}^{\theta_{12}} \Delta\theta_{12} + \mathcal{S}_{\beta}^{\theta_{13}} \Delta\theta_{13} + \mathcal{S}_{\beta}^{\theta_{23}} \Delta\theta_{23} 
 \nonumber \\
 &\quad + \mathcal{S}_{\beta}^{\delta_{\rm CP}} \Delta\delta_{\rm CP} \;.
 \label{equ:flavor_shift_generic}
\end{align}


\subsection{Numerical evaluation of flavor shifts}

To provide quantitative insight into the magnitude of the observable flavor displacements and explain our constraints from the main text, we evaluate \equ{flavor_shift_generic} using the present-day best-fit values from NuFIT 6.1~\cite{Esteban:2024eli}. The values are shown in Table~\ref{tab:results_mix_params_tev_pev} in the main text for the mixing angles and the CP-violation phase (we assume normal neutrino mass ordering, using Super-Kamiokande atmospheric data). In the SMEFT analysis, we retain the explicit dependence on the mass $m_1$ until later.

For the SMEFT analysis, we explicitly map the $t$-integrated shifts ($\Delta x \approx \dot{x} \Delta t$) to the Wilson coefficients, $C_{ij}$ in \equ{wilson_coeff_matrix_general} in the main text. Defining the global SMEFT integration factor $\tilde{\kappa} = \kappa \Delta t$ (where $\kappa$ depends on the scale $\Lambda_{\rm SMEFT}$ and is defined in Sec.~\ref{sec:rg_running-impact_smeft_coefficients} in the main text), we substitute the numerical values into the $\dot{x}$ approximations from the main text [Eqs.~(\ref{equ:th12_evol_gen})--(\ref{equ:dcp_rg_running_smeft})]. This yields parameter shifts scaling with their corresponding kinematic mass multipliers, \ie,
\begin{widetext}
\begin{align}
 \label{equ:shift_smeft_th12}
 \Delta\theta_{12} &\approx \tilde{\kappa} \left( \frac{m_1^2 + m_2^2}{\Delta m_{21}^2} \right) \big[ 0.462 C_{11} - 0.245 C_{22} - 0.217 C_{33} + 0.278 {\rm Re}(C_{12}) - 0.262 {\rm Re}(C_{13}) + 0.461 {\rm Re}(C_{23}) \big] \;, \\
 \Delta\theta_{13} &\approx \tilde{\kappa} \left( \frac{m_1^2 + m_3^2}{\Delta m_{31}^2} \right) \big[ 0.235 C_{22} - 0.235 C_{33} - 0.483 {\rm Re}(C_{12}) - 0.302 {\rm Im}(C_{12}) - 0.513 {\rm Re}(C_{13}) - 0.321 {\rm Im}(C_{13}) 
 \nonumber \\
 &\qquad\qquad\qquad\qquad + 0.028 {\rm Re}(C_{23}) + 0.294 {\rm Im}(C_{23}) \big] \;, \\
 \Delta\theta_{23} &\approx \tilde{\kappa} \left( \frac{m_2^2 + m_3^2}{\Delta m_{32}^2} \right) \big[ 0.415 C_{22} - 0.415 C_{33} + 0.381 {\rm Re}(C_{12}) + 0.405 {\rm Re}(C_{13}) + 0.050 {\rm Re}(C_{23}) \big] \;, \\
 \Delta\delta_{\rm CP} &\approx \tilde{\kappa} \left( \frac{m_1^2 + m_2^2}{\Delta m_{21}^2} \right) \big[ 0.788 {\rm Im}(C_{12}) - 0.742 {\rm Im}(C_{13}) \big] \nonumber \\
 &\quad + \tilde{\kappa} \left( \frac{m_1^2 + m_3^2}{\Delta m_{31}^2} \right) \big[ -0.981 C_{22} + 0.981 C_{33} - 2.015 {\rm Re}(C_{12}) - 2.141 {\rm Re}(C_{13}) \nonumber \\
 &\qquad\qquad\qquad\qquad + 0.118 {\rm Re}(C_{23}) + 3.224 {\rm Im}(C_{12}) + 3.424 {\rm Im}(C_{13}) - 3.144 {\rm Im}(C_{23}) \big] \;.
 \label{equ:t-shifts_params}
\end{align}
\end{widetext}

By combining these evaluated parameter dependencies with the geometric sensitivity of the PMNS matrix, \equ{geometric_sensitivity}, we compute the expected flavor shifts $\Delta f_{\beta, \oplus}$ for specific astrophysical neutrino production scenarios.


\subsection{Generic and SMEFT-induced flavor distance}
\label{app:analytical_approximation_smeft-flavor_distance}

To concisely measure how forcefully a perturbation drives the prediction away from the standard-mixing prediction, we define the Euclidean distance $D$ in flavor space between the standard-mixing and modified flavor composition at Earth:
\begin{equation}
 D = \sqrt{ (\Delta f_{e, \oplus})^2 + (\Delta f_{\mu, \oplus})^2 + (\Delta f_{\tau, \oplus})^2 } \;.
\end{equation}

In what follows, we define two explicit forms of this metric. The form $D_{\rm mix}$ isolates the generic distance strictly as a function of the fundamental mixing angle shifts ($\Delta\theta_{ij}, \Delta\delta_{\rm CP}$), whereas $D_{\rm SMEFT}$ maps these geometric sensitivities directly to the dimension-6 operators, setting $m_1 = 0.05$~eV as an illustrative hierarchical mass scheme. This dual approach demonstrates how the experimental bounds on the SMEFT coefficients are fundamentally inherited from the underlying geometric sensitivity to the mixing parameters.


\subsubsection{Case 1: No $\nu_\tau$ production ($f_{\tau, {\rm S}} = 0$, $f_{e, {\rm S}}$ free)}
\label{app:analytical_approximation_smeft-flavor_distance_generic_feS}

Imposing $f_{\tau, {\rm S}} = 0$ reduces the generic flavor composition at the sources to $\left( f_{e, {\rm S}}, f_{\mu, {\rm S}} \equiv 1 - f_{e, {\rm S}}, 0 \right)_{\rm S}$, and we treat $f_{e, {\rm S}}$ as a free parameter. Evaluating the master expression, \equ{flavor_shift_generic}, across all four mixing parameters projects the flavor shifts solely as a function of $f_{e, {\rm S}}$, \ie,

\begin{align}
 \Delta f_{e, \oplus} &\approx (0.285 - 0.623 f_{e, {\rm S}}) \Delta\theta_{12} 
 \nonumber \\ 
 &\quad + (0.074 - 0.233 f_{e, {\rm S}}) \Delta\theta_{13} 
 \label{equ:flavor_shift_electron_generic_feS}
 \;, \\
 \Delta f_{\mu, \oplus} &\approx (-0.136 + 0.298 f_{e, {\rm S}}) \Delta\theta_{12} 
 \nonumber \\
 &\quad + (0.087 - 0.157 f_{e, {\rm S}}) \Delta\theta_{13} 
 \nonumber \\
 &\quad + (0.125 - 0.510 f_{e, {\rm S}}) \Delta\theta_{23} 
 \nonumber \\
 &\quad + (-0.012 + 0.026 f_{e, {\rm S}}) \Delta\delta_{\rm CP} 
 \label{equ:flavor_shift_muon_generic_feS}
 \;, \\
 \Delta f_{\tau, \oplus} &\approx (-0.149 + 0.325 f_{e, {\rm S}}) \Delta\theta_{12} 
 \nonumber \\
 &\quad + (-0.161 + 0.390 f_{e, {\rm S}}) \Delta\theta_{13} 
 \nonumber \\
 &\quad + (-0.125 + 0.510 f_{e, {\rm S}}) \Delta\theta_{23} 
 \nonumber \\
 &\quad + (0.012 - 0.026 f_{e, {\rm S}}) \Delta\delta_{\rm CP} \;.
 \label{equ:flavor_shift_tau_generic_feS}
\end{align}

This formulation exposes a ``flavor blind spot.'' As identified in Appendix~\ref{app:analytical_approximation_smeft-flavor_shifts}, the large flavor shift driven by $\Delta\theta_{12}$ is fundamentally constrained by the $\nu_1$-$\nu_2$ population asymmetry at the sources, which  reduces to the prefactor $(0.285 - 0.623 f_{e, {\rm S}})$. If the astrophysical sources produce high-energy neutrinos with a flavor composition of $f_{e, {\rm S}} \approx 0.457$ [roughly $(1:1.2:0)_{\rm S}$ ratios], this leading-order coefficient vanishes entirely. In such case, the flavor composition at Earth is largely invariant to perturbations in $\theta_{12}$. Consequently, the measurement is rendered blind to SMEFT coefficients like $C_{11}$ that are the primary drivers of the $\theta_{12}$ RG running [see \equ{t-shifts_params}].

Mapping these geometric shifts to a flavor distance within the SMEFT framework via
\begin{equation}
 D_{\rm SMEFT} = \tilde{\kappa} \sqrt{ \mathcal{F}_e^2 + \mathcal{F}_\mu^2 + \mathcal{F}_\tau^2 } \;,
\end{equation}
the combinations of SMEFT coefficients dictating the flavor distance are:
\begin{widetext}
\begin{align}
 \mathcal{F}_e &\approx (8.87 - 19.38 f_{e, {\rm S}}) C_{11} + (-4.65 + 10.12 f_{e, {\rm S}}) C_{22} + (-4.22 + 9.27 f_{e, {\rm S}}) C_{33} \nonumber \\
 &\quad+ (5.23 - 11.32 f_{e, {\rm S}}) {\rm Re}(C_{12}) - (5.14 - 11.35 f_{e, {\rm S}}) {\rm Re}(C_{13}) + (8.85 - 19.36 f_{e, {\rm S}}) {\rm Re}(C_{23}) \nonumber \\
 &\quad- (0.07 - 0.21 f_{e, {\rm S}}) {\rm Im}(C_{12}) - (0.07 - 0.22 f_{e, {\rm S}}) {\rm Im}(C_{13}) + (0.07 - 0.21 f_{e, {\rm S}}) {\rm Im}(C_{23}) \;, \\
 \mathcal{F}_\mu &\approx (-4.23 + 9.27 f_{e, {\rm S}}) C_{11} + (2.50 - 5.76 f_{e, {\rm S}}) C_{22} + (1.73 - 3.51 f_{e, {\rm S}}) C_{33} \nonumber \\
 &\quad- (2.45 - 5.04 f_{e, {\rm S}}) {\rm Re}(C_{12}) + (2.50 - 5.83 f_{e, {\rm S}}) {\rm Re}(C_{13}) - (4.20 - 9.17 f_{e, {\rm S}}) {\rm Re}(C_{23}) \nonumber \\
 &\quad- (0.83 - 1.77 f_{e, {\rm S}}) {\rm Im}(C_{12}) + (0.39 - 0.88 f_{e, {\rm S}}) {\rm Im}(C_{13}) + (0.19 - 0.38 f_{e, {\rm S}}) {\rm Im}(C_{23}) \;, \\
 \mathcal{F}_\tau &\approx (-4.64 + 10.11 f_{e, {\rm S}}) C_{11} + (2.15 - 4.35 f_{e, {\rm S}}) C_{22} + (2.49 - 5.76 f_{e, {\rm S}}) C_{33} \nonumber \\
 &\quad- (2.78 - 6.28 f_{e, {\rm S}}) {\rm Re}(C_{12}) + (2.64 - 5.52 f_{e, {\rm S}}) {\rm Re}(C_{13}) - (4.66 - 10.19 f_{e, {\rm S}}) {\rm Re}(C_{23}) \nonumber \\
 &\quad+ (0.90 - 1.98 f_{e, {\rm S}}) {\rm Im}(C_{12}) - (0.32 - 0.66 f_{e, {\rm S}}) {\rm Im}(C_{13}) - (0.26 - 0.59 f_{e, {\rm S}}) {\rm Im}(C_{23}) \;.
\end{align}
\end{widetext}
Assuming only a single SMEFT coefficient is varied at a time (as is done to produce our single-parameter constraints in the main text), all cross-terms mixing multiple $C_{ij}$ coefficients vanish from $D$. The Euclidean distance simplifies to $D = \tilde{\kappa} \sqrt{ \sum d_{ij}^2 C_{ij}^2 }$, where the polynomials $d_{ij}(f_{e, {\rm S}})$ dictate the parameter sensitivity.


\subsubsection{Case 2: Full pion decay ($f_{e, {\rm S}} = 1/3$)}
\label{app:analytical_approximation_smeft-flavor_distance_pion_decay}

For the nominal expectation of neutrino production via full pion decay, the flavor composition at the sources is $\left( \frac{1}{3}, \frac{2}{3}, 0 \right)_{\rm S}$.  Substituting $f_{e, {\rm S}} = 1/3$ in the generic $f_{\tau, {\rm S}} = 0$ expressions above isolates the flavor shifts specific to this production case:
\begin{align}
 \Delta f_{e, \oplus} &\approx 0.078 \Delta\theta_{12} - 0.004 \Delta\theta_{13} \;, \\
 \Delta f_{\mu, \oplus} &\approx -0.037 \Delta\theta_{12} + 0.035 \Delta\theta_{13} 
 \nonumber \\
 &\quad - 0.045 \Delta\theta_{23} - 0.003 \Delta\delta_{\rm CP} \;, \\
 \Delta f_{\tau, \oplus} &\approx -0.041 \Delta\theta_{12} - 0.031 \Delta\theta_{13} 
 \nonumber \\
 &\quad + 0.045 \Delta\theta_{23} + 0.003 \Delta\delta_{\rm CP} \;.
\end{align}

Assuming a single generic high-$Q$ parameter is varied at a time (as is done to produce the single-parameter constraints in the main text), the generic flavor distance simplifies to an uncoupled quadratic sum, \ie,
\begin{align}
 D_{\rm mix} \approx \Big[ &0.009 (\Delta\theta_{12})^2 + 0.002 (\Delta\theta_{13})^2 \nonumber \\
 &+ 0.004 (\Delta\theta_{23})^2 + 0.00002 (\Delta\delta_{\rm CP})^2 \Big]^{1/2} \;.
\end{align}
This expression establishes the intrinsic sensitivity hierarchy for the standard parameters. Because $f_{e, {\rm S}} = 1/3$ sits perilously close to the $f_{e, {\rm S}} \approx 0.457$ blind spot, the $\Delta\theta_{12}$ multipliers are heavily suppressed. Consequently, the absolute coefficients inside $D_{\rm mix}$ are uniformly small.  \textbf{\textit{This explains why we are able to place only weak constraints on the generic high-$Q$ mixing parameters under our projections of flavor measurement centered on the standard-mixing expectation from full pion decay (Table~\ref{tab:results_mix_params_tev_pev} in the main text).}}

Mapping the aforementioned generic shifts to the SMEFT framework, the coefficient combinations for the flavor distance metric reduce to:
\begin{widetext}
\begin{align}
 \mathcal{F}_e &\approx 2.40 C_{11} - 1.28 C_{22} - 1.13 C_{33} + 1.46 {\rm Re}(C_{12}) - 1.36 {\rm Re}(C_{13}) + 2.40 {\rm Re}(C_{23}) \;, \\
 \mathcal{F}_\mu &\approx -1.14 C_{11} + 0.58 C_{22} + 0.56 C_{33} - 0.77 {\rm Re}(C_{12}) + 0.56 {\rm Re}(C_{13}) - 1.14 {\rm Re}(C_{23}) 
 \nonumber \\
 &\quad - 0.24 {\rm Im}(C_{12}) + 0.10 {\rm Im}(C_{13}) + 0.06 {\rm Im}(C_{23}) \;, \\
 \mathcal{F}_\tau &\approx -1.26 C_{11} + 0.70 C_{22} + 0.57 C_{33} - 0.68 {\rm Re}(C_{12}) + 0.80 {\rm Re}(C_{13}) - 1.26 {\rm Re}(C_{23}) 
 \nonumber \\
 &\quad + 0.24 {\rm Im}(C_{12}) - 0.10 {\rm Im}(C_{13}) - 0.06 {\rm Im}(C_{23}) \;.
\end{align}
\end{widetext}
Assuming a single SMEFT coefficient is varied at a time (again, as is done to produce our SMEFT constraints in the main text), the corresponding flavor distance is:
\begin{align}
 D_{\rm SMEFT} 
 &\approx 
 \tilde{\kappa} \Big[ 8.66 C_{11}^2 + 2.46 C_{22}^2 + 1.90 C_{33}^2 
 \\
 &\quad \quad + 3.17 {\rm Re}(C_{12})^2 + 0.11 {\rm Im}(C_{12})^2 
 \nonumber \\
 &\quad\quad + 2.79 {\rm Re}(C_{13})^2 + 0.02 {\rm Im}(C_{13})^2
 \nonumber \\
 &\quad\quad + 8.65 {\rm Re}(C_{23})^2 + 0.01 {\rm Im}(C_{23})^2 \Big]^{1/2}
 \nonumber \;.
\end{align}

This expression determines our sensitivity to the SMEFT parameters. While $C_{11}$ and ${\rm Re}(C_{23})$ possess relative multipliers (about 8.66) larger than the other operators, the $\nu_1$-$\nu_2$ population asymmetry at the source evaluates to a mere $\approx 0.077$ (compared to $0.285$ for the muon-damped case below). Because the geometric multipliers converting the generic $\Delta\theta_{12}$ shift into observable flavor shifts are drastically stunted, a massive injection of new physics via the SMEFT coefficients produces only a tiny displacement on the flavor composition at Earth. 

Consequently, the requisite new-physics values needed to push the prediction outside the allowed boundary---even for our ambitious, high-precision year-2050 projections with drastically shrunken likelihood contours---blow up to unobservably large numbers. \textbf{\textit{This explains why we are unable to place constraints on the SMEFT coefficients under our projections of flavor measurement centered on the standard-mixing expectation from full pion decay (Table~\ref{tab:results_smeft_full} in the main text).}}


\subsubsection{Case 3: Muon-damped pion decay ($f_{e, {\rm S}} = 0$)}
\label{app:analytical_approximation_smeft-flavor_distance_muon_damped}

For neutrino production via muon-damped pion decay, the flavor composition at the sources is $\left( 0, 1, 0 \right)_{\rm S}$.  Substituting $f_{e, {\rm S}} = 0$ in the generic $f_{\tau, {\rm S}} = 0$ expressions of Appendix~\ref{app:analytical_approximation_smeft-flavor_distance_generic_feS} yields the flavor shifts
\begin{align}
 \Delta f_{e, \oplus} &\approx 0.285 \Delta\theta_{12} + 0.074 \Delta\theta_{13} \;, \\
 \Delta f_{\mu, \oplus} &\approx -0.136 \Delta\theta_{12} + 0.087 \Delta\theta_{13} 
 \nonumber \\
 &\quad + 0.125 \Delta\theta_{23} - 0.012 \Delta\delta_{\rm CP} \;, \\
 \Delta f_{\tau, \oplus} &\approx -0.149 \Delta\theta_{12} - 0.161 \Delta\theta_{13} 
 \nonumber \\
 &\quad - 0.125 \Delta\theta_{23} + 0.012 \Delta\delta_{\rm CP} \;.
\end{align}
The corresponding generic flavor distance reduces to:
\begin{align}
 D_{\rm mix} \approx \Big[ &0.122 (\Delta\theta_{12})^2 + 0.039 (\Delta\theta_{13})^2 \nonumber \\
 &+ 0.031 (\Delta\theta_{23})^2 + 0.0003 (\Delta\delta_{\rm CP})^2 \Big]^{1/2} \;.
\end{align}

Mapping these generic shifts to the SMEFT framework, the coefficient combinations for the SMEFT flavor distance are:
\begin{widetext}
\begin{align}
 \mathcal{F}_e 
 &\approx 8.87 C_{11} - 4.65 C_{22} - 4.22 C_{33} + 5.23 {\rm Re}(C_{12}) - 5.14 {\rm Re}(C_{13}) + 8.85 {\rm Re}(C_{23}) 
 \nonumber \\
 &\quad - 0.07 {\rm Im}(C_{12}) - 0.07 {\rm Im}(C_{13}) + 0.07 {\rm Im}(C_{23}) \;, \\
 \mathcal{F}_\mu &\approx -4.23 C_{11} + 2.50 C_{22} + 1.73 C_{33} - 2.45 {\rm Re}(C_{12}) + 2.50 {\rm Re}(C_{13}) - 4.20 {\rm Re}(C_{23}) 
 \nonumber \\
 &\quad - 0.83 {\rm Im}(C_{12}) + 0.39 {\rm Im}(C_{13}) + 0.19 {\rm Im}(C_{23}) \;, \\
 \mathcal{F}_\tau &\approx -4.64 C_{11} + 2.15 C_{22} + 2.49 C_{33} - 2.78 {\rm Re}(C_{12}) + 2.64 {\rm Re}(C_{13}) - 4.66 {\rm Re}(C_{23}) 
 \nonumber \\
 &\quad + 0.90 {\rm Im}(C_{12}) - 0.32 {\rm Im}(C_{13}) - 0.26 {\rm Im}(C_{23}) \;.
\end{align}
\end{widetext}
Assuming a single SMEFT coefficient is varied at a time, the corresponding flavor distance reduces to
\begin{align}
 \label{equ:distance_smeft_muon_damped}
 D_{\rm SMEFT} 
 &\approx 
 \tilde{\kappa} \Big[ 117.99 C_{11}^2 + 32.47 C_{22}^2 + 26.95 C_{33}^2 
 \\
 &\quad\quad+ 41.05 {\rm Re}(C_{12})^2 + 1.50 {\rm Im}(C_{12})^2
 \nonumber \\
 &\quad\quad+ 39.66 {\rm Re}(C_{13})^2  + 0.26 {\rm Im}(C_{13})^2 
 \nonumber \\
 &\quad\quad + 117.65 {\rm Re}(C_{23})^2 + 0.11 {\rm Im}(C_{23})^2 \Big]^{1/2} \nonumber \;.
\end{align}
As with the full pion decay case, activation of the leading positive SMEFT coefficients forces $\Delta f_e > 0$ and $\Delta f_{\mu, \tau} < 0$, steering the predicted flavor composition at Earth toward the $\nu_e$ corner of the flavor triangle.

The above evaluation demonstrates that the underlying parameter hierarchy remains the same as in the full-pion-decay case: $C_{11}$ and ${\rm Re}(C_{23})$ retain dominant sensitivity, followed sequentially by the off-diagonal real coefficients, the remaining diagonal coefficients, and the heavily suppressed imaginary terms. However, because the initial mass-eigenstate asymmetry in \equ{flavor_shift_approx} is larger for muon-damped sources compared to full-pion-decay sources, the absolute magnitude of every geometric multiplier inside both $D_{\rm mix}$ and $D_{\rm SMEFT}$ is significantly amplified. \textbf{\textit{This explains why, unlike the full-pion-decay case, we are able to place tight constraints on both the generic high-$Q$ parameters and the SMEFT coefficients under our projections of flavor measurement centered on the standard-mixing expectation from muon-damped pion decay.}} 


\subsubsection{Case 4: Neutron decay ($f_{e, {\rm S}} = 1$)}
\label{app:analytical_approximation_smeft-flavor_distance_beta_decay}

The beta-decay of neutrons or neutron-rich isotopes in astrophysical sources produces a pure-$\bar{\nu}_{e}$ initial flux, corresponding to a flavor composition of $\left( 1, 0, 0 \right)_{\rm S}$.  Because the mass difference between neutrons and protons is small, however, these neutrinos are less energetic than neutrinos produced via pion decay. This renders neutron decay unlikely to be the dominant production mechanism for high-energy astrophysical neutrinos, which is why it does not feature in the main text.

Nevertheless, substituting $f_{e, {\rm S}} = 1$, which resides maximally far from the $f_{e, {\rm S}} \approx 0.457$ blind spot, in the generic $f_{\tau, {\rm S}} = 0$ expressions of Appendix~\ref{app:analytical_approximation_smeft-flavor_distance_generic_feS} yields the absolute maximum sensitivity available on the flavor triangle. The generic flavor shift expressions evaluate to:
\begin{align}
 \Delta f_{e, \oplus} &\approx -0.338 \Delta\theta_{12} - 0.159 \Delta\theta_{13} \\
 \Delta f_{\mu, \oplus} &\approx 0.162 \Delta\theta_{12} - 0.070 \Delta\theta_{13} 
 \nonumber \\
 &\quad - 0.385 \Delta\theta_{23} + 0.014 \Delta\delta_{\rm CP} \\
 \Delta f_{\tau, \oplus} &\approx 0.176 \Delta\theta_{12} + 0.229 \Delta\theta_{13} 
 \nonumber \\
 &\quad + 0.385 \Delta\theta_{23} - 0.014 \Delta\delta_{\rm CP} \;.
\end{align}
The corresponding generic flavor distance is:
\begin{align}
 D_{\rm mix} \approx \Big[ &0.171 (\Delta\theta_{12})^2 + 0.083 (\Delta\theta_{13})^2 \nonumber \\
 &+ 0.296 (\Delta\theta_{23})^2 + 0.0004 (\Delta\delta_{\rm CP})^2 \Big]^{1/2} \;.
\end{align}

Mapping this configuration to the SMEFT coefficients, they become
\begin{widetext}
\begin{align}
 \mathcal{F}_e &\approx -10.51 C_{11} + 5.47 C_{22} + 5.05 C_{33} - 6.09 {\rm Re}(C_{12}) + 6.21 {\rm Re}(C_{13}) - 10.51 {\rm Re}(C_{23}) 
 \nonumber \\
 &\quad + 0.14 {\rm Im}(C_{12}) + 0.15 {\rm Im}(C_{13}) - 0.14 {\rm Im}(C_{23}) \;, \\
 \mathcal{F}_\mu &\approx 5.04 C_{11} - 3.26 C_{22} - 1.78 C_{33} + 2.59 {\rm Re}(C_{12}) - 3.33 {\rm Re}(C_{13}) + 4.97 {\rm Re}(C_{23}) 
 \nonumber \\
 &\quad + 0.94 {\rm Im}(C_{12}) - 0.49 {\rm Im}(C_{13}) - 0.19 {\rm Im}(C_{23}) \;, \\
 \mathcal{F}_\tau &\approx 5.48 C_{11} - 2.21 C_{22} - 3.27 C_{33} + 3.51 {\rm Re}(C_{12}) - 2.88 {\rm Re}(C_{13}) + 5.54 {\rm Re}(C_{23}) 
 \nonumber \\
 &\quad - 1.09 {\rm Im}(C_{12}) + 0.34 {\rm Im}(C_{13}) + 0.33 {\rm Im}(C_{23}) \;.
\end{align}
\end{widetext}
Assuming a single SMEFT coefficient is varied at a time, the flavor distance reduces to:
\begin{align}
 D_{\rm SMEFT} 
 &\approx 
 \tilde{\kappa} \Big[ 165.89 C_{11}^2 + 45.43 C_{22}^2 + 39.36 C_{33}^2 
 \\
 &\quad\quad
 + 56.12 {\rm Re}(C_{12})^2 + 2.09 {\rm Im}(C_{12})^2 
 \nonumber \\
 &\quad\quad 
 + 57.94 {\rm Re}(C_{13})^2 + 0.38 {\rm Im}(C_{13})^2 
 \nonumber \\
 &\quad\quad+ 165.85 {\rm Re}(C_{23})^2 + 0.17 {\rm Im}(C_{23})^2 \Big]^{1/2} \nonumber \;.
\end{align}
In this configuration, positive activation of the leading coefficients ($C_{11}, {\rm Re}(C_{23})$) depletes the electron fraction at Earth ($\Delta f_e < 0$) while enhancing the muon and tau fractions, repelling the flavor composition away from the pure $\nu_e$ vertex of the flavor triangle, opposite to what we found in the full and muon-damped pion decay cases.


\subsection{Connecting to experimental constraints}

While the flavor distance $D$ ranks the bare capability of a generic parameter or SMEFT coefficient to deform the high-energy flavor ratios, the true experimental bounding power relies on the experimental flavor-composition measurement likelihood, $\mathcal{L}$, used in our statistical procedure (Sec.~\ref{sec:statistics} in the main text).

Concretely, our statistical parameter constraints depend on the \textit{direction} of the flavor-shift vector ($\Delta f_e, \Delta f_\mu, \Delta f_\tau$) relative to the allowed experimental flavor regions represented by $\mathcal{L}$, and not just on the magnitude, $D$, of the  shift. A high-$Q$ parameter or SMEFT coefficient that drives the flavor composition longitudinally along the major axis of the allowed flavor contour [see, \eg, \figu{ternary_theory} in the main text] will be constrained more weakly than a sub-dominant parameter whose trajectory displaces the flavor composition transversally across the narrowest boundary of the contour. 

Further, in the SMEFT analysis, the bare analytical approximations derived in this appendix represent the flavor displacement at a discrete momentum scale $Q$, captured by the fixed integration interval $\Delta t$. In reality, the parameter limits in the main text depend on flavor ratios integrated over the range of momenta accessible by high-energy astrophysical neutrinos, \equ{flavor_ratios_q_avg} in the main text. This $Q$-averaging procedure convolves the $Q$-dependent flavor shifts with the neutrino $Q$-distributions (see Appendix~\ref{app:momentum_distribution} for details). Consequently, the idealized, rigid geometric trajectories predicted by the single-$Q$ formulation in this appendix are physically smeared across the momentum range.

Therefore, our constraints on the new-physics parameters represent a convolution of the intrinsic hierarchical sensitivity evaluated herein with both the spectral $Q$-averaging of the flavor composition and the specific shape of the experimental likelihood of flavor measurements.


\section{Single-parameter SMEFT-induced flavor composition regions}
\label{app:smeft_regions_single_parameter}

\renewcommand{\theequation}{H\arabic{equation}}
\renewcommand{\thefigure}{H\arabic{figure}}
\renewcommand{\thetable}{H\arabic{table}}
\setcounter{figure}{0} 
\setcounter{table}{0} 

Figure~\ref{fig:ternary_theory_smeft_single_parameter} shows the predicted regions of flavor composition at Earth induced by RG running of the mixing parameters under our dimension-6 SMEFT scheme.  Unlike \figu{ternary_theory_smeft} in the main text (also \figu{ternary_smeft_pion_vs_muon}), the regions in \figu{ternary_theory_smeft_single_parameter} are generated by varying a single SMEFT coefficient at a time.  

The stark contrast between the single-parameter and all-parameter flavor regions stems from dimensional projection and operator interference within the SMEFT parameter space. Varying a single coefficient $C_{ij}$ in isolation forces the system along a highly restricted trajectory, allowing it to reach extreme values that project onto the two-dimensional flavor triangle as extended ``spikes'' in \figu{ternary_theory_smeft_single_parameter}. Conversely, when all parameters vary simultaneously, the RG evolution generates extensive operator mixing and cross-terms. These competing effects often result in cancellations or destructive interference. Upon projecting this full high-dimensional volume down to the physical flavor space, the extreme isolated topologies wash out, yielding the comparatively smaller, smoother, and more centralized accessible region shown in \figu{ternary_theory_smeft}. Further, the density of our Monte Carlo scanning averages out the remaining high-dimensional edges, reinforcing this centralization.

\begin{figure*}[b!]
 \centering
 \includegraphics[width=0.495\textwidth]{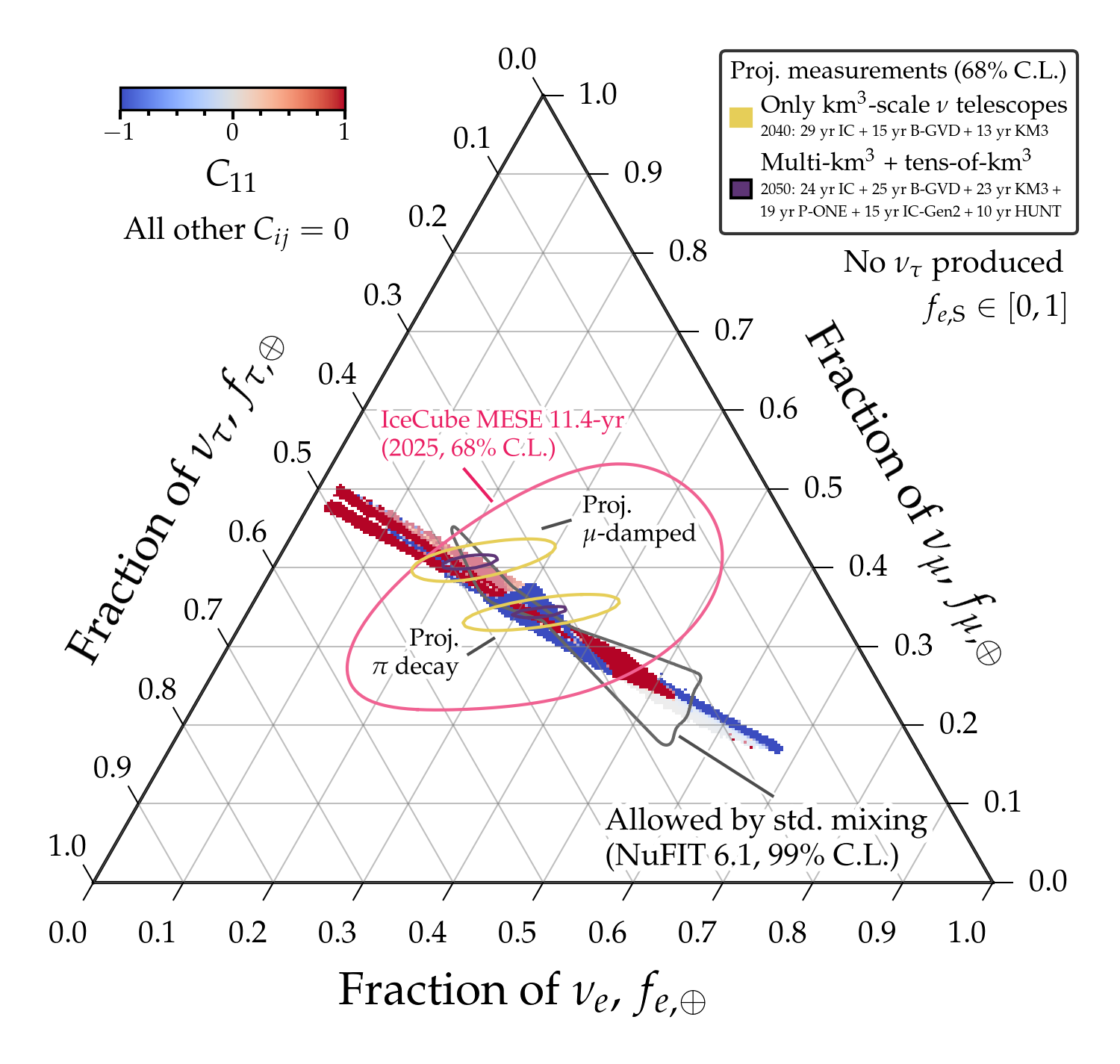}
 \includegraphics[width=0.495\textwidth]{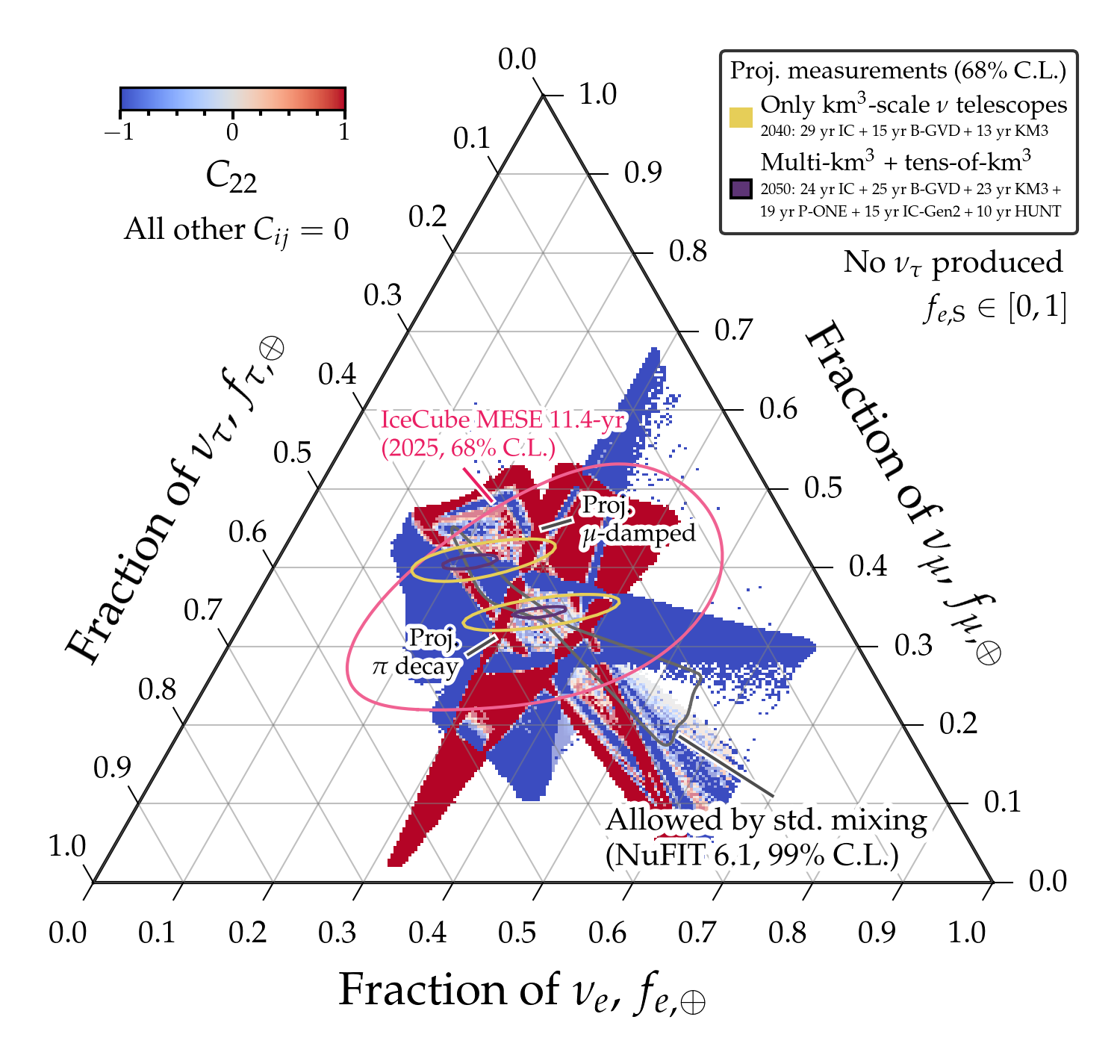}
 \includegraphics[width=0.495\textwidth]{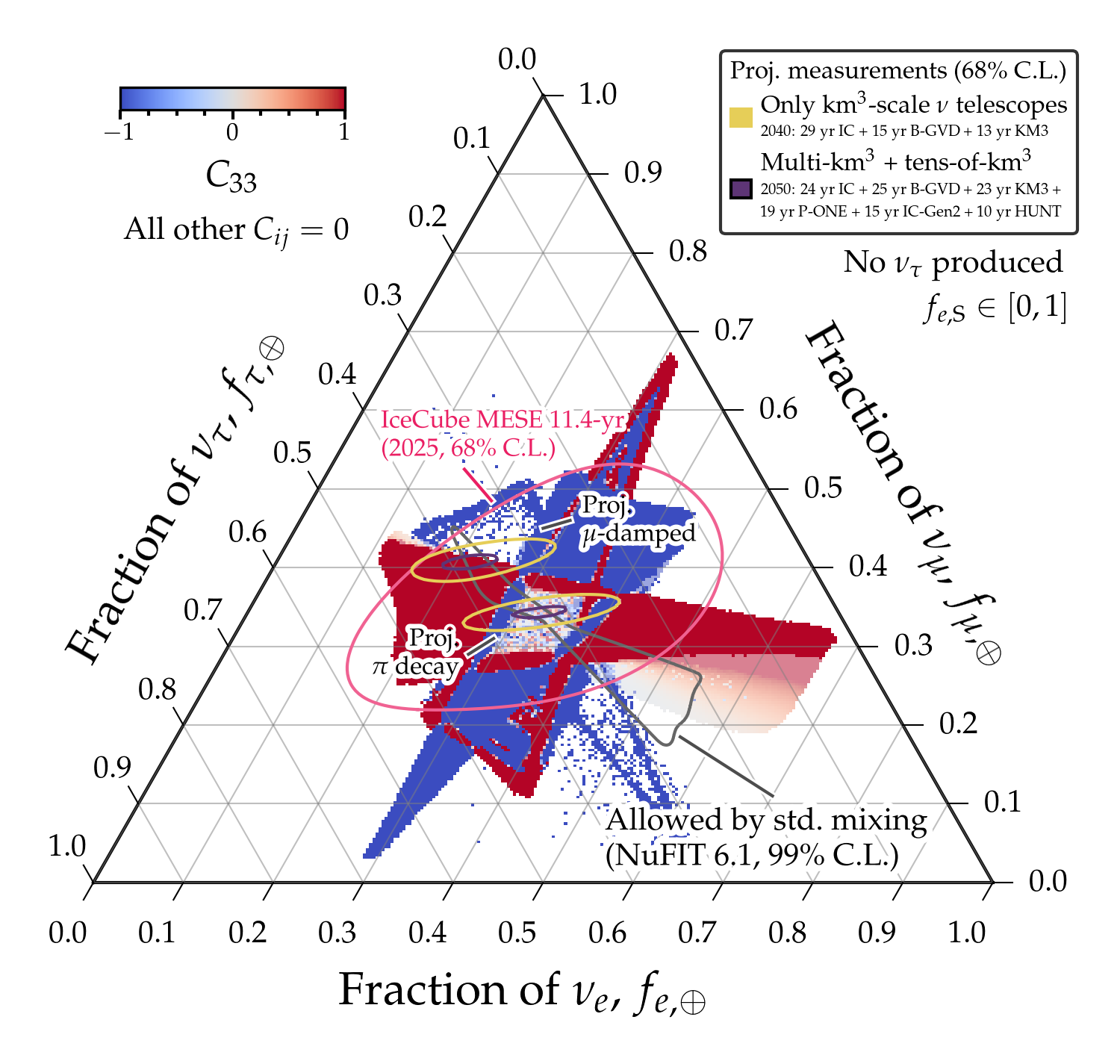}
 \includegraphics[width=0.495\textwidth]{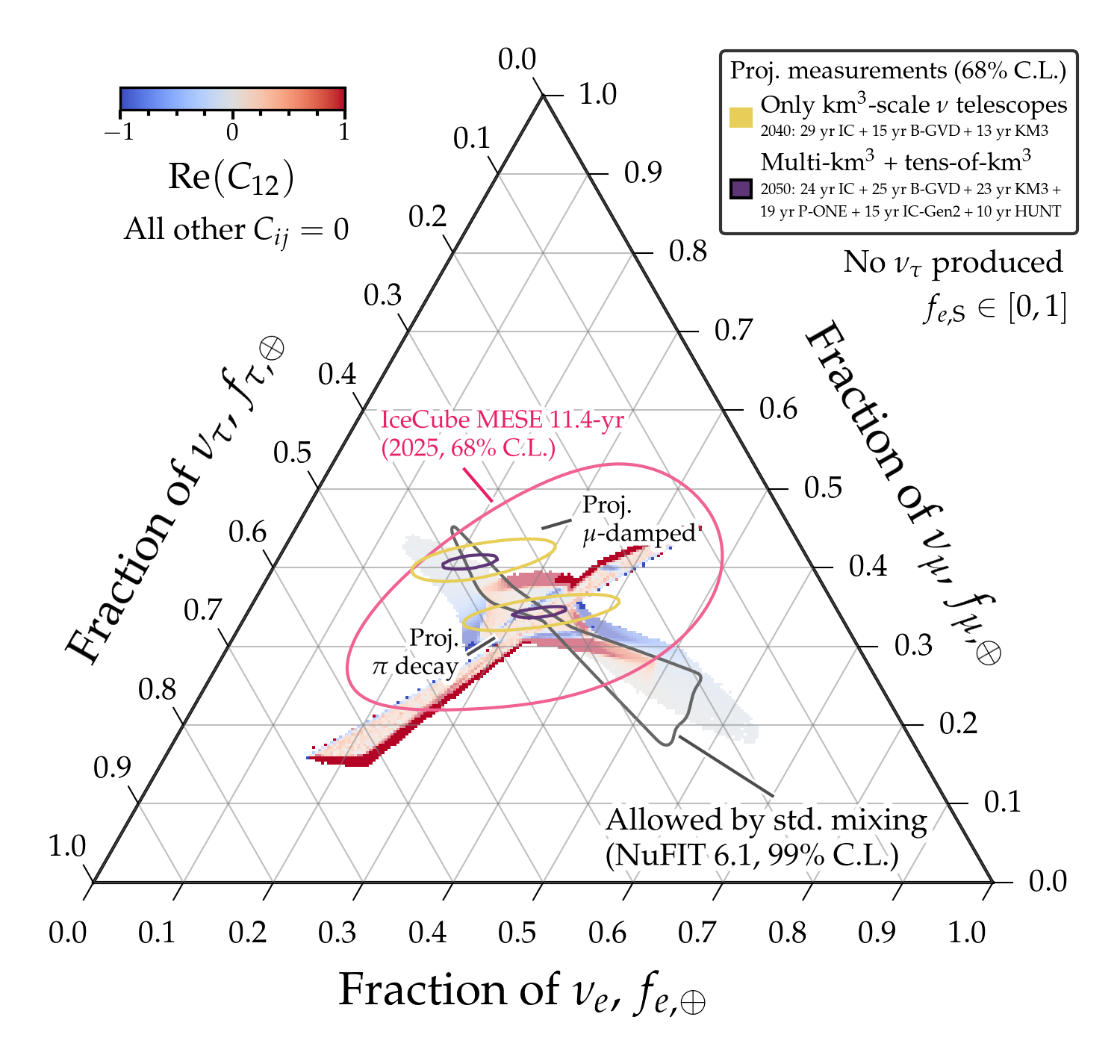}
 \caption{\textbf{Dimension-6-SMEFT allowed regions of flavor composition at Earth.} The regions are obtained by varying a single SMEFT coefficient at a time, while keeping all other coefficients fixed to zero. This figure should be compared to the all-coefficient \figu{ternary_theory_smeft} in the main text.  See Appendix~\ref{app:smeft_regions_single_parameter} for details.}
 \label{fig:ternary_theory_smeft_single_parameter}
\end{figure*}

\begin{figure*}[t!]
 \centering
 \includegraphics[width=0.495\textwidth]{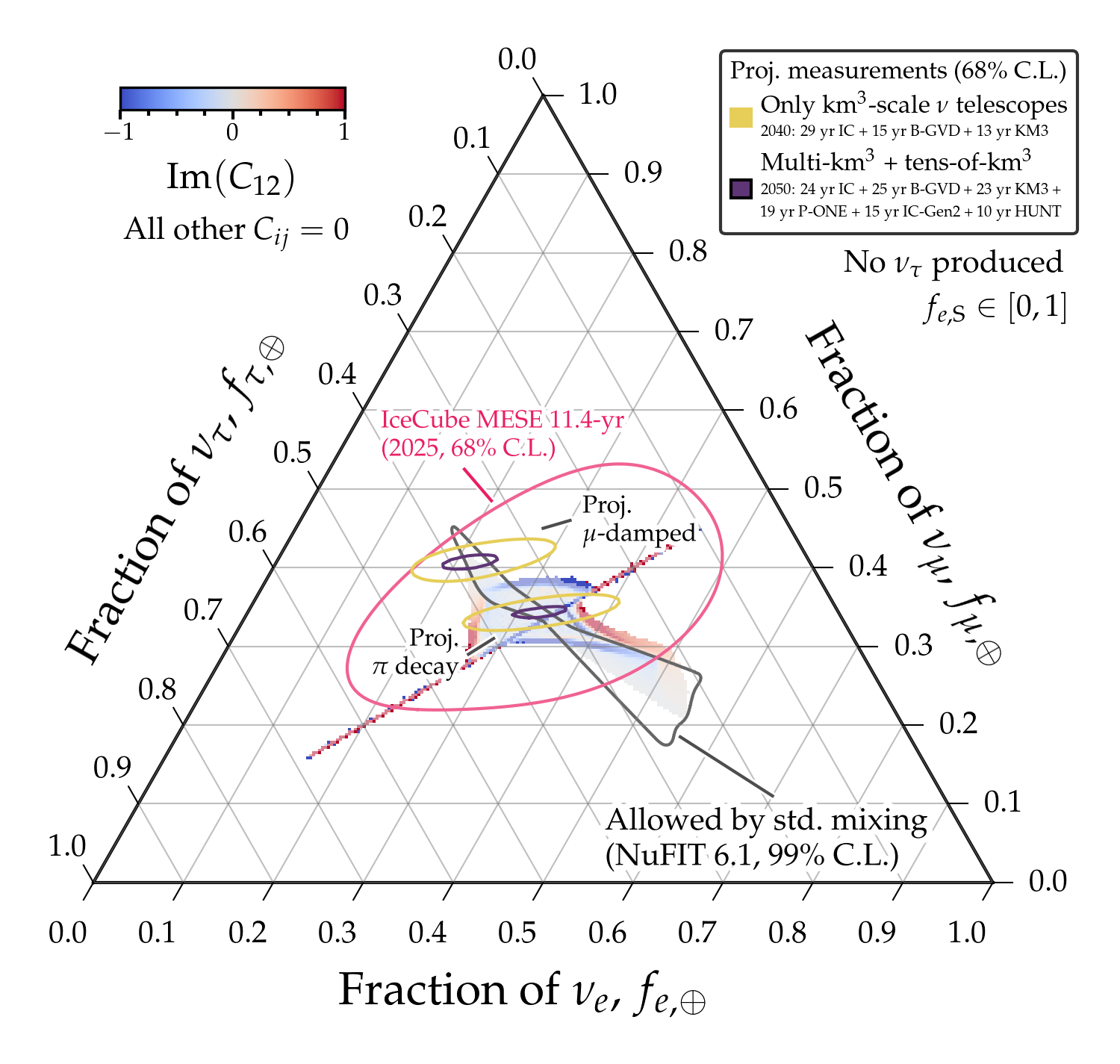}\hfill
 \includegraphics[width=0.495\textwidth]{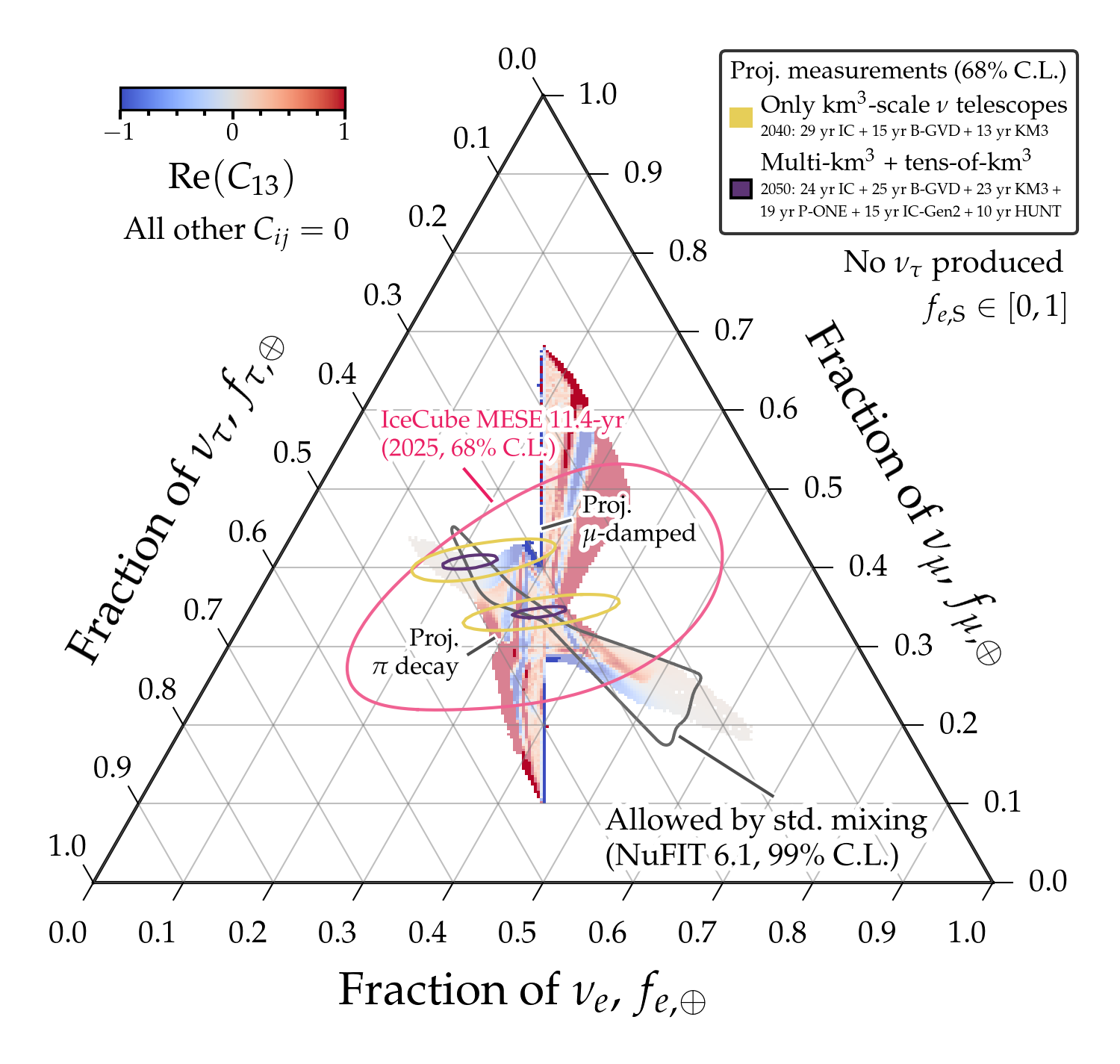}\\[-3ex]
 \includegraphics[width=0.495\textwidth]{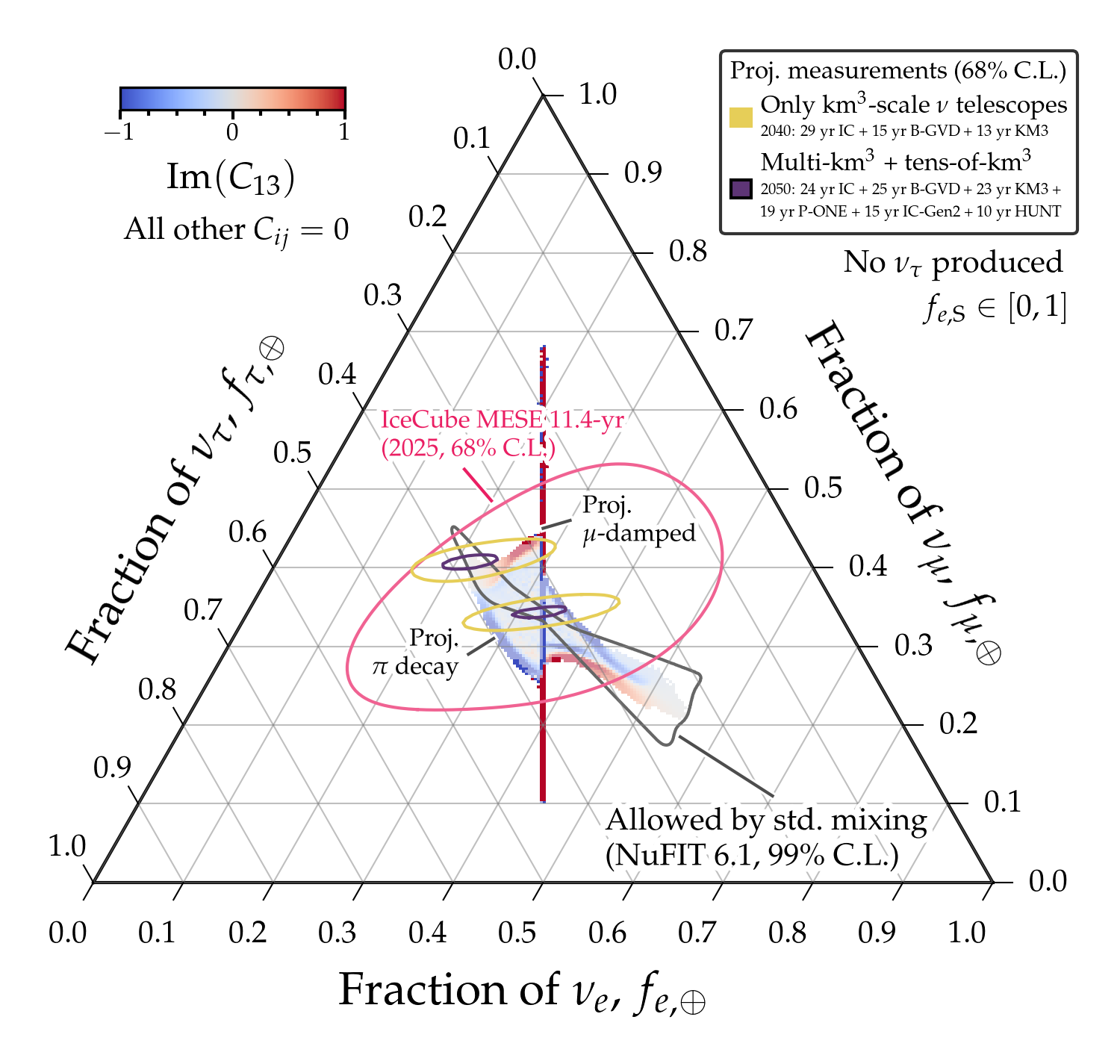}\hfill
 \includegraphics[width=0.495\textwidth]{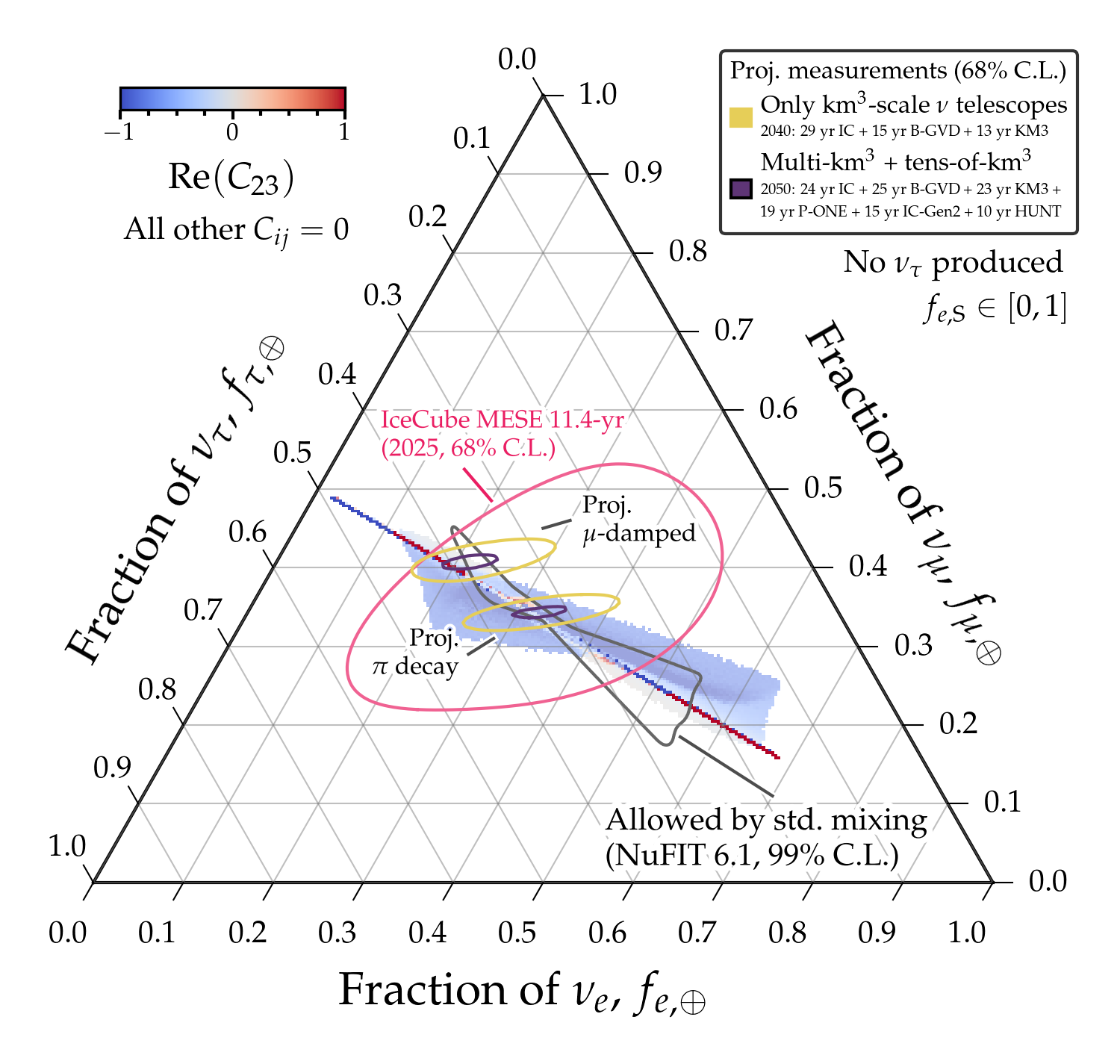}\\[-3ex]
 \includegraphics[width=0.495\textwidth]{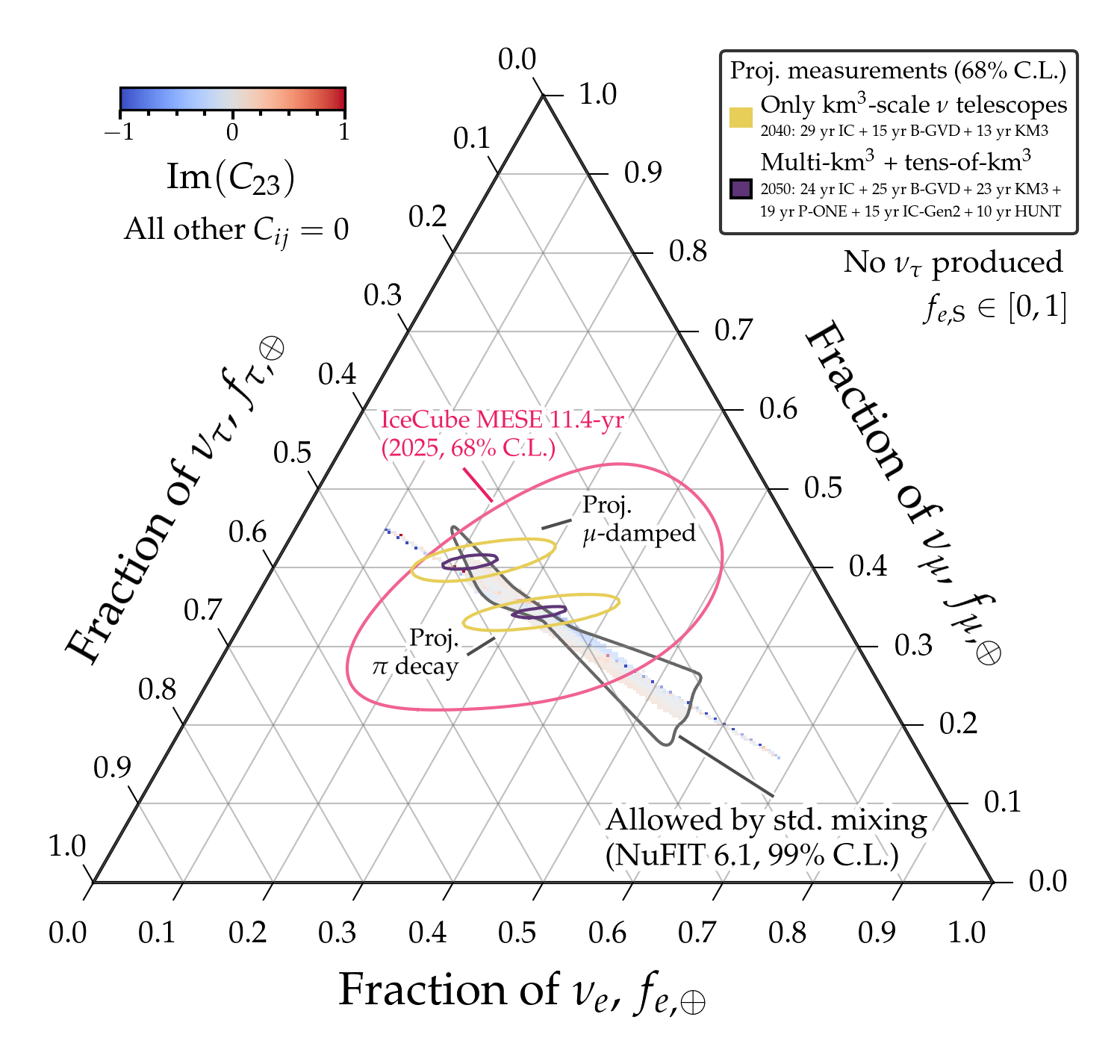}
 \addtocounter{figure}{-1}   
 \renewcommand{\thefigure}{H\arabic{figure} (cont.)}
 \vspace*{-0.5cm}
 \caption{\textbf{Dimension-6-SMEFT allowed regions of flavor composition at Earth.} Flavor regions for the remaining SMEFT coefficients.}
 \label{fig:ternary_theory_smeft_single_parameter_cont}
\end{figure*}


\end{document}